\documentclass[hidelinks, 11 pt, a4paper]{article}
\usepackage{amsfonts,amsmath,amssymb,amsthm}
\usepackage{subfigure}
\usepackage[pdftex]{graphicx}
\usepackage{natbib}
\usepackage[hypertexnames=false]{hyperref}
\usepackage{setspace}
\usepackage{pdfsync}
\usepackage{lscape}
\usepackage{verbatim}
\usepackage[hmargin=1.25in,vmargin=1.2in]{geometry}
\usepackage{bibunits}
\usepackage{footmisc}
\usepackage{rotating}
\usepackage{titling}
\usepackage{mathrsfs,dsfont}
\usepackage[dvipsnames,svgnames, x11names]{xcolor}
\usepackage[capposition=top]{floatrow}
\usepackage{booktabs}
\usepackage{multirow}
\usepackage[normalem]{ulem}

\hypersetup{colorlinks, linkcolor = {teal}, citecolor = {blue}, urlcolor ={blue!80!black}}

\makeatletter
\renewcommand\paragraph{\@startsection{paragraph}{4}{\z@}%
            {-2.5ex\@plus -1ex \@minus -.25ex}%
            {1.25ex \@plus .25ex}%
            {\normalfont\normalsize\bfseries}}
\makeatother
\defaultbibliographystyle{ims}
\defaultbibliography{references}

\usepackage{pgfplots}
\usepackage{tikz}
\usepackage{tikz-3dplot}
\usetikzlibrary{patterns,shapes, calc, positioning}
\usetikzlibrary{decorations.text}

\DeclareGraphicsExtensions{.jpg}

\def\qed{\rule{2mm}{2mm}}
\def\indep{\perp \!\!\! \perp}

\newtheorem{theorem}{Theorem}[section]
\newtheorem{lemma}{Lemma}[section]
\newtheorem{definition}{Definition}[section]
\newtheorem{example}{Example}[section]
\newtheorem{corollary}{Corollary}[section]
\newtheorem{remark}{Remark}[section]
\newtheorem{assumption}{Assumption}[section]

\let\oldmarginpar\marginpar
\renewcommand{\marginpar}[2][rectangle,draw,fill=black, text=white,text width= 2cm,rounded corners]{
    \oldmarginpar{
    \tiny \tikz \node at (0,0) [#1]{#2};}
    }

\newenvironment{packed_enum}{\begin{enumerate} \setlength{\itemsep}{1pt}\setlength{\parskip}{0pt}\setlength{\parsep}{0pt}}{\end{enumerate}}

\newcommand{\tr}{Q_0}
\newcommand{\dom}{\bar Q}
\newcommand{\restr}{\mathcal Q}
\newcommand{\Id}{\Upsilon}
\newcommand{\Eff}{\mathcal I}
\newcommand{\Rf}{\mathcal R_{T}}
\newcommand{\Rs}{\mathcal R_{t}}

\pgfplotsset{compat=1.18} 

\begin{document}

\title{Identification and Estimation in a Class of Potential Outcomes Models }

{\author{Manu Navjeevan \\ 
UCLA \\ mnavjeevan@g.ucla.edu
\and
Rodrigo Pinto \\ 
UCLA \\ rodrig@econ.ucla.edu
\and
Andres Santos\\ 
UCLA\\ andres@econ.ucla.edu}}

\date{October, 2023}

\maketitle

\begin{abstract}
This paper develops a class of potential outcomes models characterized by three main features: 
(i) Unobserved heterogeneity can be represented by a vector of potential outcomes and a ``type" describing the manner in which an instrument determines the choice of treatment;
(ii) The availability of an instrumental variable that is conditionally independent of unobserved heterogeneity; and
(iii) The imposition of convex restrictions on the distribution of unobserved heterogeneity.
The proposed class of models encompasses multiple classical and novel research designs, yet possesses a common structure that permits a unifying analysis of identification and estimation.
In particular, we establish that these models share a common necessary and sufficient condition for identifying certain causal parameters. 
Our identification results are constructive in that they yield estimating moment conditions for the parameters of interest.
Focusing on a leading special case of our framework, we further show how these estimating moment conditions may be modified to be doubly robust.
The corresponding double robust estimators are shown to be asymptotically normally distributed, bootstrap based inference is shown to be asymptotically valid, and the semi-parametric efficiency bound is derived for those parameters that are root-$n$ estimable.
We illustrate the usefulness of our results for developing, identifying, and estimating causal models through an empirical evaluation of the role of mental health as a mediating variable in the Moving To Opportunity experiment.
\end{abstract}

\begin{center}
\end{center}

\thispagestyle{empty}

\newpage
\pagenumbering{arabic}

\section{Introduction}

Potential outcomes models have become the leading framework for identifying and estimating causal effects in applications with heterogeneous treatment responses.
Originally developed for randomized experiments \citep{neyman1923} and observational studies \citep{rubin1974estimating}, these models have also proven transformative in shaping our understanding of instrumental variable approaches for addressing selection.
In this regard, fundamental contributions were made by \cite{imbens1994identification} and \cite{heckman2005structural} who highlighted the importance to identification of restricting the manner in which an instrument can impact treatment decisions.
A subsequent literature has built on their foundational work by developing identifying restrictions for a wide range of empirically relevant settings, including applications involving ordered, unordered, and multiple instruments, as well as the presence of mediating variables.

In this paper, we propose and develop a class of potential outcomes models that unifies and expands upon these identification strategies.
The main assumptions imposed by our framework are: 
(i) Unobserved heterogeneity can be represented by a vector of potential outcomes and a ``type" describing the manner in which the instruments determines the treatment decision;
(ii) The instrumental variables are conditionally independent of unobserved heterogeneity; and 
(iii) The distribution of unobserved heterogeneity belongs to a convex set.
The third requirement can include, for instance, support restrictions on the unobserved heterogeneity.
These encompass, among others, the monotonicity condition of \cite{angrist1995two}, the partial monotonicity requirement of \cite{mogstad2021causal}, and the revealed preference based restrictions of \cite{kline2016evaluating} and \cite{pinto2021beyond}.
Additional examples of convex identifying restrictions that go beyond support conditions include the sequential exogeneity requirement of \cite{imai2010identification} and the comparative compliers requirement of \cite{mountjoy2022community}.

Within the proposed class of models, we study the identification and estimation of parameters that may be expressed as the expectation (or limit of expectations) of identified functions of the unobserved heterogeneity and covariates.
These parameters include, for example, local average treatment effects, marginal treatment effects, and conditional expectations of covariates given types as in  \cite{abadie2003semiparametric}.
Our main identification result is the characterization of \emph{necessary and sufficient} conditions for such parameters to be identified.
In particular, we establish that identification is equivalent to the function whose expectation we aim to identify belonging to the closure of the range of an identified linear map $\Id$.
Intuitively, identification is tantamount to the existence of a sequence of functions $\{\kappa_j\}$ of  observable variables such that $\Id(\kappa_j)$ suitably approximates the function of unobserved heterogeneity whose expectation we wish to identify.
Critically, the map $\Id$ is the same across all the models in our framework, but the sense in which $\Id(\kappa_j)$ must converge depends on the restrictions being imposed -- i.e.\ stronger restrictions yield weaker topologies and hence the identification of additional parameters.
Our characterization of identification is additionally constructive in that it implies that if the parameter of interest is identified, then it must equal the limit of the expectations of the corresponding approximating functions $\{\kappa_j\}$ of observable variables.

The constructive nature of our identification results further suggests an estimation strategy: Simply estimate the corresponding approximating sequence $\{\kappa_j\}$ and compute its sample average.
Establishing asymptotic normality when the nuisance parameters $\{\kappa_j\}$ are estimated  via machine learning methods, however, often requires characterizing orthogonal scores that themselves depend on the specific restrictions being imposed.
In our estimation analysis, we therefore focus on a leading special case of our framework in which the identifying restrictions being imposed do not depend on the distribution of observable variables.\footnote{As we illustrate in our empirical analysis, estimation under other restrictions is also possible.}
For this class of applications, we derive a double robust moment condition and follow ideas in \cite{smucler2019unifying}, \cite{chernozhukov2022locally}, and \cite{chernozhukov2022automatic} by employing $\ell_1$-regularization to estimate nuisance parameters.
We show that the resulting estimators are asymptotically normally distributed and that bootstrap based inference is asymptotically valid even if the estimator converges at a slower than root-$n$ rate.
We additionally derive the semiparametric efficiency bound for these parameters and characterize the conditions under which it is finite.
As we illustrate in the context of \cite{mogstad2021causal}, the latter result has important implications for root-$n$ estimability in applications with continuous instruments.

Our results are not only useful in the context of existing models, but can also be instrumental in developing, identifying, and estimating novel causal models.
We highlight the utility of our analysis in this regard with an empirical analysis of the Moving to Opportunity (MTO) experiment.
Specifically, we evaluate a conjecture by \cite{ludwig2008can} who suggested that improved mental health may play an important role in the causal mechanism through which moving from high to low-poverty neighborhoods impacts economic outcomes.
Guided by our necessary and sufficient conditions for identification, we devise a model that enables us to identify and estimate the mediating effects of improved mental health for different subpopulations.
Overall, we find evidence in support of the causal channel in which mental health mediates the effect of neighborhood relocation on labor market outcomes.

This paper contributes to a vast literature on potential outcomes models.
Our analysis appears to be the first to establish the unifying role that a common linear map $\Id$ plays in determining identification across a variety of models and assumptions.
Through this common structure, our analysis delivers \emph{necessary and sufficient} conditions for identification -- a result that complements the literature, which has largely focused on \emph{sufficient} conditions for identification.\footnote{Corollary C-1 in \cite{heckman2018unordered}, for example, provides necessary and sufficient conditions for linear restrictions implied by the model to deliver identification. Our results in contrast provide necessary and sufficient conditions for identification that reflect all the restrictions of the model.}
In some applications, our results yield conditions under which existing sufficient conditions for identification are in fact necessary -- e.g., those in \cite{abadie2003semiparametric} and \cite{heckman2018unordered}.
In other applications, our results yield novel characterizations of what parameters within our framework are identified -- e.g., as in \cite{mogstad2021causal}.
Our identification analysis is further related to work providing analytical \citep{manski:1990, manski:2003, heckman2001instrumental} or computational \citep{balke1997bounds,mogstad2018using} bounds for partially identified parameters.
By establishing that point identification is determined by a single linear equation, our analysis effectively yields a simple way to derive conditions under which these bounds collapse to a point in the models that fall within our framework.

Our estimation results rely on a double robust moment equation that coincides with that of \cite{tan2006regression} and \cite{singh2022double} for the model of \cite{imbens1994identification}.
More generally, however, our results yield the first double robust moment equation for a variety of models and parameters -- e.g., under a monotonicity assumption we obtain doubly robust estimators for \cite{heckman2005structural} that do not rely on the propensity score.
The semiparametric efficiency bound derived in this paper similarly significantly extends the existing efficiency literature to a variety of models and parameters.
Our analysis corrects some approaches in the literature by relying on results by \cite{le1988preservation} that enable us to construct the tangent set generated by parametric submodels of the unobserved heterogeneity; see Remark \ref{rm:Tcorrect}.
This construction further enables us to connect to results in \cite{van1991differentiable} and characterize when the efficiency bound is finite.
The latter result is, to our knowledge, novel in all the models we consider.

The remainder of the paper is organized as follows.
In Section 2 we formally introduce the class of models we study and discuss multiple illustrative examples.
Section 3 highlights the empirical implications of our results through an analysis of the MTO experiment.
Finally, Sections 4 and 5 contain all theoretical results while Section 6 briefly concludes.
All mathematical derivations are included in the Appendix.

\section{The Model} \label{sec:model}

We consider applications in which we observe a scalar outcome $Y\in \mathbf Y \subseteq \mathbf R$, a discrete treatment $T\in \mathbf T\equiv \{t_1,\ldots, t_d\}$, an instrument $Z \in \mathbf Z$, and covariates $X\in \mathbf X$.
We model unobserved heterogeneity through a vector of potential outcomes $Y^\star \equiv (Y^\star(t_1),\ldots, Y^\star(t_d))$ and a type $T^\star : \mathbf Z \to \{t_1,\ldots, t_d\}$ that describes the  manner in which the instrument determines a unit's treatment decision.
The observed treatment and outcome are given by the treatment choice induced by the instrument and the potential outcome corresponding to the chosen treatment -- i.e.\ $T=T^\star(Z)$ and $Y = Y^\star(T)$.

In order to introduce the assumptions that characterize our model, we first define
\begin{equation*}
L^p(Q) \equiv \{f: \|f\|_{Q,p} < \infty\} \hspace{0.5 in} \|f\|_{Q,p}^p \equiv \int |f|^p dQ;
\end{equation*}
i.e.\ $L^p(Q)$ denotes the set of functions that have a finite $p^{th}$ moment under $Q$.
Also recall that a distribution $Q$ is absolutely continuous with respect to (w.r.t.) a distribution $Q^\prime$, denoted $Q\ll Q^\prime$, if $Q$ assigns zero probability to any event to which $Q^\prime$ assigns zero probability.
Importantly, whenever $Q$ is absolutely continuous w.r.t.\ $Q^\prime$ it admits a density w.r.t.\ $Q^\prime$ that we denote by $dQ/dQ^\prime$.
Finally, we let $\tr$ denote the true unknown distribution of $(Y^\star,T^\star,Z,X)$ and $P$ the identified distribution of $(Y,T,Z,X)$.

Given the introduced notation, we impose the following two assumptions:

\begin{assumption}\label{ass:setup}
(i) $(Y^\star, T^\star,Z,X)\sim \tr$ with $Y^\star \equiv (Y^\star(t_1),\ldots, Y^\star(t_d))\in \mathbf Y^\star \subseteq \mathbf R^d$, $Z\in \mathbf Z$, $X\in \mathbf X$, and $T^\star\in \mathbf T^\star$ with $\mathbf T^*$ a set of functions from $\mathbf Z$ to $\mathbf T\equiv \{t_1,\ldots, t_d\}$; (ii) We observe $T = T^\star(Z)$, $Y = Y^\star(T)$, $Z$, and $X$ with $(Y,T,Z,X)\sim P$.
\end{assumption}

\begin{assumption}\label{ass:model}
(i) $(Y^\star, T^\star)\indep Z|X$ under $\tr$;
(ii) $\tr \ll \mu$ for some identified separable probability measure $\mu$;
(iii) $d\tr/d\mu$ belongs to a set $\restr \subseteq L^1(\mu)$ for $\restr$ a closed convex subset of Banach Space $(\mathbf Q,\|\cdot\|_{\mathbf Q})$ with $\|\cdot\|_{\mathbf Q}$ (weakly) stronger than $\|\cdot\|_{\mu,1}$.
\end{assumption}

Assumption \ref{ass:setup} formalizes the data generating process, but has by itself no identifying power. 
The main conditions powering our identification results are imposed in Assumptions \ref{ass:model}.
In particular, Assumption \ref{ass:model}(i) requires that $Z$ be exogenous in the sense that it be statistically independent of the unobserved heterogeneity $(Y^\star,T^\star)$ conditional on $X$.
Assumption \ref{ass:model}(ii) in turn encodes restrictions on the support of the unobserved heterogeneity.
For instance, since $\tr$ must assign zero probability to any event to which $\mu$ assigns zero probability, we may employ $\mu$ to rule out certain realizations of $T^*$ -- e.g., to rule out ``defiers" in \cite{imbens1994identification}.
Finally, Assumption \ref{ass:model}(iii) enables us to accommodate additional convex restrictions on the density of $\tr$.
These restriction may include both regularity conditions that ensure the parameter of interest is well defined (see Example \ref{ex:rc} below) as well as more substantive identifying assumptions (see Section \ref{sec:mto} below).
We note that while not stated explicitly, we may set $\restr$ and $\mathbf Q$ to be identified instead of known.

The unknown true distribution $\tr$ of $(Y^\star,T^\star,Z,X)$ induces, through Assumption \ref{ass:setup}, the identified distribution $P$ of the observable variables $(Y,T,Z,X)$.
Absent restrictive assumptions, $\tr$ is not identified in our model because there are alternative distributions $Q$ for $(Y^\star,T^\star,Z,X)$ that induce the distribution $P$.
In what follows, we refer to any such distribution $Q$ as being observationally equivalent to $\tr$.
While it may not be possible to identify $\tr$, it is still possible to restrict it to the identified set
\begin{equation*}\label{sec2:eq2}
\Theta_0 \equiv \{Q : Q \text{ is obs.\ equiv.\ to } \tr, ~ (Y^\star,T^\star)\indep Z|X \text{ under } Q, ~ Q\ll \mu, ~ \frac{dQ}{d\mu} \in \restr\};
\end{equation*}
i.e., the identified set $\Theta_0$ is the set of distributions for $(Y^\star,T^\star,Z,X)$ that induce $P$ and additionally satisfy the requirements imposed on $\tr$ in Assumption \ref{ass:model}.

Our primary goal is to study the identification and estimation of features of the true distribution $\tr$.
Concretely, we study the identification and estimation of functionals of $\tr$ that, for some identified sequence of functions $\{\ell_j\}$, have the structure
\begin{equation}\label{sec2:eq3}
\lambda_{Q} \equiv \lim_{j\to \infty} E_Q[\ell_j(Y^\star,T^\star,X)].
\end{equation}
Here, the $Q$ subscript is meant to emphasize that the expectation is taken with respect to a distribution $Q$ that may not equal $\tr$. 
For instance, a leading example is to let $\ell_j$ equal a known function $f$ for all $j$, in which case identification of $\lambda_{\tr}$ is tantamount to the identification of the expectation of $f(Y^\star,T^\star,X)$
under the true distribution $\tr$.

\subsection{Examples}\label{subsec:ex}

In order to fix ideas, we next introduce examples that highlight the flexibility of our setup. 
We will return to some of them throughout the paper to illustrate our results.

Our first examples are based on the most studied models in the literature. 

\begin{example}\label{ex:rc} \rm
Following \cite{rosenbaum1983central}, suppose we observe an outcome $Y\in \mathbf R$, a binary treatment $T\in \{0,1\}$, covariates $X\in \mathbf X$, and that potential outcomes $Y^\star \equiv (Y^\star(0),Y^\star(1))$ are independent of $T$ conditional on $X$.
To map this setting into our framework we let $Z = T$, select $\mu$ to satisfy the restriction
\begin{equation*}
\mu(T^\star(Z)= Z) =1,    
\end{equation*}
and note that Assumption \ref{ass:model}(i) is then equivalent to the unconfoundedness assumption $(Y^\star(0),Y^\star(1))\indep T |X$.
The classical parameter of interest in the literature is the average treatment effect (ATE), which corresponds to setting $\ell(Y^\star,T^*,X) = Y^\star(1)-Y^\star(0)$ in \eqref{sec2:eq3}. 
Ensuring the ATE is well defined requires us to impose that $Y^\star(0)$ and $Y^\star(1)$ have a first moment, which can be accomplished through Assumption \ref{ass:model}(iii). \qed
\end{example}

\begin{example}\label{ex:ia} \rm
Consider a special case of \cite{imbens1994identification} in which we observe an outcome $Y$, a binary treatment $T\in \{0,1\}$, and a binary instrument $Z\in \{0,1\}$.
In this context, $T^\star$ is a random function mapping $\mathbf Z \equiv \{0,1\}$ to $\mathbf T \equiv \{0,1\}$.
Following \cite{imbens1994identification} we may employ Assumption \ref{ass:model}(ii) to impose that the instrument does not induce individuals out of treatment by setting $\mu$ to satisfy
\begin{equation*}
\mu(T^\star(1) \geq T^\star(0)) = 1;
\end{equation*}
i.e.\ $\mu$ assigns zero probability to ``defiers."
Functionals with the structure in \eqref{sec2:eq3} include the local average treatment effect (LATE) or, more generally, functionals of the marginal distributions of $Y^\star$ conditional on ``compliers" \citep{imbens1997estimating, abadie2003semiparametric}.
We also note that Assumptions \ref{ass:setup} and \ref{ass:model} can accommodate extensions to ordered discrete treatments \citep{angrist1995two} or alternatives restrictions on $T^\star$ such as the ``extensive margin compliers only" requirement in \cite{rose:shemtov:emco}. \qed
\end{example}

\begin{example}\label{ex:mte} \rm
\cite{heckman1999local} study a generalized Roy model in which a unit selects whether to adopt a binary treatment $T\in\{0,1\}$ according to
\begin{equation}\label{ex:mte1}
T = 1\{f(X,Z) \geq \xi \}    
\end{equation}
for $f$ an unknown continuous function, $\xi$ unobservable, and $(Y^\star(0),Y^\star(1),\xi)\indep Z|X$.
Assuming that $Z$ is a scalar and $f(X,\cdot)$ is monotonically increasing, we may map this model into our framework by letting $T^\star \equiv 1\{f(X,\cdot) \geq \xi\}$ and setting $\mu$ to satisfy\footnote{The upper semicontinuity of $T^*$ is a consequence of the continuity of $f(X,\cdot)$. The general case in which $Z$ is not scalar and $f(X,\cdot)$ is not monotonic corresponds to imposing that $T^\star(z) \geq T^\star(z^\prime)$ whenever $p(z,X) \geq p(z^\prime,X)$ $\mu$-almost surely for $p(Z,X) \equiv P(T=1|Z,X)$.}
\begin{equation*}
\mu(\lim_{z\downarrow z^\prime} T^*(z) = T^*(z^\prime) \text{ and } T^\star (z) \geq T^\star(z^\prime) \text{ for all } z\geq z^\prime) = 1.
\end{equation*}
Common parameters of interest in this literature, such as the policy relevant treatment effect (PRTE) of \cite{heckman2005structural}, can be expressed as
\begin{equation}\label{ex:mte3}
E[h(Y^\star,\xi,X)]
\end{equation}
where $h$ is an identified function. 
Because $T^\star$ is not necessarily an invertible function of $\xi$, the parameter in \eqref{ex:mte3} may not map 
into the functionals in \eqref{sec2:eq3} that we study.
However, under regularity conditions, it is possible to show that a necessary condition for \eqref{ex:mte3} to be identified is that $h$ must depend on $\xi$ only through $T^\star$.\footnote{Formally, we must have $h(Y^\star,\xi,X) = E[h(Y^\star,\xi,X)|Y^\star,T^\star,X]$ with probability one.}
Hence, our characterization of identification of \eqref{sec2:eq3} also characterizes identification of \eqref{ex:mte3} and our estimation results apply to \eqref{ex:mte3} whenever it is identified.
We also note that our framework can accommodate other structural equations models, such as those in \cite{lee2018identifying}. \qed    
\end{example}

Our next three examples illustrate the ability of our framework to accommodate multivalued treatments, vector valued instruments, and mediating variables.

\begin{example}\label{ex:disc} \rm
\cite{kline2016evaluating} employ the Head Start Impact Study to evaluate the cost-effectiveness of the Head Start Program.
In their analysis, $Z\in \{0,1\}$ denotes whether an individual was offered to attend a Head Start school and the treatment $T$ can take three values: Attend a Head Start School $(h)$, attend other schools $(c)$, or receive home care $(n)$.
Here, $T^\star$ maps $\{0,1\}$ to $\{h,c,n\}$ and we can therefore characterize $T^\star$ as a vector $T^\star = (T^\star(0),T^\star(1))$ taking values in $\{h,c,n\}\times \{h,c,n\}$.
The main identification assumption imposed by \cite{kline2016evaluating} is that receiving an offer to attend a Head Start school can only (weakly) induce individuals to attend a Head Start school.
Formally, they require that $T^\star = (T^\star(0),T^\star(1))$ belong to the set
\begin{equation*}
    \mathbf R^\star \equiv  \left\{ \left(\begin{array}{c} n\\ h \end{array}\right), \left(\begin{array}{c} c\\ h \end{array}\right), \left(\begin{array}{c} n\\ n \end{array}\right), \left(\begin{array}{c} c\\ c \end{array}\right), \left(\begin{array}{c} h\\ h \end{array}\right) \right \} 
\end{equation*}
with probability one, which can be mapped into our framework by setting $\mu$ to satisfy $\mu(T^\star \in \mathbf R^\star) = 1$.
More generally, in applications with a discrete valued instrument $Z$ we may always impose support restrictions on $T^\star$ by demanding that $\mu(T^\star \in \mathbf R^\star) = 1$ for some finite set of vectors $\mathbf R^\star \equiv \{t^\star_1,\ldots, t^\star_r\}$.
Through this observation, our framework can accommodate the unordered monotonicity condition of \cite{heckman2018unordered}, the analysis of the Moving to Opportunity experiment by  \cite{pinto2021beyond}, and the double threshold crossing model of survey non-response by  \cite{dutz2021selection}. \qed   
\end{example}

\begin{example}\label{ex:pm} \rm
\cite{mogstad2021causal} propose a partial monotonicity condition that can deliver a causal interpretation for the two stage least squares (TSLS) estimand in applications with vector valued instruments.
For instance, in an empirical re-examination of \cite{carneiro2011estimating}, the authors consider a setting in which $T\in \{0,1\}$ indicates whether an individual attended college, $Y$ represents log average hourly wage, and $Z = (C,W)$ where $C\in \{0,1\}$ indicates whether a college is present in the county of residence at age 14 and $W$ denotes average log earnings in the county of residence at age 17.
\cite{mogstad2021causal} further suppose that increasing $C$ induces individuals into treatment, while increasing $W$ induces individuals out of treatment.
Formally, their requirement may be mapped into our framework by selecting $\mu$ to satisfy
\begin{equation}\label{ex:pm1}
\mu(T^*(1,w) \geq T^*(0,w) \text{ and } T^*(c,w) \leq T^*(c,w^\prime) \text{ for all } c\in \{0,1\},~ w\geq w^\prime) = 1.
\end{equation}
Under an appropriate choice of sequence $\{\ell_j\}$, parameters such as \eqref{sec2:eq3} can then include, for example, analogues to the marginal treatment effect (MTE) of \cite{heckman2005structural}.
We also note that restrictions analogous to \eqref{ex:pm1} where employed in the empirical study of the returns to two-year colleges by \cite{mountjoy2022community}. \qed    
\end{example}

\begin{example}\label{ex:med} \rm
Mediation analysis aims to identify how a treatment can affect an outcome through intermediate variables called mediators. 
\cite{angrist2022marginal}, for instance, argue that engagement in the first year of college is an important mediator through which student grants impact graduation rates.
Letting $D\in\{0,1\}$ indicate whether a student is awarded a grant, $M\in \{e,ne\}$ denote whether she was engaged $(e)$ or not $(ne)$, and $Y\in \{0,1\}$ indicate whether she graduated within six years, we may map their study into our framework by letting $T = (D,M)$ and $Z = D$. 
Potential outcomes $Y^*$ are then indexed by $t = (d,m) \in \{0,1\}\times \{e,ne\}$, while $T^*$ is a function mapping $\mathbf Z \equiv \{0,1\}$ to $\mathbf T \equiv \{0,1\}\times \{e,ne\}$.
Assumptions \ref{ass:model}(i)(ii) can then be employed to impose identifying restrictions such as the sequential ignorability requirement of \cite{imai2010identification}, while parameters with the structure in \eqref{sec2:eq3} include the direct and indirect effects of \cite{pearl2001} and \cite{robins2003semantics}.
Finally, we note our framework can also accommodate IV mediation models, such as those of \cite{imai2013experimental} and \cite{frolich2017direct}. \qed
\end{example}

\section{Moving to Opportunity} \label{sec:mto}

As a preview of our theoretical results, we first illustrate their ability to develop, identify, and estimate a causal model in the context of an empirical analysis of the MTO experiment.
MTO was a housing experiment in which households living in high-poverty neighborhoods were offered vouchers that incentivized them to relocate to low-poverty neighborhoods.
The experiment targeted disadvantaged families residing in impoverished housing projects from June 1994 to July 1998 \citep{Orr_etal_2003}. 
Approximately 75\% of these households relied on welfare support, 92\% were female-headed, and only one-third of adult family members had attained a high school diploma. 

The MTO literature has found significant impacts on adult mental health, psychological well-being, and risky behavior  \citep{katz2001moving, Kling_etal_2005, kling2007experimental} 
as well as on economic outcomes for compliers moving from high to low-poverty neighborhoods \citep{Clampet_Massey_2008,pinto2021beyond}.\footnote{The evidence on economic impacts from moving from high to medium-poverty neighborhoods is less conclusive, with treatment on the treated estimates often being insignificant \citep{ludwig2013long}.} 
We revisit MTO to investigate a conjecture by \citet{ludwig2008can}, who hypothesize that relocation to low-poverty neighborhoods can improve mental health and empower previously marginalized women to obtain steady employment.
Specifically, we employ our theoretical results to obtain the first estimates of the role improved mental health plays as a mediator in the causal channel through which neighborhood relocation affects economic outcomes.

To map this application into our framework, we let $Z \in \{0,1\}$ indicate whether a voucher is offered, $Y$ denote an economic outcome of interest, $D\in \{0,1\}$ indicate whether the household relocated to a low-poverty neighborhood, and $M\in \{0,1\}$ indicate whether the head of household reported having positive mental health.\footnote{Specifically, $M=1$ if the head out household reported feeling calm during the past thirty days.}
We further set $T = (D,M)$ and let potential outcomes $Y^*(t)$ depend on $t=(d,m)$ to reflect that mental health and relocating neighborhoods can both affect economic outcomes.
For our covariates $X$, we follow the literature in employing experimental site indicators and variables pertaining to household and neighborhood characteristics.
We report additional implementation details for this empirical study in Appendix A.4.

\subsection{Learning About Types}\label{subsec:mtot}

We begin by studying functionals of the distribution of types $T^*$, which here describe the heterogeneous manner in which $Z$ affects mental health and the relocation decision.
In particular, for identified functions $\ell$, we first estimate expectations with the structure
\begin{equation}\label{eq:tmto1}
E_{\tr}[\ell(T^*,X)].    
\end{equation}
A leading special case of such expectations is the probability that $T^*$ equals a point $t^*$ in its support, which corresponds to setting $\ell(T^*,X) = 1\{T^* = t^*\}$.

By selecting $\mu$ in Assumption \ref{ass:model}(ii) to restrict the support of $T^*$, our model enables us to restrict how a voucher offer affects relocation decisions and mental health.
Under such support restrictions, our identification results imply that \eqref{eq:tmto1} is identified if and only if there exists a function $\kappa$ of $(T,Z,X)$ satisfying the equation
\begin{equation}\label{eq:tmto2}
\sum_{t\in \mathbf T} \sum_{z\in \mathbf Z} 1\{t^*(z) = t\}\kappa(t,z,X)P(Z=z|X) = \ell(t^*,X)
\end{equation}
for every $t^*$ in the support of $T^*$ and all $X$; see Corollary \ref{cor:discint}.
Moreover, any $\kappa$ satisfying equation \eqref{eq:tmto2} can be employed to identify the expectation of $\ell(T^*,X)$ through the equality
\begin{equation}\label{eq:tmto3}
E_{\tr}[\ell(T^*,X)] = E_P[\kappa(T,Z,X)].    
\end{equation}
Guided by this result, we impose three requirements that deliver identification of the distribution of $(T^*,X)$: 
(i) A voucher offer (weakly) incentivizes households to relocate;
(ii) Moving to a low-poverty neighborhood (weakly) improves mental health; and
(iii) A voucher offer affects mental health only through the relocation decision.
Formally, we impose these restrictions by letting $T^*(z) \equiv (D^*(z),M^*(z))$ with $D^*$ and $M^*$ describing how relocation and mental health  respond to a voucher offer, and setting
\begin{equation}\label{eq:tmto4}
    \mu(D^*(1) \geq D^*(0) \text{ and } M^* = F^*\circ D^* \text{ with } F^*(1) \geq F^*(0)) = 1.
\end{equation}

\begin{table}[t!] 
\begin{center}
\caption{Type Probabilities and Conditional Expectation of Baseline Variables }						\label{tab:type}
\resizebox{\textwidth}{!}{
\begin{tabular}{c cc ccc cc}
\hline \hline
& \multicolumn{7}{c}{Type Definitions and Probabilities} \\ \cmidrule{2-8}
                     & NN       & NA    &  CN    & CC    & CA    &  AN    & AA  \\ \midrule
\multicolumn{1}{r}{($D^*(0),M^*(0)$)}    & (0,0)    & (0,1) &  (0,0) & (0,0) & (0,1) &  (1,0) & (1,1) \\
\multicolumn{1}{r}{($D^*(1),M^*(1)$)}    & (0,0)    & (0,1) &  (1,0) & (1,1) & (1,1) &  (1,0) & (1,1) \\ \\
\multicolumn{1}{r}{\small{Probability Point Estimate	}}&	0.258	&	0.253	&	0.194	&	0.065	&	0.203	&	0.014	&	0.013	\\	
\multicolumn{1}{r}{	\small{s.e.}	}&\small{(0.014)}&\small{(0.014)}&\small{(0.013)}&\small{(0.023)}&\small{(0.022)}&\small{(0.004)}&\small{(0.003)}\\	\\
& \multicolumn{7}{c}{Expected Value of Covariate Conditional on Types} \\ \cmidrule{2-8}
    & NN       & NA   &  CN    & CC    & CA    &  AN    & AA  \\ \midrule
\multicolumn{1}{r}{\small{Household member }}&	0.183	&	0.145	&	0.156	&	0.388	&	0.076	&	0.091	&	0.077	\\	
\multicolumn{1}{r}{\small{has a disability	}}&\small{(0.020)}&\small{(0.019)}&\small{(0.022)}&\small{(0.156)}&\small{(0.039)}&\small{(0.066)}&\small{(0.054)}\\	\\
\multicolumn{1}{r}{\small{	No teens in the}}&	0.535	&	0.566	&	0.656	&	0.736	&	0.653	&	0.602	&	0.902	\\	
\multicolumn{1}{r}{\small{household	}}&\small{(0.028)}&\small{(0.029)}&\small{(0.030)}&\small{(0.176)}&\small{(0.050)}&\small{(0.121)}&\small{(0.071)}\\	\\
\multicolumn{1}{r}{\small{	Applied for a Section}}&	0.394	&	0.408	&	0.491	&	0.277	&	0.491	&	0.162	&	0.399	\\	
\multicolumn{1}{r}{\small{Eight Voucher	}}&\small{(0.027)}&\small{(0.029)}&\small{(0.033)}&\small{(0.182)}&\small{(0.052)}&\small{(0.088)}&\small{(0.125)}\\	\\
\multicolumn{1}{r}{\small{	Moved 3+ times }}&	0.083	&	0.080	&	0.120	&	0.130	&	0.079	&	0.002	&	0.049	\\	
\multicolumn{1}{r}{\small{in past 5 years	}}&\small{(0.014)}&\small{(0.016)}&\small{(0.021)}&\small{(0.097)}&\small{(0.027)}&\small{(0.013)}&\small{(0.044)}\\	\\
\multicolumn{1}{r}{\small{	No friends in the}}&	0.355	&	0.400	&	0.410	&	0.699	&	0.360	&	0.341	&	0.659	\\	
\multicolumn{1}{r}{\small{neighborhood	}}&\small{(0.026)}&\small{(0.029)}&\small{(0.033)}&\small{(0.204)}&\small{(0.051)}&\small{(0.115)}&\small{(0.118)}\\	\\
\multicolumn{1}{r}{\small{	Neighborhood is 	}}&	0.462	&	0.429	&	0.534	&	0.592	&	0.514	&	0.471	&	0.496	\\	
\multicolumn{1}{r}{\small{unsafe at night	}}&\small{(0.027)}&\small{(0.029)}&\small{(0.033)}&\small{(0.181)}&\small{(0.052)}&\small{(0.119)}&\small{(0.132)}\\	\\
\multicolumn{1}{r}{\small{	Gangs/Drugs are primary}}&	0.768	&	0.741	&	0.841	&	0.500	&	0.848	&	0.806	&	0.818	\\	
\multicolumn{1}{r}{\small{reason to move	}}&\small{(0.022)}&\small{(0.024)}&\small{(0.023)}&\small{(0.178)}&\small{(0.042)}&\small{(0.091)}&\small{(0.105)}\\	
\hline 
\hline
\end{tabular} }
\end{center}
\begin{flushleft}
\scriptsize{
First panel reports the seven support points of $T^*$ and their respective estimated probabilities.
Second panel reports estimates for the expectation of baseline variables conditioned on types.
All estimates account for the person-level weight for the adult survey of the interim analyses. 
Standard errors are displayed in parentheses.}		
\end{flushleft}\vspace{-0.2 in}
\end{table}

The imposed restrictions limit the support of $T^*$ to seven possible types. 
These types, displayed in the first panel of Table \ref{tab:type}, are characterized by the possible realizations of $D^*$ and $M^*$ -- i.e.\ whether they are never takers, compliers, or always takers with regards to relocation and mental health status.
For instances, types CN, CA, and CC always relocate when offered a voucher.
Relocation, however, does not change the mental health status of types CN and CA, but improves the mental health status of type CC.

It is straightforward to verify that, for any function $\ell$, equation \eqref{eq:tmto2} admits a solution and hence that the expectation of $\ell(T^*,X)$ is identified.
In particular, it follows that the probability of each type is identified and that we may apply our asymptotically normal estimator based on \eqref{eq:tmto3} to estimate it; see Theorem \ref{th:typenorm}.
The first panel of Table \ref{tab:type} reports our estimates of the type probabilities. 
While all types in our model occur with a strictly positive probability, the vast majority of households either do not relocate with a voucher offer (types NN and NA) or only relocate when given a voucher (types CN, CC, CA). In contrast, only $2.6\%$ of households would relocate to low-poverty neighborhoods without a voucher offer (types AN and AA). 
We also note that only $4.7\%$ of households experience an improvement in mental health upon relocating (type CC).

Since the type probabilities are identified, for any type $t^*$ we may set $\ell(T^*,X)= X1\{T^* = t^*\}/\tr(T^*=t^*)$, in which case \eqref{eq:tmto1} equals the expected value of baseline variables conditional on type.
The second panel of Table \ref{tab:type} presents estimates for such identified type characteristics.
Interestingly, the observed characteristics of double compliers (type CC) substantially differ from those of other types. 
The CC households are more likely to include a disabled family member yet are less likely to have teenagers. 
They seldom apply for Section 8, exhibit higher neighborhood mobility than other types, and are less likely to report having friends in the neighborhood. 
Although they are slightly more likely to feel unsafe in the neighborhood, they do not cite gang or drug-related issues as the primary reason for seeking to move to relocate.

\subsection{Learning About Outcomes} \label{subsec:mtoo}

We next turn to estimating treatment effects in our model.
A natural starting point is to examine the LATE of \cite{imbens1994identification}, which in our context equals:
\begin{equation*}
\text{LATE} \equiv E_{\tr}[Y^*(1,M^*(1)) - Y^*(0,M^*(0))|T^* \in \{CN,CC,CA\}].
\end{equation*}
The LATE informs us about the treatment effect of relocating to a low-poverty neighborhood for the subgroup of individuals who decide to relocate in response to being offered a voucher.
However, the LATE is a weighted average of causal effects across types with different mental health statuses and is, as a result, not suitable for assessing the mediating role of mental health. 
Specifically, the LATE is a weighted average of:
\begin{align}\label{eq:omto2}
\text{CDE}_0 & \equiv E_{\tr}[Y^*(1,0) - Y^*(0,0)|T^* = CN] \notag \\
\text{CDE}_1 & \equiv E_{\tr}[Y^*(1,1) - Y^*(0,1)|T^* = CA] \notag \\
\text{CTE} & \equiv E_{\tr} [Y^*(1,1) - Y^*(0,0)|T^* = CC];
\end{align}
i.e., the LATE aggregates the ``controlled direct effects" of relocating while keeping mental health status constant ($\text{CDE}_0$ and $\text{CDE}_1$) and the ``controlled total effect" of simultaneously relocating and improving mental health ($\text{CTE}$).

Because the marginal distribution of $T^*$ is identified, the identification of $\text{CDE}_0$, $\text{CDE}_1$, and $\text{CTE}$ reduces to the identification of expectations with the structure
\begin{equation}\label{eq:omto3}
E_{\tr}[\rho(Y^*(t))\ell(T^*,X)]    
\end{equation}
for identified $\rho$ and $\ell$.
Applying our identification results to this context immediately implies that restriction \eqref{eq:tmto4} fails to identify $\text{CDE}_0$, $\text{CDE}_1$, and $\text{CTE}$; see Corollary \ref{cor:discintout}.
We therefore introduce an ``exogeneity of irrelevant mediator choices" (EIMC) assumption: 
Potential outcomes corresponding to high (resp.\ low) poverty neighborhood and poor (resp.\ good) mental health are conditionally independent of what mental health would have been in a low (resp.\ high) poverty neighborhood.
Formally, EIMC requires that
\begin{align*}
Y^*(0,0)\perp M^*(1) & | D^*(1)>D^*(0), M^*(0) = 0,X  \\
Y^*(1,1)\perp M^*(0) & | D^*(1)>D^*(0), M^*(1) = 1 ,X 
\end{align*}
which we note can be imposed in our model through the set $\restr$ in Assumption \ref{ass:model}(iii).

Our identification results imply that EIMC and restriction \eqref{eq:tmto4} secure the identification of $\text{CDE}_0,$ $\text{CDE}_1,$ and $\text{CTE}$; see Theorem \ref{th:outmain}.
More generally, our results yield that expectations with the structure in \eqref{eq:omto3} are identified if and only if there is a $\kappa$ solving
\begin{equation}\label{eq:omto6}
E_{\tr}[\sum_{z\in \mathbf Z} 1\{T^*(z)=t\}\kappa(z,X)P(Z=z|X)|V^*(t),X] = E_{\tr}[\ell(T^*,X)|V^*(t),X],
\end{equation}
where $V^*(t) = T^*$ if $t\in\{(0,1),(1,0)\}$, $V^*((0,0)) = T^*1\{T^*\notin\{CN,CC\}\}$, and $V^*((1,1)) = T^*1\{T^*\notin\{CA,CC\}\}$.\footnote{Here, with some abuse of notation, we understand $T^*\times 1$ to equal $T^*$ and $T^*\times 0$ to equal 0.}
Moreover, any function $\kappa$ satisfying equation \eqref{eq:omto6} can be employed to identify the expectation of $\ell(T^*,X)$ through the equality
\begin{equation}\label{eq:omto7}
E_{\tr}[\rho(Y^*(t))\ell(T^*,X)] = E_P[\rho(Y)1\{T=t\}\kappa(Z,X)].    
\end{equation}
We highlight that the identifying equations in \eqref{eq:tmto2} and \eqref{eq:omto6} are both linear, but \eqref{eq:omto6} requires us to ``equal" $\ell(T^*,X)$ in a weaker sense than \eqref{eq:tmto2}.
This contrast reflects a deeper observation, established in Theorem \ref{th:idom}, that identification is driven by a common linear map $\Id$ and a topology that reflects the strength of the identifying assumptions.

\begin{table}[t!]
\begin{center}
\caption{Treatment Effects Estimates}														
\label{tab:out}
\begin{tabular}{c  cc cc c}
\hline \hline
                            &                &                &              &              & \small{Implied} \\ 
Outcome                     & $\text{CDE}_0$ & $\text{CDE}_1$ & $\text{CTE}$ &  \text{LATE} & \text{LATE}     \\  \midrule
													

\multicolumn{1}{r}{\small{	Household is Economically}}&	0.024	&	0.048	&	0.059	&	0.035	&	0.039	\\	
\multicolumn{1}{r}{\small{Self-Sufficient	}}&\small{(0.050)}&\small{(0.063)}&\small{(0.078)}&\small{(0.039)}&\small{ }\\	\\
													
\multicolumn{1}{r}{\small{	Sampled Adult}}&	$0.127^{*}$	&	-0.003	&	0.132	&	0.066	&	0.070	\\	
\multicolumn{1}{r}{\small{is Employed	}}&\small{(0.066)}&\small{(0.079)}&\small{(0.122)}&\small{(0.051)}&\small{ }\\	\\
													
\multicolumn{1}{r}{\small{	Sampled Adult not }}&	-0.042	&	-0.074	&	-$0.412^{**}$	&	-$0.113^{**}$	&	-0.106	\\	
\multicolumn{1}{r}{\small{in Labor Force	}}&\small{(0.071)}&\small{(0.072)}&\small{(0.144)}&\small{(0.049)}&\small{ }\\	\\
													
\multicolumn{1}{r}{\small{	Household Total }}&	0.498	&	2.928	&	2.479	&	1.811	&	1.840	\\	
\multicolumn{1}{r}{\small{Income	}}&\small{(2.262)}&\small{(2.191)}&\small{(3.014)}&\small{(1.523)}&\small{ }\\
													
\hline
\hline
\end{tabular}
\end{center}
\begin{flushleft}
\scriptsize{
Estimates for the treatment effects $\text{CDE}_0,$  $\text{CDE}_1,$  $\text{CTE}$ (as in \eqref{eq:omto2}), and LATE for different economic outcomes.
The last column evaluates LATE as a weighted average of $\text{CDE}_0,$  $\text{CDE}_1,$ and $\text{CTE}.$
All estimates account for the person-level weight for the adult survey of the interim analyses.
Standard errors are displayed in parentheses.
}		
\end{flushleft}\vspace{-0.3 in}
\end{table}

Table \ref{tab:out} reports treatment effects estimates based on an orthogonal score of \eqref{eq:omto7}; see Appendix A.4 for details.
We examine four different outcomes:
(i) Household is self-sufficient; \footnote{Defined as total household income in 2001 being above the poverty line and the household not currently being a recipient of welfare programs, namely, AFDC/TANF, food stamps, SSI, or Medicaid.}
(ii) Adult participant is employed;
(iii) Adult participant is not in the labor force; and
(iv) Household Total Income.
LATE estimates suggest that moving from high to low-poverty neighborhoods is associated with improved self-sufficiency, a higher likelihood of being employed, and increased income.
The estimates for $\text{CDE}_0$ and $\text{CDE}_1$ indicate that these positive effects from relocation are largely present even if mental heath status is unchanged.
Parameter $\text{CTE}$ encompasses two effects: the impact of moving to a low-poverty neighborhood and the effect of enhanced mental health. 
In full support of  the conjecture by \cite{ludwig2008can}, we see that mental health plays an important role in mediating the effects of neighborhood relocation on labor force participation.
Table \ref{tab:out} additionally reports the LATE implied by our estimates for $\text{CDE}_0$, $\text{CDE}_1$, $\text{CTE}$, and type probabilities.
The implied and estimated LATEs closely align, providing credence to our decomposition of LATE into direct and total effects.

\begin{figure}[t!]
\caption{Quantile Treatment Effects for Total Household Income}
\label{fig:QuantilesY5}
\begin{center}
\resizebox{\textwidth}{!}{
\begin{tabular}{c c}
\input{Results/Quantiles/QDE0Y5}
&
\input{Results/Quantiles/QDE1Y5}\\
&\\
\input{Results/Quantiles/CTEY5}
&
\input{Results/Quantiles/LATEY5}\\
\end{tabular}}
\end{center}
\begin{flushleft}
\scriptsize{
Graph A: Difference in the quantiles of $Y^*(1,0)$ and $Y^*(0,0)$ conditional on type CN.
Graph B: Difference in the quantiles of $Y^*(1,1)$ and $Y^*(0,1)$ conditional on type CA. 
Graph C: Difference in the quantiles of $Y^*(1,1)$ and $Y^*(0,0)$ conditional on type CC.
Graph D: Difference in the quantiles of $Y^*(1,M^*(1))$ and $Y^*(0,M^*(0))$ conditional type $\{CA,CN,CC\}$.
All estimates consider the person-level weight from the adult survey in the interim analyses. Income is measured in thousands of dollars per year.}\vspace{-0.3 in}
\end{flushleft}
\end{figure}

We further investigate treatment impacts across the outcome distribution by computing Quantile Treatment Effects (QTEs) analogues to the average effects estimated in Table \ref{tab:out} -- e.g., the QTE for CTE consists of comparing the quantiles of $Y^*(1,1)$ against those of $Y^*(0,0)$ for type CC.
Figure \ref{fig:QuantilesY5} reports our QTE estimates for total household income with 95$\%$ pointwise confidence regions.
Overall, we find the estimates for the QTEs corresponding to relocating while keeping mental health constant (${\rm CDE}_0$ and ${\rm CDE}_1$) are decreasing though mostly statistically insignificant.
In contrast, we find that the QTEs corresponding to both relocating and improving mental health (CTE) are positive and statistically significant across the quantiles we examine.
Reflecting the low proportion of type CC in the population, the LATE QTEs exhibit mixed findings, being decreasing and statistically significant for lower quantiles only.


\section{Identification} \label{sec:id}

We next turn to our theoretical results starting, in this section, by developing a characterization of point identification for the functionals that we study.


\subsection{Two Key Lemmas}\label{subsec:2lemma}

We begin by introducing two lemmas that play a fundamental role in our characterization of identification.
The first result is technical in nature, but crucial for our analysis.

\begin{lemma}\label{th:idom}
If Assumptions \ref{ass:setup} and \ref{ass:model} hold, then $\Theta_0$ is convex and there is a $\dom \in \Theta_0$ such that all $Q\in \Theta_0$ are absolutely continuous with respect to $\dom$.
\end{lemma}

In words, Lemma \ref{th:idom} establishes the existence of a distribution $\dom$ that is both in the identified set $\Theta_0$ for the true distribution $\tr$ and ``larger" than any other distribution in $\Theta_0$.
Intuitively, by ``larger" we mean that the support of any distribution in the identified set must be contained in the support of the distribution $\dom$. 
We note that there may be multiple measures $\dom$ satisfying the conclusion of Lemma \ref{th:idom}.
However, such measures are equivalent in the sense that they must be mutually absolutely continuous -- i.e.\ they must assign zero probability to the same sets.
Hence, whether $\dom$ assigns probability zero (or one) to a set is a property that is identified from the distribution of the data.

Our second lemma is the cornerstone of our identification analysis.
In order to formally state this key result, we first introduce a linear operator $\Id$ that maps functions of $(Y,T,Z,X)$ to functions of $(Y^*,T^*,X)$. 
Specifically, for any $f\in L^1(P)$ we set 
\begin{equation*}
\Id(f) \equiv \sum_{t\in \mathbf T} E_{P_{Z|X}}[f(Y^\star(t),t,Z,X)1\{T^\star(Z) = t\}],
\end{equation*}
where $P_{Z|X}$ denotes the conditional distribution of $Z$ given $X$ and the notation $E_{P_{Z|X}}$ emphasizes the expectation is taken with respect to $Z$ while $(Y^\star,T^\star,X)$ are kept ``fixed."
Under our assumptions, it is possible to show that $\Id(f)$ in fact satisfies
\begin{equation}\label{sec4:eq2}
    \Id(f) = E_Q[f(Y,T,Z,X)|Y^*,T^*,X]
\end{equation}
for any $Q$ in the identified set $\Theta_0$ -- i.e.\ $\Id$ maps functions of observables into functions of unobservables by taking conditional expectations given $(Y^*,T^*,X)$.

The next lemma combines the map $\Id$ with the measure $\dom$ to obtain a sufficient condition for the expectation of a function $\ell$ of $(Y^\star,T^\star,X)$ to be identified.

\begin{lemma}\label{lm:warmup}
Let Assumptions \ref{ass:setup} and \ref{ass:model} hold, and $\ell$ satisfy $\dom(\Id(\kappa) = \ell)=1$ for some $\kappa \in L^1(P)$.
Then it follows that $\ell \in L^1(Q)$ for all $Q\in \Theta_0$ and in addition
\begin{equation}\label{lm:warmupdis}
E_Q[\ell(Y^\star,T^\star,X)] = E_{P}[\kappa(Y,T,Z,X)]. 
\end{equation}
\end{lemma}

The conclusion of Lemma \ref{lm:warmup} is straightforward to obtain after noting that the conditions imposed on $\kappa$ and the equality in \eqref{sec4:eq2} ensure, for any $Q\in \Theta_0$, that
$$E_Q[\kappa(Y,T,Z,X)|Y^*,T^*,X] = \ell(Y^*,T^*,X)$$
from whence result \eqref{lm:warmupdis} is immediate by the law of iterated expectations.
The principal implication of Lemma \ref{lm:warmup} is a recipe for identification and estimation of the expectation of a function $\ell$ of $(Y^*,T^*,X)$.
In particular, Lemma \ref{lm:warmup} suggests estimating the expectation of $\ell$ by employing sample moments based on an estimator of a function $\kappa$ solving the equation $\Id(\kappa) = \ell$. 
In implementing this approach, it is often fruitful to rely on our next corollary, which obtains an alternative representation for $\kappa$ in terms of the density
\begin{equation*}
\pi \equiv \frac{dP_{Z|X}}{d\mu_{Z|X}}.    
\end{equation*}


\begin{corollary}\label{cor:warmup}
Let Assumptions \ref{ass:setup} and \ref{ass:model} hold, and suppose $\nu \in L^1(P)$ is such that
\begin{equation}\label{cor:warmupdisp1}
\mu(\sum_{t\in \mathbf T} E_{\mu_{Z|X}}[\nu(Y^*(t),t,Z,X)1\{T^*(Z) = t\}] = \ell(Y^*,T^*,X)) = 1.
\end{equation}
If $\mu(\pi(Z,X)> \delta) = 1$ for some $\delta > 0$, then the function $\kappa \equiv \nu/\pi$ satisfies $\kappa \in L^1(P)$ and $\dom(\Id(\kappa) = f) = 1$, and therefore $E_{\tr}[\ell(Y^\star,T^\star,X)] = E_P[\kappa(Y,T,Z,X)]$.
\end{corollary}

Under the requirement that $\pi$ be bounded away from zero, Corollary \ref{cor:warmup} shows that we may find a function $\kappa$ solving $\Id(\kappa) = \ell$ by taking the ratio of a function $\nu$ satisfying \eqref{cor:warmupdisp1} and the density $\pi$.
This characterization is particularly useful in applications in which the measure $\mu$ is known (instead of identified), as is the case in the majority of the examples discussed in Section \ref{subsec:ex}.
Specifically, if $\mu$ is known, then the functions $\nu$ satisfying \eqref{cor:warmupdisp1} are known in that they may be computed analytically or numerically. 
In particular, it follows that $\kappa = \nu/\pi$ is known up to the identified density $\pi$.
We will extensively employ these observations when developing our estimators in Section \ref{sec:est}.

\begin{remark}\rm
Revisiting Example \ref{ex:rc} can be instructive in illustrating the content of Corollary \ref{cor:warmup}.
In this example, based on \cite{rosenbaum1983central}, we imposed $\mu(T^*(Z)=Z)=1$ and set $\ell(Y^*,T^*,X) = Y^*(1)-Y^*(0)$.
In order to identify the ATE, Corollary \ref{cor:warmup} suggests finding a function $\nu$ satisfying equation \eqref{cor:warmupdisp1}.
To this end, we select $\mu$ to satisfy $\mu(Z=0|X)=\mu(Z=1|X)=1/2$, which implies that
\begin{equation*}
\nu(Y,T,Z,X) = 2Y(1\{Z=1\} - 1\{Z=0\})
\end{equation*}
solves \eqref{cor:warmupdisp1}.
Moreover, under such a choice of $\mu$, $\pi$ satisfies $\pi(z,X) = P(Z=z|X)/2$ for $z\in\{0,1\}$.
Therefore, computing $\kappa = \nu/\pi$ and employing Corollary \ref{cor:warmup} yields that
\begin{equation*}
E_{\tr}[Y^*(1)-Y^*(0)] = E_P[\kappa(Y,T,Z,X)] = E_P[\frac{Y1\{Z=1\}}{P(Z=1|X)} - \frac{Y1\{Z=0\}}{P(Z=0|X)}],
\end{equation*}
which recovers the canonical propensity score reweighing moment for identifying the ATE.
Similarly, applying Corollary \ref{cor:warmup} to the model in \cite{imbens1994identification} recovers the the ``$\kappa$-weights" identifying equations of \cite{abadie2003semiparametric}. \qed
\end{remark}

\subsection{Main Result}\label{subsec:idmain}

Lemma \ref{lm:warmup} establishes that a sufficient condition for the identification of the expectation of a function $\ell$ of $(Y^\star,T^\star,X)$ is the existence of a function $\kappa$ of $(Y,T,Z,X)$ satisfying $\Id(\kappa) = \ell$ in an appropriate sense. 
The conclusion of Lemma \ref{lm:warmup} is additionally constructive in that it suggests an estimator for the parameter of interest.
However, our analysis so far leaves two important questions unanswered.
First: Is it possible to employ a similar approach to identify and estimate the more general class of parameters that interest us?
Second: Is such an approach applicable whenever the parameter of interest is identified?
In other words, are our sufficient conditions for identification also necessary?
We next provide affirmative answers to these questions.

Specifically, we next return to the general class of functionals with the structure
\begin{equation}\label{sec4:eq4}
\lambda_{Q} \equiv \lim_{j\to \infty} E_Q[\ell_j(Y^\star,T^\star,X)]
\end{equation}
and provide necessary and sufficient conditions for the identification of $\lambda_{\tr}$.
In particular, we will show that $\lambda_{\tr}$ is identified \emph{if and only if} the functional $Q\mapsto \lambda_{Q}$ is in the ``closure" of the set of functions that equal $\Id(\kappa)$ for some $\kappa$ -- notice the distinction with Lemma \ref{lm:warmup}, which requires $\ell$ to exactly equal $\Id(\kappa)$ for some $\kappa$.
Intuitively, we will establish that for a functional of $\tr$ to be identified, it must be the ``limit" of functionals of $\tr$ whose identification can be shown through Lemma \ref{lm:warmup}.
Such a characterization of identification is additionally constructive in that it suggests estimating $\lambda_{\tr}$ by employing the sample averages of a sequence of functions $\{\kappa_j\}$ for which $\Id(\kappa_j)$ ``converges" to the desired functional.
Formalizing this discussion, however, first requires to clarify the sense (i.e.\ topology) in which we mean ``converge," ``closure," and ``limit."
We next turn to this task, which requires us to introduce additional assumptions and notation.

Our next assumption introduces the final regularity conditions for our model.

\begin{assumption}\label{ass:genpar}
(i) $\{\ell_j\}_{j=1}^\infty$ is an identified sequence satisfying $\{\ell_j\}_{j=1}^\infty \subset L^1(Q)$ for all $Q\in \Theta_0$;
(ii) $\lambda_Q$ (as in \eqref{sec4:eq4}) is well defined and satisfies $|\lambda_Q|<\infty$ for all  $Q\in \Theta_0$;
(iii) $dQ/d\mu$ belongs to the interior of $\restr$ in $\mathbf Q$ for some $Q\in \Theta_0$.
\end{assumption}

Assumption \ref{ass:genpar}(i) formalizes the requirement that the functions $\{\ell_j\}$ be identified and integrable with respect to every $Q\in \Theta_0$.
The latter requirement can be ensured, for example, by imposing suitable regularity conditions through $\restr$ in Assumption \ref{ass:model}(iii).
In turn, Assumption \ref{ass:genpar}(ii) formalizes the structure of the parameter of interest by imposing that the limit in \eqref{sec4:eq4} exists and is finite for any $Q\in \Theta_0$ -- a requirement that can again be ensured through the specification of $\restr$.
Finally, Assumption \ref{ass:genpar}(iii) will help us establish that our sufficient conditions for identification are also necessary.
Intuitively, Assumption \ref{ass:genpar}(iii) requires that the restrictions imposed through $\restr$ do not bind at some $Q\in \Theta_0$ and as a result cannot point identify $\lambda_{\tr}$.

As we have informally discussed, the identification of $\lambda_{\tr}$ hinges on whether the functional $Q\mapsto \lambda_Q$ is in, an appropriate sense, the closure of the set of functions that equal $\Id(\kappa)$ for some $\kappa$.
To formally introduce the relevant topology, we first define
$$\langle f,g\rangle_Q \equiv \int fg dQ$$
for any $f,g$ such that $|fg|\in L^1(Q)$ and let $Q_V$ denote the marginal distribution of a random variable $V$ under $Q$ -- e.g., $Q_{T^*}$ denotes the marginal distribution of $T^\star$ under $Q$.
It is also useful to note that, for any suitably ``smooth" $s\in L^\infty(\dom_{Y^\star T^\star X})$, the limit
\begin{equation}\label{sec4:eq5}
\lim_{j \to \infty} \langle s, \ell_j\rangle_{Q}
\end{equation}
will often exist for any $Q\in \Theta_0$.
For instance, if we let ${\bf 1}$ be the function that is constant at one and evaluate \eqref{sec4:eq5} at $s = {\bf 1}$, then we recover $\lambda_Q$.
Given this observation, we set
\begin{equation*}
\mathcal S_Q \equiv \{s \in L^\infty(Q_{Y^*T^*X}) : |\lim_{j\to \infty} \langle s,\ell_j\rangle_Q| < \infty \text{ and } s \frac{dQ}{d\mu} \in \mathbf Q \},  
\end{equation*}
where we tacitly understand every $s\in \mathcal S_Q$ to be such that the limit in \eqref{sec4:eq5} exists.
Additionally, we let $\text{span}\{A\}$ denotes the linear span of a set $A$, and introduce a vector space $\mathcal L$ of linear functionals defined on the space $\bigcap_{Q\in \Theta_0} L^1(Q)$ by setting
\begin{equation*}
\mathcal L \equiv \text{span}\{L: \bigcap_{Q\in \Theta_0} L^1(Q) \to \mathbf R \text{ s.t. } L = \langle \cdot ,s\rangle_Q \text{ for some } s\in \mathcal S,  ~ Q \in \Theta_0\}.    
\end{equation*}

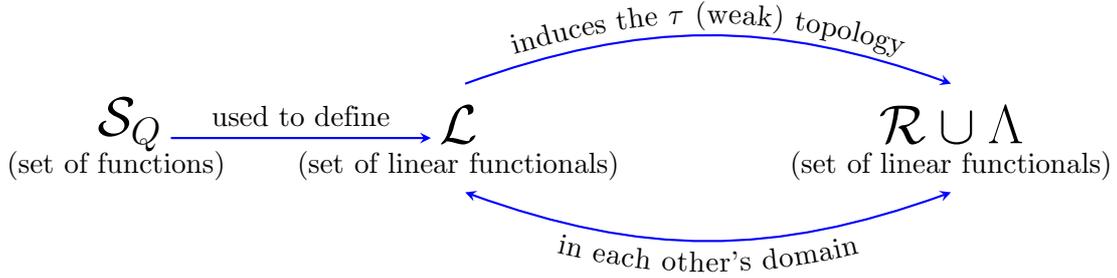
\begin{figure}[t!]
\centering
\begin{tikzpicture}[scale=1.8]
\draw (-3,0) node {\huge $\mathcal S_Q$}; 
\draw (-3.1,-0.3) node { (set of functions)}; 
\draw[blue,thick, ->, {-stealth[scale=10]},postaction={decorate,decoration={raise=1ex,text along path,text align=center,text={used to define}}}] (-2.7,-0.1)--(-0.8,-0.1); 
\draw (-0.6,0) node {\huge $\mathcal L$}; 
\draw (-0.6,-0.3) node { (set of linear functionals)}; 
\draw (3,0) node {\huge $\mathcal R \cup \Lambda $};
\draw (3,-0.3) node { (set of linear functionals)}; 
\draw[blue, <->, thick, {stealth[scale=10]-stealth[scale=10]},postaction={decorate,decoration={raise=-2.5ex,text along path,text align=center,text={in each other's domain}}}] (-0.55,-0.5) to [bend right = 20] (3,-0.5);
\draw[blue, ->, thick,  {-stealth[scale=10]}, postaction={decorate,decoration={raise=1ex,text along path,text align=center,text={induces the {$\tau$} (weak) topology}}}] (-0.55,0.3) to [bend left = 20] (3,0.3);
\end{tikzpicture}
\caption{Diagram of the definition of the $\tau$ topology dictating point identification. \\ \vspace{0.2 in} } \label{fig1} 
\end{figure}

By Lemma \ref{lm:warmup}, a function $\Id(\kappa)$ is in the domain of all the linear functionals $L\in \mathcal L$ for any $\kappa \in L^1(P)$. 
Through duality, however, it is more instructive to identify such functions with a set of linear functionals on $\mathcal L$. 
We therefore define the set $\mathcal R$ by
\begin{equation*}
\mathcal R \equiv \{L^\prime : \mathcal L \to \mathbf R \text{ s.t. } L^\prime(L) = L(\Id(\kappa)) \text{ for some } \kappa \in L^1(P) \} .  
\end{equation*}
Similarly, $\{\ell_j\}$ generates a linear functional on $\mathcal L$, which we denote by $\Lambda$ and equals
\begin{equation*}
\Lambda(L) \equiv \lim_{j\to \infty} L(\ell_j),
\end{equation*}
and we note that any $L\in \mathcal L$ can also be viewed as a functional on $\{\mathcal R \cup\Lambda\}$ through the relation $L^\prime \mapsto L^\prime(L)$.
Given the introduced concepts, we can finally define the topology that dictates identification.
Specifically, we let $\tau$ denote the weak topology on $\{\mathcal R \cup \Lambda\}$ that is generated by the functionals $L \in \mathcal L$ -- i.e.\ $\tau$ is the weakest topology on $\{\mathcal R \cup \Lambda\}$ that makes all $L\in \mathcal L$ continuous; see Figure \ref{fig1} for a diagram summarizing this construction.

The next theorem is our main identification result.

\begin{theorem}\label{th:genmain}
If Assumptions \ref{ass:setup}, \ref{ass:model}, and \ref{ass:genpar} hold, then it follows that $\lambda_{\tr}$ is identified if and only if $\Lambda$ belongs to the the $\tau$-closure of $\mathcal R$.
\end{theorem}

Intuitively, Theorem \ref{th:genmain} establishes that $\lambda_{\tr}$ is identified if and only if there is a sequence $\{L_j^\prime\} \subseteq \mathcal R$ converging to $\Lambda$ in the $\tau$ topology.\footnote{We discuss sequences for ease of exposition. However, we note that our formal arguments rely on nets because the $\tau$ topology may not be first countable and therefore not be metrizable.} 
To the best of our knowledge, the characterization of all the functionals that are identified is novel in the context of all the examples in Section \ref{subsec:ex}.
To gain some insight into why $\Lambda$ belonging to the $\tau$-closure of $\mathcal R$ is a sufficient condition for identification, let $L_{\tr} \equiv \langle \cdot,{\bf 1}\rangle_{\tr}$ and note
\begin{equation}\label{sec4:eq10}
\lambda_{\tr} = \lim_{j\to \infty} \langle \ell_j,{\bf 1}\rangle_{\tr}  = \Lambda(L_{\tr}) .   
\end{equation}
Moreover, since $L_j^\prime \in \mathcal R$ implies $L_j^\prime(L) = L(\Id(\kappa_j))$ for some $\kappa_j$, it follows from $\{L_j^\prime\}$ converging to $\Lambda$ in the $\tau$ topology that there is a sequence $\{\kappa_j\}$ satisfying
\begin{equation}\label{sec4:eq11}
\Lambda(L_{\tr}) = \lim_{j\to \infty} L_j^\prime(L_{\tr}) = \lim_{j\to \infty} \langle \Id(\kappa_j),{\bf 1}\rangle_{\tr} = \lim_{j\to \infty} E_P[\kappa_j(Y,T,Z,X)],
\end{equation}
where the final equality follows from Lemma \ref{lm:warmup}.
Importantly, results \eqref{sec4:eq10} and \eqref{sec4:eq11} not only establish the identification of $\lambda_{\tr}$, but also suggest an estimation strategy: Simply employ the sample moments based on estimates of the approximating sequence $\{\kappa_j\}$.

More surprisingly, Theorem \ref{th:genmain} also establishes that the existence of the desired sequence $\{\kappa_j\}$ is in fact a necessary condition for the identification of $\lambda_{\tr}$.\footnote{Theorem \ref{th:genmain} only implies the existence of a net, but we again discuss sequences for ease of exposition.}
As a result, it is without loss of generality to estimate $\lambda_{\tr}$ by employing the discussed estimation strategy that is motivated by results \eqref{sec4:eq10} and \eqref{sec4:eq11}.
Moreover, while our preceding discussion suggests that we need only find a sequence $\{\kappa_j\}$ satisfying \eqref{sec4:eq11}, Theorem \ref{th:genmain} states that the identification of $\lambda_{\tr}$ is in fact only possible if the stronger requirement that $L(\Id(\kappa_j))\to\Lambda(L)$ for all $L\in \mathcal L$ is satisfied.
As our next corollary illustrates, the latter observation can be helpful in characterizing the desired sequence $\{\kappa_j\}$.

\begin{corollary}\label{cor:int}
Let Assumptions \ref{ass:setup}, \ref{ass:model} hold with $\restr = \mathbf Q = L^\infty(\mu)$, $\mu \ll \dom$ with $d\mu/d\dom$ bounded, and $\lambda_Q \equiv E_Q[\ell(Y^*,T^*,X)]$ for some identified $\ell \in L^1(\mu_{Y^*T^*X})$. Then:\vspace{-0.15 in}
\begin{packed_enum}
    \item[(i)] $\lambda_{\tr}$ is identified if and only if $\lim_{j\to \infty} \|\ell - \Id(\kappa_j)\|_{\mu,1}= 0$ for some $\{\kappa_j\}\subseteq L^1(P)$. Moreover, any such sequence $\{\kappa_j\}$ satisfies $\lambda_{\tr} = \lim_{j\to \infty} E_P[\kappa_j(Y,T,Z,X)]$.
    \item[(ii)] Suppose in addition that $\mu(\pi(Z,X) > \delta) = 1$ for some $\delta >0$. Then, $\lambda_{\tr}$ is identified if and only if there is a sequence $\{\nu_j\}\subseteq L^1(P)$ satisfying
    \begin{equation}\label{eq:cor:intdisp}
    \lim_{j\to \infty} E_{\mu}[|\ell(Y^*,T^*,X) - \sum_{t\in \mathbf T} E_{\mu_{Z|X}}[\nu_j(Y^*(t),t,Z,X)1\{T^*(Z)=t\}]|] = 0.
    \end{equation}
    Moreover, for any such $\{\nu_j\}$, $\kappa_j = \nu_j/\pi$ satisfies $\lambda_{\tr} =\lim_{j\to \infty} E_P[\kappa_j(Y,T,Z,X)]$.
\end{packed_enum}
\end{corollary}

Corollary \ref{cor:int} specializes Theorem \ref{th:genmain} to the case in which the functional of interest is the expectation of a function $\ell$ of $(Y^*,T^*,X)$ and Assumption \ref{ass:model}(iii) only imposes that the density of $\tr$ be bounded.
Within this context, Corollary \ref{cor:int}(i) shows that $\lambda_{\tr}$ is identified if and only if $\ell$ is the limit of a sequence of functions $\{\Id(\kappa_j)\}$ in the $\|\cdot\|_{\mu,1}$-norm.
Paralleling Corollary \ref{cor:warmup}, Corollary \ref{cor:int}(ii) additionally provides conditions under which it is without loss of generality to set $\kappa_j = \nu_j/\pi$ for any sequence $\{\nu_j\}$ satisfying \eqref{eq:cor:intdisp}.
Corollary \ref{cor:int}(ii) has two important implications for applications in which $\mu$ is known and therefore the sequence $\{\nu_j\}$ is known and computable analytically or numerically; see Remark \ref{rm:num}.
First, Corollary \ref{cor:int}(ii) provides us with a simple characterization of $\{\kappa_j\}$ in terms of the identified density $\pi$ that we will use in estimation.
Second, condition \eqref{eq:cor:intdisp} allows us to assess whether the restrictions of our model (as embodied in $\mu$) point identify a functional of interest or not; see our discussion of Example \ref{ex:pm} below. 

\subsection{Special Case: Types}\label{sec:idtype} 

Functionals of the joint distribution of $(T^\star, X)$ are often of interest in their own right or as building blocks towards estimating other parameters.
In this section, we specialize our analysis to such functionals by considering parameters with the structure
\begin{equation}\label{sec4:eq12}
\lambda_{Q} \equiv \lim_{j\to \infty} E_Q[\ell_j(T^\star,X)],
\end{equation}
which we refer to as functionals about ``types."
While Theorem \ref{th:genmain} of course continues to apply to this context, the fact that $\{\ell_j\}$ now only depends on $(T^\star,X)$ will allow us to sharpen our identification results.
In particular, we will show that $\lambda_{\tr}$ is identified if and only if it can be identified from the joint distribution of $(T,Z,X)$ -- i.e.\ from ``first stage" information.
As a result, in estimating functionals about types we may simplify estimation by only employing the sample for $(T,Z,X)$ (instead of $(Y,T,Z,X)$).

In order to formally state the conditions for our result, we first define the measure
\begin{equation*}
\dom^{\rm it} \equiv \dom_{Y^\star|X} \dom_{T^\star ZX},
\end{equation*}
which shares the same marginal distributions for $(Y^*,X)$ and $(T^*,X)$ as $\dom$, but is such that $Y^*$ is independent of $T^*$ conditionally on $X$.
Given this notation we impose: 
\begin{assumption}\label{ass:4types}
(i) $\dom^{\rm it} \ll \dom$;
(ii) $(d\dom^{\rm it}_{Y^\star T^\star X}/d\dom_{Y^\star T^\star X})s \in \mathcal S_{\dom}$ for all $s\in L^\infty(\dom_{T^\star X}) \cap \mathcal S_{\dom}$; 
(iii) $ (dQ_{T^\star X}/d\dom_{T^\star X})s \in \mathcal S_{\dom}$ for all $Q\in \Theta_0$ and $s\in L^\infty(Q_{T^\star X}) \cap \mathcal S_Q$; 
(iv) For any $Q\in \Theta_0$ and $s\in \mathcal S_Q$ we have that $E_Q[s(Y^\star,T^\star,X)|T^\star,X]\in \mathcal S_Q$. 
\end{assumption}

Assumption \ref{ass:4types}(i) essentially requires the support of $Y^\star$ conditional on $(T^\star,X)$ under $\dom$ to not depend on $T^\star$.
We view Assumption \ref{ass:4types}(i) as the key requirement ensuring that the identification of functionals about types can be characterized by the distribution of $(T,Z,X)$.
Assumptions \ref{ass:4types}(ii) and \ref{ass:4types}(iii) impose restrictions on the densities $d\dom^{\rm it}/d\dom$ and $dQ_{T^\star X}/d\dom_{T^\star X}$ (for $Q\in \Theta_0$), while Assumption \ref{ass:4types}(iv) requires that conditional expectations of functions in $\mathcal S_Q$ belong to $\mathcal S_Q$ as well.
Assumptions \ref{ass:4types}(ii)-(iv) can in many applications be verified by appropriately selecting $\restr$ and $\mathbf Q$ in Assumption \ref{ass:model}(iii); see, e.g., Corollary \ref{cor:discint} and Section \ref{sec:idtype:ex} below.

The next theorem is our main result on identification of functionals about types.

\begin{theorem}\label{th:typemain}
Let Assumptions \ref{ass:setup}, \ref{ass:model}, \ref{ass:genpar}, \ref{ass:4types} hold, $\lambda_Q$ be as in \eqref{sec4:eq12}, and define
$$\Rf \equiv \{L^\prime : \mathcal L \to \mathbf R \text{ s.t. } L^\prime(L) = L(\Id(\kappa)) \text{ for some } \kappa\in L^1(P_{TZX})\}.$$
Then, it follows that $\lambda_{\tr}$ is identified if and only if $\Lambda$ belongs to the the $\tau$-closure of $\Rf$.
\end{theorem}

Theorem \ref{th:typemain} establishes that functionals about types are identified if and only if they are identified from the distribution of $(T,Z,X)$.
Formally, Theorem \ref{th:typemain} shows that $\lambda_{\tr}$ is identified if and only if it belongs to the $\tau$-closure of the subset $\mathcal R_T \subseteq \mathcal R$ (instead of the $\tau$-closure of $\mathcal R$ as in Theorem \ref{th:genmain}).
In particular, $\mathcal R_T$ is generated by first stage information in that it consists of functionals corresponding to $\Id(\kappa)$ for some $\kappa$ depending on $(T,Z,X)$ only.
The main implication of this result is that, when estimating functionals about types, we may search for the desired approximating sequence $\{\Id(\kappa_j)\}$ for $\Lambda$ by considering functions $\{\kappa_j\}$ that depend on $(T,Z,X)$ only.

Theorem \ref{th:typemain} further yields an analogue to Corollary \ref{cor:int}.
In particular, under parallel conditions to those imposed in Corollary \ref{cor:int}, it is possible to show that the expectation of a function $\ell$ of $(T^*,X)$ is identified if and only if $\ell$ can be approximated by a sequence $\{\Id(\kappa_j)\}$ with $\kappa_j$ depending only on $(T,Z,X)$.
For conciseness, however, we do not formally state such a result.
Instead, in our next corollary we highlight the implications of Theorem \ref{th:typemain} in the empirically salient case of discrete instruments.

\begin{corollary}\label{cor:discint}
Let Assumption \ref{ass:setup}, \ref{ass:model} hold, $\restr = \mathbf Q = L^1(\mu)$, $\dom^{\rm it} \ll \dom$ with $d\dom^{\rm it}/d\dom$ bounded,  $\ell \in L^1(\dom_{T^\star X})$ be identified, $Z$ be discrete, $P(Z=z|X) \geq \varepsilon > 0$ a.s.\ for all $z\in \mathbf Z$, and $\dom(T^\star = t^\star|X) \geq \varepsilon > 0$ a.s.\ for any $t^\star\in \mathbf T^\star$ with $\dom(T^\star = t^\star) > 0$.
Then:\vspace{-0.15 in} 
\begin{packed_enum}
    \item[(i)] $\lambda_{\tr} \equiv E_{\tr}[\ell(T^\star,X)]$ is identified if and only if  $\dom(\Id(\kappa) = \ell) = 1$ for some $\kappa\in L^1(P_{TZX})$.
    Moreover, any such $\kappa$ satisfies $\lambda_{\tr} = E_P[\kappa(T,Z,X)]$.
    \item[(ii)] Suppose in addition that $\mu(\pi(Z,X) > \delta) = 1$ for some $\delta > 0$ and $\mu \ll \dom$. 
    Then $\lambda_{\tr}$ is identified if and only if there exists a $\nu\in L^1(P_{TZX})$ satisfying
    \begin{equation}\label{cor:discintdisp}
    \mu(\ell(T^*,X) = \sum_{t\in \mathbf T} E_{\mu_{Z|X}}[\nu(t,Z,X)1\{T^*(Z)=t\}]) = 1.
    \end{equation}
    Moreover, for any such function $\nu$, $\kappa = \nu/\pi$ satisfies $\lambda_{\tr} = E_P[\kappa(T,Z,X)]$.
\end{packed_enum}
\end{corollary}

Corollary \ref{cor:discint}(i) specializes our analysis to the case of discrete instruments and no additional identifying assumptions being imposed in Assumption \ref{ass:model}(ii) -- a setting that covers many of the examples in Section \ref{subsec:ex}.
In this context, Corollary \ref{cor:discint}(i) establishes that the expectation of a function $\ell$ of $(T^*,X)$ is identified if and only if $\ell$ equals $\Id(\kappa)$ for some function $\kappa$ of $(T,Z,X)$. 
Under its assumptions, Corollary \ref{cor:discint}(i) therefore delivers a converse to Lemma \ref{lm:warmup}. 
In turn, Corollary \ref{cor:discint}(ii) parallels Corollary \ref{cor:int} in providing conditions that can be helpful in assessing whether the desired $\kappa$ exists and estimating it if it does.
Such a characterization is particularly useful when $\mu$ is known, in which case the validity of \eqref{cor:discintdisp} for some $\nu$ is independent of the distribution of the data.

\subsubsection{Examples Revisited}\label{sec:idtype:ex} 

We next revisit Examples \ref{ex:disc} and \ref{ex:pm} to illustrate the implications of our results in models with discrete and continuous instruments respectively.

\noindent {\bf Example \ref{ex:disc} (cont.)} 
The main features of this example, based on \cite{kline2016evaluating}, are that $Z$ and $T^*$ are discrete and $\mu$ imposed $\mu(T^*\in \mathbf R^*) = 1$ for some set $\mathbf R^* \equiv \{t_1^*,\ldots, t_r^*\}$.
Denoting the support of $Z$ by $\mathbf Z = \{z_1,\ldots, z_q\}$, we then let
\begin{equation*}
\omega_j(t^*) \equiv ( 1\{t^*(z_j) = t_1\}, \cdots, 1\{t^*(z_j) = t_d\})
\end{equation*}
and note Corollary \ref{cor:discint}(ii) implies the expectation of $\ell(T^*,X)$ is identified if and only if
\begin{equation}\label{ex:disc3}
\min_{\{s_j\}_{j=1}^q \subset \mathbf  R^d } \sum_{i=1}^r ( \ell(t^*_i,X) - \sum_{j=1}^q s_j^\prime \omega_j(t^*_i))^2 = 0   
\end{equation}
with probability one (over $X$).
Moreover, provided condition \eqref{ex:disc3} holds, we can find a function $\kappa$ of $(T,Z,X)$ whose expectation equals the expectation of $\ell(T^*,X)$ by setting
\begin{equation*}
(\kappa(t_1,z_j,X), \ldots, \kappa(t_d,z_j,X)) \equiv \frac{s_j(X)^\prime }{P(Z=z_j|X)} 
\end{equation*}
for any $(s_1(X),\ldots, s_q(X))$ minimizing \eqref{ex:disc3}.
For instance, specializing \eqref{ex:disc3} to \cite{kline2016evaluating} implies that the distribution of $(T^*,X)$ is identified in that application.
More generally, the preceding discussion highlights that identifying and estimating a functional about types reduces to a simple numerical problem when $Z$ is discrete. \qed

\noindent {\bf Example \ref{ex:pm} (cont.)} 
In this example, based on \cite{mogstad2021causal}, $Z = (C,W)$ with $C$ binary, $W$ a scalar, and $\mu$ imposed that $T^*(c,w)$ be increasing in $c$ and decreasing in $w$.
As in Example \ref{ex:mte}, we also require $T^*(c,w)$ to be lower semicontinuous in $w$ and for simplicity assume that $W$ is continuously distributed with compact support $[\underline{\bf w},\overline{\bf w}]$. 
Under these restrictions, each $T^*$ can be identified with a unique pair $(K_0^*,K_1^*)$ satisfying\begin{equation*}
T^*(c,w) = \sum_{i=0}^1 1\{c=i,K_i^* > w\}
\end{equation*}
and $K_i^*\in [\underline{\bf w},\overline{\bf w}] \cup \{\infty\}$ -- note that, for $c = i$, $K_i^* = \underline {\bf w}$ and $K_i^* = \infty$ corresponds to ``never-takers" and ``always-takers."
We therefore study the identification of the distribution of $(K_0^*,K_1^*,X)$ and note that the restriction that $T^*(c,w)$ be increasing in $c$ is equivalent to imposing $\mu(K_1^*\geq K_0^*)=1$.
It is convenient to let $K_i^*$ be continuously distributed on $(\underline{\bf w},\overline{\bf w}]$ under $\mu$, though we allow $\mu$ to possibly assign positive mass to $\{\underline {\bf w}\}$ and $\{\infty\}$. 
Under conditions paralleling those in Corollary \ref{cor:int}(ii), Theorem \ref{th:typemain} here implies that the expectation of a function $\ell$ of $(K_0^*,K_1^*,X)$ is identified if and only if 
\begin{equation}\label{ex:pm4}
\lim_{j\to \infty} E_\mu[|\sum_{i=0}^1 \int_{K_i^*}^{\bar {\bf w}} \nu_j(0,i,w,X)dw + \int_{\underline{\bf w}}^{K_i^*}\nu_j(1,i,w,X)dw - \ell(K_0^*,K_1^*,X)|] = 0 
\end{equation}
for some sequence $\{\nu_j(T,C,W,X)\}$.
Moreover, provided condition \eqref{ex:pm4} holds, the expectation of $\ell(K_0^*,K_1^*,X)$ equals the limit of the expectations of the functions
\begin{equation*}
\kappa_j(t,c,w,X) = \frac{\nu_j(t,c,w,X)}{f_{W|CX}(w|c,X)P(C=c|X)}
\end{equation*}
where $f_{W|CX}$ denotes the conditional density of $W$ given $(C,X)$.
The characterization of identification obtained in \eqref{ex:pm4} in fact implies that the expectation of a function $\ell$ of $(K_0^*,K_1^*,X)$ is identified if and only if $\ell$ belongs to the $\|\cdot\|_{\mu,1}$-closure of the set
\begin{equation}\label{ex:pm6}
\mathcal T \equiv \{f : f(K_0^*,K_1^*,X) = g_0(K_0^*,X) + g_1(K_1^*,X) \text{ for some } g_0,g_1\}.
\end{equation}
For instance, since $1\{K_1^* > a_1, ~ K_0^* \leq a_0\} = 1\{K_1^* > a_1\}-1\{K_0^* > a_0\}$ for any $a_0\geq a_1$ under $\|\cdot\|_{\mu,1}$ due to $\mu(K_1^*\geq K_0^*) = 1$, it follows that the probability of the event $\{K_1^* > a_1, ~ K_0^* \leq a_0\}$ is identified.
Conversely, $1\{K_1^* > a_1,~ K_0^* \leq a_0\}$ does not belong to the $\|\cdot\|_{\mu,1}$-closure of $\mathcal T$ when $a_1 > a_0$, and hence the probability of the event $\{K_1^* > a_1,~ K_0^* \leq a_0\}$ is identified if and only if $a_0 \geq a_1$. \qed

\subsection{Special Case: Outcomes}\label{sec:idout} 

We conclude our discussion of identification by specializing our analysis to functionals of the distribution of $(Y^*(t),T^*,X)$ for some $t\in \mathbf T$.
In particular, we focus on parameters that for some identified function $\rho$ and sequence $\{\ell_j\}$ have the structure
\begin{equation}\label{sec4:eq14}
\lambda_Q \equiv \lim_{j\to \infty} E_Q[\rho(Y^\star(t))\ell_j(T^\star,X)],
\end{equation}
which we refer to as functionals about ``outcomes."
Our primary motivation for studying these functionals is that they include features of the conditional distribution of a potential outcome given types and covariates as a special case.

Intuitively, identification of functionals of the distribution of $(Y^*(t),T^*,X)$ should only be possible from the distribution of observations for which treatment assignment $T$ equals $t$.
As a result, identification will now require us to approximate the sequence $\{\ell_j\}$ by employing only the subset of observations for which $T$ equals $t$ -- contrast with Theorem \ref{th:typemain} which instead employs all treatment values.
In order to introduce the assumptions that enable us to formalize this intuition, we first define the measure
\begin{equation*}
\dom^{\rm io} \equiv \dom_{Y^\star(t_1)|X}\cdots \dom_{Y^\star(t_d)|X} \dom_{T^\star Z X};
\end{equation*}
i.e., for any $t\in \mathbf T$, $\dom^{\rm io}$ shares the same marginal distributions for $(Y^*(t),X)$ and $(T^*,X)$ as $\dom$, but is such that all coordinates of $Y^*$ and $T^*$ are mutually independent conditionally on $X$.
We also define a function $\phi_{\dom,\rho}$ of $(Y^*(t),X)$ to be given by\footnote{If $\text{Var}_{\dom}\{\rho(Y^\star(t))|X\} = 0$, then $\rho(Y^\star(t)) = E_{\dom}[\rho(Y^\star(t))|X]$ and, setting 0/0 = 0, we therefore let $\phi_{\dom,\rho}(Y^\star(t),X) = 0$ whenever $\text{Var}_{\dom}\{\rho(Y^\star(t))|X\} = 0$.}
\begin{equation*}
\phi_{\dom,\rho}(Y^\star(t),X) \equiv \frac{\rho(Y^\star(t)) - E_{\dom}[\rho(Y^\star(t))|X]}{\text{Var}_{\dom}\{\rho(Y^\star(t))|X\}}.
\end{equation*}

Given the introduced notation, we impose the following assumptions.

\begin{assumption}\label{ass:4outbasic}
(i) $\rho$ and $\{\ell_j\}$ are identified, $\rho \in L^\infty(\dom)$, and $\{\ell_j\}\subset  L^1(Q)$ for all $Q\in \Theta_0$;
(ii) $\lambda_Q$ (as in \eqref{sec4:eq14}) is well defined and satisfies $|\lambda_Q|<\infty$ for all $Q\in \Theta_0$.
\end{assumption}

\begin{assumption}\label{ass:4out}
(i) $E_Q[\rho(Y^\star(t))|T^\star,X] \in \mathcal S_Q$ and $E_Q[s(Y^\star,T^\star,X)|T^\star,X] \in \mathcal S_Q$ for any $Q\in \Theta_0$ and $s\in \mathcal S_Q$;
(ii) $\dom^{\rm io} \ll \dom$ and $d\dom^{\rm io}/d\dom\in L^\infty(\dom)$; 
(iii) $\phi_{\dom,\rho}\in L^\infty(\dom)$, ${\rm Var}_{\dom}\{\rho(Y^\star(t))|X\} > 0$ a.s.\ under $\dom$, and $s (d\dom^{\rm io}_{Y^\star T^\star X}/d\dom_{Y^\star T^\star X})\phi_{\dom,\rho}  \in \mathcal S_{\dom}$ for all $s\in L^\infty(\dom_{T^\star X})\cap \mathcal S_{\dom}$;
(iv) $(dQ_{T^\star X}/d\dom_{T^\star X})s\in \mathcal S_{\dom}$ for all $Q\in \Theta_0$ and $s\in L^\infty(Q_{T^\star X})\cap \mathcal S_Q$.
\end{assumption}

Assumption \ref{ass:4outbasic} ensures that $\lambda_Q$ is well defined and $\{\ell_j \rho\}$ is integrable under any $Q\in \Theta_0$ -- note that Assumption \ref{ass:4outbasic} essentially imposes that Assumptions \ref{ass:genpar}(i)(ii) hold with $\{\ell_j \rho\}$ in place of $\{\ell_j\}$.
In turn, Assumption \ref{ass:4out} is similar in spirit to the conditions we imposed in Assumption \ref{ass:4types} to establish our results concerning functionals about types.
Specifically, we note Assumptions \ref{ass:4out}(i)(iv) imposes restrictions on conditional expectations and densities that parallel those of Assumptions \ref{ass:4types}(ii)-(iv), while Assumption \ref{ass:4out}(ii) imposes a key support requirement that parallels Assumption \ref{ass:4types}(i).
Finally, Assumption \ref{ass:4out}(iii) requires that $\text{Var}_{\dom}\{\rho(Y^\star(t))|X\}$ be positive, which implies the parameter of interest indeed concerns features of the outcomes distribution -- e.g.  if $\rho$ were constant, then \eqref{sec4:eq14} would fall within the framework of Section \ref{sec:idtype}.

Our next theorem characterizes the identification of functionals about outcomes.

\begin{theorem}\label{th:outmain}
Let Assumptions \ref{ass:setup}, \ref{ass:model}, \ref{ass:genpar}(iii), \ref{ass:4outbasic}, \ref{ass:4out} hold, $\lambda_Q$ be as in \eqref{sec4:eq14}, define $L^1(P_{tZX}) \equiv \{f\in L^1(P) : f(T,Z,X) = 1\{T=t\}g(Z,X)$ for some $g\in L^1(P)\}$ and
$$\Rs \equiv \{L^\prime : \mathcal L \to \mathbf R \text{ s.t. } L^\prime(L) = L(\Id(\kappa)) \text{ for some } \kappa \in L^1(P_{tZX})\}.$$
Then, it follows that $\lambda_{\tr}$ is identified if and only if $\Lambda$ belongs to the $\tau$-closure of $\Rs$.
\end{theorem}

Theorem \ref{th:outmain} establishes that functionals about outcomes are identified if and only if they are identified from the distribution of observations with treatment assignment $T$ equal to $t$.
We emphasize the contrast with Theorem \ref{th:typemain}, which showed identification of functionals about types is equivalent to $\Lambda$ being in the $\tau$-closure of $\Rf$ (instead of $\Rs$ in Theorem \ref{th:outmain}). 
In particular, since $\Rs \subseteq \Rf$, it follows that identification of a functional about outcomes for a given sequence $\{\ell_j\}$ implies the identification of the corresponding functional about types.
More generally, since $\Lambda$ depends only on the sequence $\{\ell_j\}$, the identification of $\lambda_{\tr}$ for \emph{some} $\rho$ implies that $\Lambda$ is in the $\tau$-closure of $\Rs$ and therefore that $\lambda_{\tr}$ is in fact identified for \emph{all} suitable $\rho$.

Our next corollary illustrates these implications in the case of discrete instruments.

\begin{corollary}\label{cor:discintout}
Let Assumptions \ref{ass:setup}, \ref{ass:model} hold, $\restr = \mathbf Q = L^1(\mu)$, $\dom^{\rm io} \ll \dom$ with $d\dom^{\rm io}/d\dom$ bounded, $\ell \in L^1(\dom_{T^\star X})$ be identified,  $\rho$ be bounded, identified, and  $\text{\rm Var}_{\dom}\{\rho(Y^\star(t))|X\} \geq \varepsilon > 0$ a.s.. 
If $Z$ is discrete, $P(Z=z|X) \geq \varepsilon > 0$ a.s.\ for all $z\in \mathbf Z$, and $\dom(T^\star = t^\star|X) \geq \varepsilon > 0$ a.s.\ for any $t^\star\in \mathbf T^\star$ with $\dom(T^\star = t^\star) > 0$, then the following are equivalent:
\vspace{-0.15 in}
\begin{packed_enum}
    \item[(i)] $E_{\tr}[\rho(Y^\star(t)) \ell(T^\star,X)]$ is identified.
    \item[(ii)] $E_{\tr}[f(Y^\star(t)) \ell(T^\star,X)]$ is identified for any bounded $f$.
    \item[(iii)] $\dom(\Id(\kappa) = \ell) = 1$ for some $\kappa(T,Z,X) = 1\{T=t\}g(Z,X)$ with $g\in L^1(P_{ZX})$, and therefore $E_{\tr}[f(Y^*(t))\ell(T^*,X)] = E_P[f(Y)\kappa(T,Z,X)]$ for any bounded $f$.
\end{packed_enum}
\end{corollary}

Through the equivalence of (i) and (ii), Corollary \ref{cor:discintout} formalizes that the identification of $\lambda_{\tr}$ for some $\rho$ implies the identification of $\lambda_{\tr}$ for all $\rho$.
Corollary \ref{cor:discintout} additionally establishes that identification of an expectation about outcomes requires that there be a $\kappa$ solving $\ell = \Id(\kappa)$.
Unlike Corollary \ref{cor:discint}, however, identification of functionals about outcomes further requires $\kappa$ to only employ observations corresponding to treatment status $t$ -- i.e., $\kappa$ must satisfy $\kappa(T,Z,X) = 1\{T=t\}g(Z,X)$ for some $g$.
Paralleling Corollaries \ref{cor:int} and \ref{cor:discint}, it is further possible to show that identification of $\lambda_{\tr}$ is also equivalent to the existence of a function $\nu \in L^1(P_{ZX})$ satisfying
\begin{equation}\label{out:eq2}
\ell(T^*,X) = E_{\mu_{Z|X}}[\nu(Z,X)1\{T^*(Z)=t\}]
\end{equation}
with $\mu$-probability one.
Such a result is again particularly helpful when $\mu$ is known, in which  it is straightforward to asses whether $\lambda_{\tr}$ is identified (through \eqref{out:eq2}) and estimate the desired $\kappa$ through the relation $\kappa(T,Z,X) = 1\{T=t\}\nu(Z,X)/\pi(Z,X)$.

\subsubsection{Examples Revisited}

We conclude our discussion on identification by revisiting Examples \ref{ex:disc} and \ref{ex:pm}.

\noindent {\bf Example \ref{ex:disc} (cont.)} In this context, Corollary \ref{cor:discintout} implies that the expectation of a function with the structure $\rho(Y^*(t))\ell(T^*,X)$ is identified if and only if
\begin{equation}\label{ex:disc5}
\min_{\{s_j\}_{j=1}^q \subset \mathbf R} \sum_{i=1}^r (\ell(t_i^*,X) - \sum_{j=1}^qs_j1\{t_i^*(z_j)=t\})^2 = 0
\end{equation}
with probability one (over $X$).
Moreover, provided condition \eqref{ex:disc5} holds, the expectation of $\rho(Y^*(t))\ell(T^*,X)$ is equal to the expectation of $\rho(Y)\kappa(T,Z,X)$ where
\begin{equation*}
\kappa(T,z_j,X) = 1\{T = t\}\frac{s_j(X)}{P(Z=z_j|X)}    
\end{equation*}
for any $(s_1(X),\ldots, s_q(X))$ minimizing \eqref{ex:disc5}.
These results highlight that identifying a functional about outcomes reduces to a simple numerical problem when $Z$ is discrete. \qed

\noindent {\bf Example \ref{ex:pm} (cont.)} In this application, under conditions paralleling those in Corollary \ref{cor:int}, Theorem \ref{th:outmain} implies that the expectation of a function $\rho(Y^*(0))\ell(K^*_0,K^*_1,X)$ is identified if and only if there exists a sequence $\{\nu_j(C,W,X)\}$ satisfying
\begin{equation}\label{ex:pm7}
\lim_{j\to \infty} E_{\mu}[|\ell(K_0^*,K_1^*,X)-\sum_{c=0}^1 \int_{K_c^*}^{\bar {\bf w}} \nu_j(c,w,X)dw|] = 0.
\end{equation}
Moreover, provided condition \eqref{ex:pm7} holds, the expectation of $\rho(Y^*(0))\ell(K^*_0,K^*_1,X)$ is identified as the limit of the expectations of $\rho(Y)\kappa_j(T,C,W,X)$ with
\begin{equation*}
\kappa_j(t,c,w,X) \equiv \frac{1\{t=0\}\nu_j(c,w,X)}{f_{W|CX}(w|c,X)P(C=c|X)}.
\end{equation*}
More generally, our analysis yields that the expectation of $\rho(Y^*(t))\ell(K_0^*,K_1^*,X)$ is identified if and only if $\ell$ belongs to the $\|\cdot\|_{\mu,1}$-closure of the set $\mathcal T_t$, where
\begin{align*}
\mathcal T_0 & \equiv \{f: f(K_0^*,K_1^*,X) = g_0(K_0^*,X) + g(K_1^*,X) \text{ with } g_0(\infty,X) = g_1(\infty,X)=0\}\\
\mathcal T_1 & \equiv \{f: f(K_0^*,K_1^*,X) = g_0(K_0^*,X) + g(K_1^*,X) \text{ with } g_0(\underline{\bf w},X) = g_1(\underline{\bf w},X)=0\}  .
\end{align*}
For instance, setting $\ell(K_0^*,K_1^*) = 1\{K_0^* \leq a_0, K_1^* > a_1\}$ with $a_0,a_1\in [\underline{\bf w},\bar {\bf w}]\cup\{\infty\}$ we can conclude that the expectation of $\rho(Y^*(t))1\{K_0^* \leq a_0, K_1^* > a_1\}$ is identified for both $t=0$ and $t=1$ if and only if $a_0 \geq a_1$.
In particular, it follows that parameters such as
\begin{equation}\label{ex:pm8p5}
E_{\tr}[Y^*(1)-Y^*(0)|K_0^*\leq a_0, K_1 = a_1] \text{ and } E_{\tr}[Y^*(1)-Y^*(0)|K_0^* = a_0, K_1 > a_1] 
\end{equation}
are identified for any points $a_0,a_1$ satisfying $\underline{\bf w}\leq a_1 \leq a_0 < \infty$. \qed

\section{Estimation} \label{sec:est}

In our analysis so far, we have allowed features of our model (e.g., the measure $\mu$ and functions $\{\ell_j\}$) to depend on the distribution $P$ of the data.
To construct an estimator, however, we need to incorporate additional information on the exact manner in which these features depend on $P$.  
For concreteness, in what follows we therefore focus on a leading special case in which $\mu$ and $\{\ell_j\}$ are known instead of identified -- a setting that encompasses the majority of our examples in Section \ref{subsec:ex}.

Our estimation strategy is based on two observations that follow from our identification analysis.
First, if the parameter of interest is point identified, then it must equal the limit of expectations of a sequence of unknown functions $\{\kappa_j\}$.
Second, provided $\{\ell_j\}$ and $\mu$ are known, the functions $\{\kappa_j\}$ can often be set to equal $\kappa_j = \nu_j/\pi$ for known $\{\nu_j\}$ and $\pi = dP_{Z|X}/d\mu_{Z|X}$; see, e.g., Corollaries \ref{cor:int}, \ref{cor:discint}, and \ref{cor:discintout}.
These observations enable us to devise double robust identifying moment conditions that readily yield asymptotically normal estimators.
We next construct such estimators for functionals about types and about outcomes and characterize their semiparametric efficiency bound.

\subsection{Estimation: Types}\label{subsec:inftype}

Recall that functionals about types, as studied in Section \ref{sec:idtype}, have the structure
\begin{equation}\label{eq:inftype1}
\lambda_Q = \lim_{j\to\infty} E_Q[\ell_j(T^\star,X)].
\end{equation}
By Theorem \ref{th:typemain}, if $\lambda_{\tr}$ is point identified, then it must equal the limit of the expectation of functions $\{\kappa_j\}$ of $(T,Z,X)$.
Moreover, in an important class of applications, the functions $\{\kappa_j\}$ satisfy $\kappa_j = \nu_j/\pi$ for some known functions $\{\nu_j\}$ of $(T,Z,X)$.

Our estimator is based on the observation that the structure $\kappa_j = \nu_j/\pi$ with $\pi = dP_{Z|X}/d\mu_{Z|X}$ implies that for any $t\in \mathbf T$ and function $f$ we have the equality
\begin{equation*}
E_P[\kappa_j(t,Z,X)f(Z,X)] = E_{P_X}[E_{\mu_{Z|X}}[\nu_j(t,Z,X)f(Z,X)]].
\end{equation*}
Therefore, we may equivalently express the expectation of $\kappa_j(T,Z,X)$ as being equal to
\begin{align}\label{eq:inftype2}
 E_P  [\kappa_j(T,Z,X)]
=  \sum_{t\in \mathbf T} & E_P[\kappa_j(t,Z,X)(1\{T=t\}-P(T=t|Z,X))] \notag \\ & + \sum_{t\in \mathbf T} E_{P_X}[E_{\mu_{Z|X}}[\nu_j(t,Z,X)P(T=t|Z,X)]].
\end{align}
Crucially, the identifying moment in \eqref{eq:inftype2} is double robust in the sense that the equality continues to hold if for any $t\in \mathbf T$ we substitute either of the nuisance parameters $\kappa_j(t,Z,X)$ or $P(T=t|Z,X)$ with different functions of $(Z,X)$.
This double robustness readily enables estimation through a variety of plug-in machine learning methods. 
For concreteness, we follow ideas in \cite{smucler2019unifying}, \cite{chernozhukov2022locally}, and \cite{chernozhukov2022automatic} and employ an $\ell_1$-regularized double robust estimator. 

Specifically, our estimator is obtained from the following algorithm:

\noindent {\sc Step 1.} Partition $\{1,\ldots, n\}$ into $K$ subsets $\{I_k\}_{k=1}^K$, select  functions $\{b_l\}_{l=1}^p$ of $(Z,X)$ with $p$ potentially larger than $n$, and let $b(Z,X)\equiv (b_1(Z,X),\ldots, b_p(Z,X))^\prime$. The number of partitions $K$ is fixed with $n$, and usually set to five or ten. \qed

\noindent {\sc Step 2.} For each treatment value $t \in \mathbf T$ and partition $k$ compute the estimators
\begin{align}
\hat \beta_{t,k} & \in \arg\min_{\beta \in \mathbf R^p} \sum_{i \in I_{k}^c} (1\{T_i = t\} - b(Z_i,X_i)^\prime \beta)^2 + \alpha \|\beta\|_1 \label{type:alg1}\\
\hat \gamma_{t,k} & \in \arg\min_{\gamma \in \mathbf R^p} \sum_{i \in I_k^c} \frac{1}{2}(b(Z_i,X_i)^\prime \gamma)^2 - E_{\mu_{Z|X}} [\nu_j(t,Z,X_i)b(Z,X_i)^\prime \gamma] + \alpha \|\gamma\|_1, \label{type:alg2}
\end{align}
where $I_k^c = \{1,\ldots, n\}\setminus I_k$. We note that the penalty $\alpha$ need not be the same in both estimation problems, but the set of functions $\{b_l\}_{l=1}^p$ must be the same. 
The penalty $\alpha$ can be selected in a data-drive way such as, e.g., cross-validation. \qed

\noindent {\sc Step 3.} For each $k$, let $|I_k|$ denote the number of observations in $I_k$ and set $\hat \lambda_k$ to equal
$$\hat \lambda_{k} \equiv \frac{1}{|I_k|} \sum_{i \in I_k} \sum_{t\in \mathbf T} b(Z_i,X_i)^\prime \hat \gamma_{t,k}(1\{T_i = t\} - b(Z_i,X_i)^\prime \hat \beta_{t,k}) + E_{\mu_{Z|X}}[\nu_j(t,Z,X_i)b(Z,X_i)^\prime \hat \beta_{t,k}].$$
Note that in computing the estimator $\hat \lambda_{k}$ we employ estimators $\hat \gamma_{t,k}$ and $\hat \beta_{t,k}$ that are obtained from data not in partition $I_k$ (see Step 2). \qed

\noindent {\sc Step 4.} The estimator for $\lambda_{\tr}$ is given by $\hat \lambda \equiv \sum_k \hat \lambda_{k}|I_k|/n$ -- i.e.\ $\hat \lambda$ is simply the weighted average of the estimators $\{\hat \lambda_{k}\}_{k=1}^K$ obtained from each partition $I_k$. \qed

Intuitively, we may view $b(Z,X)^\prime \hat \beta_{t,k}$ and $b(Z,X)^\prime \hat \gamma_{t,k}$ as estimators for the nuisance parameters $P(T=t|Z,X)$ and $\kappa_j(t,Z,X)$ and $\hat \lambda$ as a plug-in estimator based on \eqref{eq:inftype2}.
The sample splitting in Step 1 is important for relaxing our assumptions, though we note $\hat \lambda$ will remain asymptotically normal without sample splitting provided we impose sufficiently strong sparsity requirements. 
We also note that we may substitute $b(Z,X)^\prime \hat \beta_{t,k}$ with certain nonlinear estimators, such as logistic regression, and still obtain a double robust estimator for $\lambda_{\tr}$ provided $b(Z,X)^\prime \hat \gamma_{t,k}$ is modified accordingly as well \citep{smucler2019unifying, chernozhukov2022locally}. 
Alternatively, in Step 3 we may substitute $b(Z,X)^\prime \hat \beta_{t,k}$ and $b(Z,X)^\prime \hat \gamma_{t,k}$ with any suitably convergent machine learning estimators for the nuisance parameters \citep{chernozhukov2018double}.
The resulting estimator for $\lambda_{\tr}$, however, may fail to be double robust in the sense that inference based on it can be invalid if any of the nuisance parameter estimators is inconsistent.

In order to state sufficient conditions for the asymptotic normality of our estimator $\hat \lambda$, we first need to introduce some additional notation. 
To this end, we define
\begin{align*}
\beta_t& \in \arg\min_{\beta \in \mathbf R^p} E_P[(1\{T=t\} - b(Z,X)^\prime \beta)^2] \notag \\
& \gamma_t \in\arg\min_{\gamma \in \mathbf R^p} \{\frac{1}{2}E_P[(b(Z,X)^\prime \gamma)^2] - E_{P_X}[E_{\mu_{Z|X}}[\nu_j(t,Z,X)b(Z,X)^\prime \gamma]\},
\end{align*}
which are the estimands for which $\hat \beta_{t,k}$ and $\hat \gamma_{t,k}$ will be assumed to be consistent for.
We additionally denote the estimation error for $\beta_t$ and $\gamma_t$ in the prediction norm by
\begin{equation*}
r_{t}^\beta  \equiv \max_{1\leq k \leq K} \{E_P[(b(Z,X)^\prime(\hat \beta_{t,k}-\beta_t))^2]\}^{1/2} \hspace{0.3 in} 
r_{t}^\gamma  \equiv \max_{1\leq k \leq K} \{E_P[(b(Z,X)^\prime(\hat \gamma_{t,k}-\gamma_t))^2]\}^{1/2} .
\end{equation*}
The estimands $b(Z,X)^\prime \beta_t$ and $b(Z,X)^\prime \gamma_t$ are approximations to the nuisance parameters $P(T=t|Z,X)$ and $\kappa_j(t,Z,X)$, and we denote their approximation errors by
\begin{align*}
\delta_{t}^\beta & \equiv \{E_P[(P(T=t|Z,X)-b(Z,X)^\prime \beta_t)^2]\}^{1/2} \notag \\
\delta_{t}^\gamma & \equiv \{E_P[(\kappa_j(t,Z,X)-b(Z,X)^\prime \gamma_t)^2]\}^{1/2}.
\end{align*}
Finally, it will be convenient to denote the influence function of our estimator $\hat\lambda$ by
\begin{align}\label{eq:inftype3}
\psi(T,Z,X) \equiv  \sum_{t\in \mathbf T} &  b(Z,X)^\prime \gamma_t(1\{T = t\}-b(Z,X)^\prime \beta_t) \notag \\ &+ \sum_{t\in \mathbf T} E_{\mu_{Z|X}}[\nu_j(t,Z,X)b(Z,X)^\prime \beta_t]-\lambda_{\tr},
\end{align}
and to let $\sigma^2 \equiv \text{Var}_P\{\psi(T,Z,X)\}$ denote its variance.
While we have suppressed it from the notation, it is important to note that $p$ (the dimension of $b(Z,X)$) and $j$ (as indexing $\kappa_j$) can depend on $n$, and as a result so do all the terms we have defined.

Given the introduced notation, we impose the following assumptions:

\begin{assumption}\label{ass:proptype}
(i) $\{Y_i,T_i,X_i,Z_i\}_{i=1}^n$ is i.i.d.;
(ii) There are known $\{\nu_j\}\subseteq L^\infty(P_{TZX})$ such that $\kappa_j \equiv \nu_j/\pi$ satisfies $\Id(\kappa_j)\stackrel{\tau}{\rightarrow} \Lambda$;
(iii) $\mu_{Z|X}\ll P_{Z|X}$ and $\|1/\pi\|_\infty < \infty$.
\end{assumption}

\begin{assumption}\label{ass:regtype}
(i) $\max_{t} \|b^\prime \beta_t\|_{\infty}=O(1)$ and $B \equiv \max_{t} \|b^\prime \gamma_t\|_\infty\vee \|\nu_j\|_\infty < \infty$ satisfies $B\log(n) = o(\sigma \sqrt n)$;
(ii) $r_{t}^\gamma \vee B r_{t}^\beta \vee \sqrt n r_{t}^\beta r_{t}^\gamma = o_P(\sigma)$ for all $t\in \mathbf T$;
(iii) $\sqrt n \delta_{t}^\beta \delta_{t}^\gamma = o(\sigma)$ for all $t\in \mathbf T$;
(iv) $\sqrt n|\lambda_{\tr}-E_P[\kappa_j(T,Z,X)]| = o(\sigma)$;
(v) $|I_k|\asymp n$.
\end{assumption}

Assumption \ref{ass:proptype}(ii) formalizes our conditions on $\kappa_j$ which, by our identification analysis, is equivalent to the identification of $\lambda_{\tr}$ in a variety of applications.
Assumption \ref{ass:proptype}(iii) imposes that $\pi$ be bounded away from zero. 
In turn, Assumption \ref{ass:regtype} states conditions on $ \hat \beta_{t,k}$ and $ \hat \gamma_{t,k}$  -- we impose high level conditions given the preponderance of results in the literature justifying these assumptions under lower level assumptions.
Specifically, Assumption \ref{ass:regtype}(ii) demands that $\hat \beta_{t,k}$ and $\hat \gamma_{t,k}$ be suitably convergent to their respective estimands in the prediction norm. 
Sufficient conditions for deriving convergence rates for $\hat \gamma_{t,k}$ can be found in \cite{chernozhukov2022automatic}, and for $\hat \beta_k$ in \cite{buhlmann2011statistics} and \cite{bartlett2012} with and without sparsity assumptions respectively.
Assumption \ref{ass:regtype}(iii) states our rate requirements on the approximation errors $\delta_{t}^\gamma$ and $\delta_{t}^\beta$.
The rate is double robust in that Assumption \ref{ass:regtype}(iii) can hold even if one of the estimands is not consistent for its corresponding nuisance parameter.
Finally, Assumption \ref{ass:regtype}(iv) is automatically satisfied if $\{\kappa_j\}$ does not depend on $j$ (as in Lemma \ref{lm:warmup}) and may be viewed as an undersmoothing requirement otherwise.

\begin{remark}\label{rm:num} \rm
Sufficient conditions for Assumption \ref{ass:regtype}(iv) can be analytically derived in certain applications in which $\kappa_j$ depends on $j$; see, e.g., our discussion of Example \ref{ex:pm} below.
Alternatively, a numerical bound can be obtained through the inequality
\begin{multline*}
|E_{\tr}[\ell(T^*,X) - \kappa_j(T,Z,X)]| \\ 
\leq \|\frac{d{\tr}}{d\mu}\|_{\infty} \times E_{\mu}[|\ell(T^*,X)-\sum_{t\in\mathbf T} E_{\mu_{Z|X}}[\nu_j(t,Z,X)1\{T^*(Z)=t\}]|].
\end{multline*}
In particular, if $\nu_j$ is computed through, e.g., Corollary \ref{cor:int} then we may set it to control the bias in Assumption \ref{ass:model}(iii)  given a sup-norm bound on $d\tr/d\mu$. \qed 
\end{remark}

Our next result establishes the asymptotic normality of our estimator.

\begin{theorem}\label{th:typenorm}
Let Assumptions \ref{ass:setup}, \ref{ass:model}, \ref{ass:genpar}, \ref{ass:proptype}, \ref{ass:regtype} hold, $\lambda_Q$ and $\psi$ be as defined in \eqref{eq:inftype1} and \eqref{eq:inftype3}, and $\sigma^2 \equiv \text{\rm Var}_P\{\psi(T,Z,X)\}$.
Then, there is a $\mathbb Z\sim N(0,1)$ satisfying
\begin{equation}\label{th:typenormdisp}
\frac{\sqrt n}{\sigma}(\hat \lambda - \lambda_{\tr}) = \frac{1}{\sqrt n \sigma}\sum_{i=1}^n \psi(T_i,Z_i,X_i) + o_P(1) = \mathbb Z+ o_P(1).
\end{equation}
\end{theorem}

For inference, we will rely on a multiplier bootstrap procedure that approximates the distribution in Theorem \ref{th:typenorm} and further extends to vector valued parameters and their nonlinear functionals. 
Specifically, for each $k$ we define an estimator for $\psi$ by setting
\begin{align}\label{eq:inftype4}
\hat \psi_k(T,Z,X) \equiv 
\sum_{t\in \mathbf T} & b(Z,X)^\prime \hat \gamma_{t,k}(1\{T=t\}-b(Z,X)^\prime \hat \beta_{t,k}) \notag \\ & + \sum_{t\in \mathbf T} E_{\mu_{Z|X}}[\nu_j(t,Z,X)b(Z,X)^\prime \hat \beta_{t,k}] - \hat \lambda.
\end{align}
Our ``bootstrapped" estimator $\hat \lambda^*$ is then obtained by employing $\hat \psi_k$ and an i.i.d.\ sample $\{W_i\}_{i=1}^n$ of standard normal weights independent of the data to perturb $\hat \lambda$ according to
\begin{equation}\label{eq:inftype5}
\hat \lambda^* \equiv \hat \lambda + \frac{1}{n}\sum_{k=1}^K \sum_{i\in I_k} W_i\hat \psi_k(T_i,Z_i,X_i).
\end{equation}
We employ standard normal weights $W$ to simplify our technical arguments, though under appropriate moment restrictions the proposed bootstrap remains valid provided $W$ satisfies $E[W]= 0$ and $E[W^2] = 1$ -- e.g.,\ for $W$ set to be Rademacher weights.

The next result establishes the validity of the proposed bootstrap.

\begin{theorem}\label{th:typeboot}
Let the conditions of Theorem \ref{th:typenorm} hold and $\{W_i\}_{i=1}^n$ be i.i.d.\ standard normal random variables independent of $\{Y_i,T_i,Z_i,X_i\}_{i=1}^n$. 
Then, there exists a standard normal random variable $\mathbb Z^*$ independent of $\{Y_i,T_i,Z_i,X_i\}_{i=1}^n$ and satisfying
\begin{equation}\label{th:typebootdisp}
\frac{\sqrt n}{\sigma}(\hat \lambda^* - \hat \lambda) = \frac{1}{\sqrt n \sigma}\sum_{i=1}^n W_i \psi(T_i,Z_i,X_i) + o_P(1) = \mathbb Z^* +o_P(1).
\end{equation}    
\end{theorem}

Theorems \ref{th:typenorm} and \ref{th:typeboot} justify employing the distribution of $ (\hat \lambda^*-\hat \lambda)$ conditional on the data as an approximation to the distribution of $(\hat \lambda - \lambda_{\tr})$.
For instance, in order to obtain a two sided confidence region we would: 
(i) Draw $1\leq b \leq B$ samples $\{W_i^{(b)}\}_{i=1}^n$ of the weights independently of the data; 
(ii) Employ each sample $\{W_i^{(b)}\}_{i=1}^n$ to obtain a bootstrap estimator $\hat\lambda^{*(b)}$ through  \eqref{eq:inftype5}; 
(iii) Compute the $1-\alpha$ quantile $\hat c_\alpha$ of $\{|\hat \lambda^{*(b)} - \hat \lambda|\}_{b=1}^B$; and 
(iv) Set the two sided confidence region to equal $\hat \lambda \pm \hat c_\alpha$.

\subsubsection{Examples Revisited}

We next illustrate our results in the context of Examples \ref{ex:disc} and \ref{ex:pm}, focusing our discussion on the computation of the terms in our algorithm that are model specific.

\noindent {\bf Example \ref{ex:disc} (cont.)} 
Suppose $\ell$ is a known function of $(T^*,X)$ and recall that we showed the expectation of $\ell(T^*,X)$ is identified if and only if with probability one
\begin{equation}\label{ex:disc7}
\min_{\{s_j\}_{j=1}^q\subset \mathbf R^d} \sum_{i=1}^r (\ell(t_i^*,X) - \sum_{j=1}^q s_j^\prime \omega_j(t_i^*))^2 = 0,
\end{equation}
where $\omega_j(t^*) \equiv (1\{t^*(z_j) = t_1\},\ldots, 1\{t^*(z_j) = t_d\})$.
To implement our estimator in this context, let $(s_1(X),\ldots, s_q(X))$ be a minimizer of \eqref{ex:disc7} and $s_{jm}(X)$ denote the $m^{th}$ coordinate of $s_j(X)$.
It is then possible to show that $\nu_j$ does not depend on $j$ and
\begin{equation*}
E_{\mu_{Z|X}}[\nu(t_m,Z,X_i)b(Z,X_i)] = \sum_{j=1}^q s_{jm}(X_i)b(z_j,X_i) .
\end{equation*}
This construction yields a double robust estimator of, e.g., $\tr(T^* \in A)$ for any $A$ for which the probability is identified (i.e.\ \eqref{ex:disc7} holds with $\ell(t^*,X) = 1\{t^*\in A\}$). \qed

\noindent {\bf Example \ref{ex:pm} (cont.)} 
We focus on discussing estimators for the expectation of a function $\ell$ of $(K_c^*,X)$ for some $c\in \{0,1\}$ -- estimators for the expectation of a function of $(K_0^*,K_1^*,X)$ then readily follow from our identification results (see \eqref{ex:pm6}).
To this end, suppose $\ell(K_c^*,X)$ is differentiable in $K_c^*$ on $(\underline{\bf w},\bar {\bf w})$ with derivative $\ell^\prime(K_c^*,X)$ and that $\ell(K_c^*,X) = 0$ whenever $K_c^*\in \{\underline{\bf w},\bar {\bf w},+\infty\}$.
It can then be shown that we may  set
\begin{equation}\label{ex:pm8}
E_{\mu_{Z|X}}[\nu(1,C,W,X)b(C,W,X_i)] = \int_{\underline{\bf w}}^{\bar{\bf w}}\ell^\prime(w,X_i)b(c,w,X_i)  dw 
\end{equation}
and $\nu(0,C,W,X) = 0$ (see \eqref{ex:pm4}).
Expectations of more general functions of $(K_c^*,X)$ can in turn be estimated by approximating them with differentiable functions.
For example, the expectation of $\ell(K_c^*) = 1\{a\leq K_c^*\leq b\}$ with $\underline{\bf w} < a < b < \bar{\bf w}$ can be approximated by the expectation of $\ell_j(K_c^*) = F((b-K_c^*)/h_j) - F((a-K_c^*)/h_j)$ for some $h_j \downarrow 0$ and $F$ the c.d.f.\ of a compactly supported mean zero continuous random variable.
In this case, we may again set $\nu(0,C,W,X) = 0$ while \eqref{ex:pm8} becomes 
\begin{equation}\label{ex:pm9}
E_{\mu_{Z|X}}[\nu_j(1,C,W,X)b(C,W,X_i)] = \int_{\underline{\bf w}}^{\bar{\bf w}}\frac{1}{h_j}(F^\prime(\frac{a-w}{h_j}) - F^\prime(\frac{b-w}{h_j}))b(c,w,X_i) dw
\end{equation}
and, under regularity conditions, $B\asymp \sigma^2 \asymp 1/h_j$ and $|\lambda_{\tr} - E_P[\kappa_j(T,Z,X)]| = O(h^2)$ so that Assumption \ref{ass:regtype} requires us to set $\log^2(n)/(nh_j) =o(1)$ and $nh_j^5 =o(1)$. \qed

\subsection{Estimation: Outcomes}\label{subsec:infout}

We next turn to developing an estimator for functionals about outcomes, as studied in Section \ref{sec:idout}.
Recall that these functionals are characterized by having the structure
\begin{equation}\label{eq:infout1}
\lambda_Q = \lim_{j\to \infty} E_Q[\rho(Y^*(t))\ell_j(T^\star,X)]
\end{equation}
for some known $\rho$ and $t\in \mathbf T$.
If $\lambda_{\tr}$ is identified, then by Theorem \ref{th:outmain} it must equal the limit of the expectation of $\{\rho \kappa_j\}$ for some sequence of functions $\{\kappa_j\}$ of $(T,Z,X)$.
While the functions $\{\kappa_j\}$ are unknown, in a leading set of applications they satisfy
\begin{equation}\label{eq:infout2}
\kappa_j(T,Z,X) = 1\{T=t\}\frac{\nu_j(Z,X)}{\pi(Z,X)}
\end{equation}
for known functions $\{\nu_j\}$; see, e.g., Corollary \ref{cor:discintout} and subsequent discussion.
Due to the similarities between the identifying equations for functionals about types and outcomes, we are able to obtain estimators for functionals about outcomes by slightly modifying our preceding analysis for types.
As a result, in what follows we keep exposition brief though note that the discussion and remarks of Section \ref{subsec:inftype} apply to this section as well.

Our estimator for functionals about outcomes is obtained though the algorithm:

\noindent {\sc Step 1.} Partition $\{1,\ldots, n\}$ into $K$ subsets $\{I_k\}_{k=1}^K$, select a set of functions $\{b_l\}_{l=1}^p$ of $(Z,X)$, and let $b(Z,X)\equiv (b_1(Z,X),\ldots, b_p(Z,X))^\prime$. \qed

\noindent {\sc Step 2.} For each partition $1\leq k\leq K$ compute the following two estimators 
\begin{align}
\hat \beta_{k} & \in \arg\min_{\beta \in \mathbf R^q} \sum_{i \in I_{k}^c} (\rho(Y_i)1\{T_i = t\} - b(Z_i,X_i)^\prime \beta)^2 + \alpha \|\beta\|_1 \label{out:alg1}\\
\hat \gamma_{k} & \in \arg\min_{\gamma \in \mathbf R^q} \sum_{i \in I_k^c}\{\frac{1}{2}(b(Z_i,X_i)^\prime \gamma)^2 - E_{\mu_{Z|X}} [\nu_j(Z,X_i)b(Z,X_i)^\prime \gamma]\} + \alpha \|\gamma\|_1, \label{out:alg2}
\end{align}
where the set of functions $\{b_l\}_{l=1}^p$ must be the same in both estimation problems. \qed

\noindent {\sc Step 3.} For each partition $1\leq k \leq K$ compute the plug-in estimator $\hat \lambda_k$ given by
$$\hat \lambda_{k} \equiv \frac{1}{|I_k|} \sum_{i \in I_k} b(Z_i,X_i)^\prime \hat \gamma_{k}(\rho(Y_i)1\{T_i = t\} - b(Z_i,X_i)^\prime \hat \beta_{k}) + E_{\mu_{Z|X}}[\nu_j(Z,X_i)b(Z,X_i)^\prime \hat \beta_{k}]$$
where $|I_k|$ denotes number of observations in the partition $I_k$. \qed

\noindent {\sc Step 4.} Compute $\hat \lambda \equiv \sum_k \hat \lambda_k |I_k|/n$ as the final estimator for $\lambda_{\tr}$. \qed

The asymptotic properties of $\hat \lambda$ can unsurprisingly be established under similar conditions to those employed in Section \ref{subsec:inftype}.
Adjusting notation, we now define the estimands
\begin{align*}
\beta & \equiv \arg\min_{\beta \in \mathbf R^p} E_P[(\rho(Y)1\{T=t\} -b(Z,X)^\prime \beta)^2] \\
\gamma & \equiv \arg\min_{\gamma \in \mathbf R^p} \{\frac{1}{2}E_P[(b(Z,X)^\prime \gamma)^2] - E_{P_X}[E_{\mu_{Z|X}}[\nu_j(Z,X)b(Z,X)^\prime \gamma]]\}
\end{align*}
and denote the convergence rates for $\hat \beta_k$ and $\hat \gamma_k$ to $\beta$ and $\gamma$ in the prediction norm by
\begin{equation*}
r^\beta  \equiv \max_{1\leq k \leq K} \{E_P[(b(Z,X)^\prime(\hat \beta_k - \beta))^2]\}^{1/2} \hspace{0.4 in}
r^\gamma  \equiv \max_{1\leq k \leq K} \{E_P[(b(Z,X)^\prime(\hat \gamma_k - \gamma))^2]\}^{1/2}.
\end{equation*}
The functions $b(Z,X)^\prime \beta$ and $b(Z,X)^\prime \gamma$ represent approximations to  $E[\rho(Y)1\{T=t\}|Z,X]$ and $\nu_j(Z,X)/\pi(Z,X)$ respectively, and we denote their approximation errors by
\begin{align*}
\delta^\beta & \equiv \{E_P[(E_P[\rho(Y)1\{T=t\}|Z,X] - b(Z,X)^\prime \beta)^2]\}^{1/2} \\
\delta^\gamma & \equiv \{E_P[(\frac{\nu_j(Z,X)}{\pi(Z,X)} - b(Z,X)^\prime \gamma)^2]\}^{1/2}.
\end{align*}
Finally, we introduce the influence function for our estimator, which here is given by
\begin{multline}\label{eq:infout3}
\psi(Y,T,Z,X) \\ 
\equiv b(Z,X)^\prime \gamma(1\{T=t\}\rho(Y) - b(Z,X)^\prime \beta) + E_{\mu_{Z|X}}[\nu_j(Z,X)b(Z,X)^\prime \beta] - \lambda_{\tr},
\end{multline}
and set $\sigma^2 \equiv \text{Var}_P\{\psi(Y,T,Z,X)\}$.
We again note that the introduced parameters are allowed to depend on $n$, though we suppressed such dependence from the notation.

The following assumptions suffice for estalibshing the asymptotic properties of $\hat \lambda$.

\begin{assumption}\label{ass:propout}
(i) $\{Y_i,T_i,X_i,Z_i\}_{i=1}^n$ is i.i.d.;
(ii) There are known $\{\nu_j\}\subseteq L^\infty(P_{ZX})$ such that $\kappa_j$ given by \eqref{eq:infout2} satisfies $\Id(\kappa_j)\stackrel{\tau}{\rightarrow} \Lambda$;
(iii) $\mu_{Z|X}\ll P_{Z|X}$ and $\|1/\pi\|_\infty < \infty$.
\end{assumption}

\begin{assumption}\label{ass:regout}
(i) $\|\rho\|_\infty < \infty$, $\|b^\prime \beta\|_{\infty}=O(1)$, and $B \equiv \|b^\prime \gamma\|_\infty\vee \|\nu_j\|_\infty < \infty$ satisfies $B\log(n) = o(\sigma \sqrt n)$;
(ii) $r^\gamma \vee B r^\beta \vee \sqrt n r^\beta r^\gamma = o_P(\sigma)$;
(iii) $\sqrt n \delta^\beta \delta^\gamma = o(\sigma)$;
(iv) $\sqrt n|\lambda_{\tr}-E_P[\rho(Y)\kappa_j(T,Z,X)]| = o(\sigma)$;
(v) $|I_k|\asymp n$.
\end{assumption}

Assumptions \ref{ass:propout} and \ref{ass:regout} are simply adaptations of Assumptions \ref{ass:proptype} and \ref{ass:regtype} to the present estimation problem.
The most substantive difference between these sets of assumptions is that Assumption \ref{ass:regout}(i) requires $\rho$ to be bounded -- a condition that enables us to establish our results employing convergence rates in the prediction norm.
While we impose this requirement for simplicity, we note that it may be relaxed by strengthening the norm under which we require $\hat \beta_k$ and $\hat \gamma_k$ to converge to $\beta$ and $\gamma$.

The next result establishes the asymptotic normality of our estimator.

\begin{theorem}\label{th:outnorm}
Let Assumptions \ref{ass:setup}, \ref{ass:model}, \ref{ass:genpar}, \ref{ass:propout}, \ref{ass:regout} hold, $\lambda_Q$ and $\psi$ be as defined in \eqref{eq:infout1} and \eqref{eq:infout3}, and $\sigma^2 \equiv \text{\rm Var}_P\{\psi(Y,T,Z,X)\}$.
Then, there is a $\mathbb Z\sim N(0,1)$ satisfying
\begin{equation}\label{th:typeoutdisp}
\frac{\sqrt n}{\sigma}(\hat \lambda - \lambda_{\tr}) = \frac{1}{\sqrt n \sigma}\sum_{i=1}^n \psi(Y_i,T_i,Z_i,X_i) + o_P(1) = \mathbb Z+ o_P(1).
\end{equation}
\end{theorem}

For inference we again rely on the multiplier bootstrap.
Specifically, for each $1\leq k\leq K$ in our partition we define an estimator for the influence function by setting
\begin{multline}\label{eq:infout4}
\hat \psi_k(Y,T,Z,X) \\ \equiv 
b(Z,X)^\prime \hat \gamma_{k}(1\{T=t\}\rho(Y)-b(Z,X)^\prime \hat \beta_{k}) + E_{\mu_{Z|X}}[\nu_j(Z,X)b(Z,X)^\prime \hat \beta_{k}] - \hat \lambda.
\end{multline}
For $\{W_i\}_{i=1}^n$ an i.i.d.\ sample of standard normal random variables independent of the data, we then obtain a ``bootstrapped" analogue $\hat \lambda^*$ to $\hat \lambda$ by setting 
$$\hat \lambda^* \equiv \hat \lambda +\frac{1}{n}\sum_{k=1}^K \sum_{i\in I_k} W_i \hat \psi_k(Y_i,T_i,Z_i,X_i).$$

Our next result establishes the validity of the proposed bootstrap procedure.

\begin{theorem}\label{th:outboot}
Let the conditions of Theorem \ref{th:outnorm} hold and $\{W_i\}_{i=1}^n$ be i.i.d.\ standard normal random variables independent of $\{Y_i,T_i,Z_i,X_i\}_{i=1}^n$. 
Then, there exists a standard normal random variable $\mathbb Z^*$ independent of $\{Y_i,T_i,Z_i,X_i\}_{i=1}^n$ and satisfying
\begin{equation}\label{th:outbootdisp}
\frac{\sqrt n}{\sigma}(\hat \lambda^* - \hat \lambda) = \frac{1}{\sqrt n \sigma}\sum_{i=1}^n W_i \psi(Y_i,T_i,Z_i,X_i) + o_P(1) = \mathbb Z^* +o_P(1).
\end{equation}    
\end{theorem}

Theorems \ref{th:outnorm} and \ref{th:outboot} justify employing the proposed bootstrap to conduct inference on functionals about outcomes.
Moreover, together with Theorems \ref{th:typenorm}, \ref{th:typeboot}, and the Delta method, they also justify employing the bootstrap to conduct inference on parameters such as, e.g.,  conditional expectations of potential outcomes given types and of types given covariates.\footnote{See Lemma \ref{lm:delta} in the Appendix for a version of the Delta method suitable for our setting.}
Specifically, such parameters have the structure
\begin{equation*}
F(\lambda_{\tr1},\ldots, \lambda_{\tr q}) 
\end{equation*}
where $F:\mathbf R^q\to \mathbf R$ is a known differentiable function and each $\lambda_{\tr j}\in \mathbf R$ is a functional about types or outcomes.\footnote{E.g., for some event $A$ set $\lambda_{\tr 1} = E_{\tr}[Y^*(t)1\{T^*\in A\}]$, $\lambda_{\tr 2} = E_{\tr}[1\{T^*\in A\}]$, and $F(\lambda_{\tr 1},\lambda_{\tr 2}) = \lambda_{\tr 1}/\lambda_{\tr 2}$ to obtain $F(\lambda_{\tr 1}/\lambda_{\tr 2}) = E_{\tr}[Y^*(t)|T^*\in A]$.}
For instance, to obtain a two sided confidence region we would: 
(i) Compute estimators $(\hat \lambda_1,\ldots, \hat \lambda_q)$ for $(\lambda_{\tr 1},\ldots, \lambda_{\tr q})$ using our results for types or outcomes; 
(ii) Draw $B$ samples $\{W_i^{(b)}\}_{i=1}^n$ of weights independent of the data; 
(iii) Employ each sample $\{W_i^{(b)}\}_{i=1}^n$ to obtain bootstrap estimators $(\hat \lambda_1^{*(b)},\ldots, \hat \lambda_q^{*(b)})$ using our results for types or outcomes; 
(iv) Set $\hat c_\alpha$ to equal the $1-\alpha$ quantile of $\{|F(\hat \lambda_1,\ldots, \hat \lambda_q) - F(\hat \lambda_1^{*(b)},\ldots,\hat \lambda_q^{*(b)})|\}_{b=1}^B$ across the $B$ samples; and
(v) Report $F(\hat \lambda_1,\ldots, \hat \lambda_q) \pm \hat c_\alpha$ as a two sided confidence region.
Similarly, our results also allow us to conduct inference on directionally (but not fully) differentiable functionals of $(\lambda_{\tr 1},\ldots, \lambda_{\tr q})$ by relying on the framework developed in \cite{fang2018inference}.

\subsubsection{Examples Revisited}


\noindent {\bf Example \ref{ex:disc} (cont.)} We previously established that, for a known function $\ell$ of $(T^*,X)$, the expectation of $\rho(Y^*(t))\ell(T^*,X)$ is identified if and only if
\begin{equation}\label{ex:disc9}
\min_{\{s_j\}_{j=1}^q\subset \mathbf R} \sum_{i=1}^r (\ell(t_i^*,X) - \sum_{j=1}^q s_j1\{t_i^*(z_j) = t\})^2 = 0
\end{equation}
with probability one (over $X$). 
In order to estimate an identified functional about outcomes (i.e.\ one for which \eqref{ex:disc9} holds), we may implement our estimator with 
\begin{equation*}
E_{\mu_{Z|X}}[\nu(Z,X_i)b(Z,X_i)] = \sum_{j=1}^q s_{j}(X_i)b(z_j,X_i) ,
\end{equation*}
where $(s_1(X),\ldots, s_q(X))$ is any minimizer of \eqref{ex:disc9}.
Hence, we may for example conduct inference on $E_{\tr}[Y^*(t)1\{T^*\in A\}]$ (provided $\ell(t^*,X) = 1\{t^*\in A\}$ satisfies \eqref{ex:disc9}) or, in combination with our results on functionals about types, on $E_{\tr}[Y^*(t)|T^*\in A]$. \qed

\noindent {\bf Example \ref{ex:pm} (cont.)} 
When illustrating the implementation of our estimator for functionals about types in this example we employed  functions $\kappa_j$ with the structure $\kappa_j(T,Z,X) = 1\{T=1\}\nu_j(Z,X)/\pi(Z,X)$.
Hence, the same $\nu_j$ can be employed to estimate functionals about $Y^*(1)$ -- e.g., to estimate $E_{\tr}[\rho(Y^*(1))1\{a\leq K_c^*\leq b\}]$ we may employ \eqref{ex:pm9}.
Similarly, to estimate $E_{\tr}[\rho(Y^*(0))1\{a\leq K_c^*\leq b\}]$ we may set
\begin{equation*}
    E_{\mu_{Z|X}}[\nu_j(C,W,X_i)b(C,W,X_i)] = \int_{\underline{\bf w}}^{\bar{\bf w}}\frac{1}{h_j}(F^\prime(\frac{b-w}{h_j}) - F^\prime(\frac{a-w}{h_j}))b(c,w,X_i) dw,
\end{equation*}
and by combining estimators we may conduct inference on average treatment effects for individuals with $K_c^*\in [a,b]$.
More generally, our results enable us to conduct inference on average treatment effects for groups determined by $(K_0^*,K_1^*)$ as in, e.g.,  \eqref{ex:pm8p5}. \qed

\subsection{Efficiency Bound}\label{subsec:eff}

We conclude this section by deriving the semiparametric efficiency bound for the estimation problems studied in Sections \ref{subsec:inftype} and \ref{subsec:infout} that required $\mu$ to be known (instead of identified).
To this end, we first introduce a series of definitions that are standard in the literature on semiparametric efficiency \citep{bickel:klaassen:ritov:wellner}.

\begin{definition} \label{def:path} \rm
A \emph{path} $\eta \mapsto Q_{\eta,g}$ is a function defined on $[0,1)$ such that $Q_{\eta,g}$ is a probability distribution on $\mathbf Y^\star\times\mathbf T^\star\times\mathbf Z\times \mathbf X$ satisfying $Q_{\eta,g} \ll \mu$ for every $\eta$ and 
\begin{equation}\label{def:path1}
\lim_{\eta \to 0}\int (\frac{1}{\eta}(\frac{dQ_{\eta,g}^{1/2}}{d\mu} - \frac{dQ_{0,g}^{1/2}}{d\mu}) - \frac{1}{2}g \frac{dQ_{0,g}^{1/2}}{d\mu})^2d\mu = 0.
\end{equation}
The function $g\in L^2(Q_{0,g})$ is called the \emph{score} of the path $\eta \mapsto Q_{\eta,g}$. \qed
\end{definition}

\begin{definition} \label{def:submodel} \rm
We say that a path $\eta \mapsto Q_{\eta,g}$ is a \emph{submodel} if: (i) $(Y^\star,T^\star)\indep Z|X$ under $Q_{\eta,g}$ for all $\eta \in [0,1)$, and (ii) $Q_{0,g} \in \Theta_0$. \qed
\end{definition}

A \emph{path} is simply a ``smooth" one dimensional parametrization of distributions for random variables $(Y^\star,T^\star,Z,X)$.
We emphasize that in Definition \ref{def:path} we are relying on the fact that $\mu$ is known and is therefore fixed along the path.
A \emph{submodel} is a path that in addition: (i) Satisfies the requirements of our model -- i.e.\ $Q_{\eta,g}\ll \mu$ and $(Y^\star,T^\star)\indep Z|X$ under $Q_{\eta,g}$; and (ii) Induces the distribution $P$ on $(Y,T,Z,X)$ at $\eta = 0$ -- i.e.\ $Q_{0,g}$ is observationally equivalent to $\tr$.
We note that we do not require that the path satisfy Assumption \ref{ass:model}(iii).
In this regard, our analysis concerns applications in which the conditions encoded in $\restr$ are not informative or we do not want to use such information in estimation. 
In applications in which $\restr$ encodes regularity conditions, such as in the majority of the examples in Section \ref{subsec:ex}, it is often possible to establish that Assumption \ref{ass:genpar}(iii) implies that Assumption \ref{ass:model}(iii) is uninformative, though such arguments rely on the specific choice of $\restr$.

For any $\eta$, a distribution $Q_{\eta,g}$ for $(Y^\star,T^\star,Z,X)$ induces a distribution for $(Y,T,Z,X)$ through the relation $(Y,T,Z,X) = (Y^\star(T),T^\star(Z),Z,X)$.
As a result, each submodel $\eta \mapsto Q_{\eta,g}$ induces a path $\eta\mapsto P_{\eta,s}$ of probability distributions for $(Y,T,Z,X)$ with a score that we denote by $s$ -- i.e.\ the map $\eta \mapsto P_{\eta,s}$ satisfies smoothness requirements analogous to those imposed in \eqref{def:path1} and by construction $P_{0,s} = P$ \citep{le1988preservation}.
The resulting set of scores $s$ that can be produced in this manner generate the so-called tangent space for our model, which plays a crucial role in characterizing semiparametric efficiency bounds -- see Theorem \ref{th:eff} in the appendix for a characterization of the tangent space that may be of independent interest.

\begin{remark}\rm \label{rm:Tcorrect}
By construction, every path $\eta \mapsto P_{\eta,s}$ we consider satisfies the restrictions of our model.
This approach contrasts with, for instance, \cite{frolich2007nonparametric} who does not impose that $\eta \mapsto P_{\eta,s}$ be generated by an underlying path $\eta \mapsto Q_{\eta,g}$ satisfying the restrictions of the model.
Nonetheless, the efficiency bound of \cite{frolich2007nonparametric} is correct because in the model he examines $P$ is just identified in the sense of \cite{chen2018overidentification}.
We emphasize, however, than in models in which $P$ is overidentified, neglecting to impose the restrictions of the model can lead to incorrect efficiency bounds. \qed
\end{remark}

Our first result derives the semiparametric efficiency bound for estimating $\lambda_{\tr}$ when
\begin{equation}\label{eq:eff1}
\lambda_{Q} = E_{Q}[\ell(Y^\star,T^\star,X)]
\end{equation}
with $\ell$ a known function.
Following \cite{bickel:klaassen:ritov:wellner}, for any submodel $\eta \mapsto Q_{\eta,g}$ inducing a path $\eta \mapsto P_{\eta,s}$ we define the \emph{information bound} for estimating $\lambda_{\tr}$ by
\begin{equation}\label{eq:eff2}
    I^{-1}(Q_{\cdot,g}) \equiv \{\frac{\partial}{\partial \eta} \lambda_{Q_{\eta,g}}\Big|_{\eta = 0}\}^2 \times \{E_P[s^2(Y,T,Z,X)]\}^{-1}.
\end{equation}
Intuitively, the information bound $I^{-1}(Q_{\cdot,g})$ is the asymptotic variance of the maximum likelihood estimator for $\lambda_{\tr}$ in the parametric submodel $\eta \mapsto Q_{\eta,g}$.
We note that in order for $I^{-1}(Q_{\cdot,g})$ to be well defined, $\eta \mapsto Q_{\eta,g}$ must be regular in the sense that it induces a path $\eta \mapsto P_{\eta,s}$ whose score has positive variance and hence has positive Fisher information.
The semiparametric efficiency bound for estimating $\lambda_{\tr}$ is then defined as
\begin{equation}\label{eq:eff3}
    I^{-1} \equiv \sup_{Q_{\cdot,g}} I^{-1}(Q_{\cdot,g}),
\end{equation}
where the supremum is taken over all submodels $\eta \mapsto Q_{\eta,g}$ for which $I^{-1}(Q_{\cdot,g})$ is well defined (i.e.\ the Fisher information of the submodel is positive).

As a final of notation we introduce a map $\Eff$ mapping functions $s$ of the observables $(Y,T,Z,X)$ to functions $\Eff(s)$ of $(Y^*,T^*,X)$ by setting 
\begin{equation}\label{def:Eff}
\Eff(s)\equiv \sum_{t\in \mathbf T} E_{P_{Z|X}}[s(Y^\star(t),t,Z,X)1\{T^\star(Z) = t\}]-  E_{P}[s(Y,T,Z,X)|X]. 
\end{equation}
The null space of $\Eff$, defined as $N(\Eff)  \equiv \{s \in L^2(P) : \|\Eff(s)\|_{\dom,2} = 0\}$, and its orthocomplement $[N(\Eff)]^\perp  \equiv \{s \in L^2(P) : \langle s, \tilde s\rangle_P = 0 \text{ for all } \tilde s \in N(\Eff)\}$, play a crucial role in our next result characterizing the semiparametric efficiency bound for $\lambda_{\tr}$.

\begin{theorem}\label{th:effmain}
Let Assumptions \ref{ass:setup} and \ref{ass:model} hold, $\mu$ be known, $\lambda_Q \equiv E_Q[\ell(Y^*,T^*,X)]$ for some known bounded $\ell$, and $\lambda_{\tr}$ be identified. Then the following hold:
\vspace{-0.1 in}
\begin{packed_enum}
    \item[(i)] Suppose $\dom(\Upsilon(\kappa) = \ell) = 1$ for some $\kappa \in L^2(P)$ and let $\varphi$ denote the projection of $\kappa$ onto  $[N(\Eff)]^\perp$. Then: $I^{-1} = {\rm Var}_P\{\varphi(Y,T,Z,X)\} + \text{\rm Var}_P\{E_P[\kappa(Y,T,Z,X)|X]\}$.
    
    \item[(ii)] Suppose Assumption \ref{ass:genpar}(iii) holds and the projection of $\ell$ onto the $\|\cdot\|_{\dom,2}$-closure of the range of $\Id: L^2(P)\to L^2(\dom)$ is bounded. If there is no $\kappa \in L^2(P)$ satisfying $\dom(\Upsilon(\kappa) = \ell) = 1$, then it follows that $I^{-1} = \infty$.
\end{packed_enum}
\end{theorem}

Theorem \ref{th:effmain}(i) characterizes the semiparametric efficiency bound for estimating $\lambda_{\tr}$ when: (i) $\ell$ is \emph{known}, and (ii) There is a $\kappa$ such that $\dom(\Id(\kappa) = \ell) = 1$ -- i.e., $\lambda_{\tr}$ falls within the scope of Lemma \ref{lm:warmup}.
By Theorem \ref{th:genmain}, we know that $\lambda_{\tr}$ may be identified even if there is no $\kappa$ solving $\dom(\Id(\kappa) = \ell) = 1$.
Subject to an additional regularity condition, however, Theorem \ref{th:effmain}(ii) establishes that such functionals have an infinite semiparametric efficiency bound\footnote{Intuitively, the additional regularity condition enables us to show that $\ell$ belonging to the $\|\cdot\|_{\dom,1}$ closure of $\Id(L^1(P))$ implies $\ell$ also belongs to the $\|\cdot\|_{\dom,2}$ closure of $\Id(L^2(P))$.} -- a conclusion that is often interpreted as equivalent to the functional not being (regularly) estimable at the root-$n$ rate \citep{chamberlain1986asymptotic}.
In summary, we can conclude that: (i) Lemma \ref{lm:warmup} characterizes a set of functionals with a finite semiparametric efficiency bound, and (ii) Theorem \ref{th:genmain} characterizes all additional functionals that are identified, though such functionals are not root-$n$ estimable.

Theorem \ref{th:effmain}(i) both recovers previously available semiparametric efficiency bounds as special cases \citep{hahn1998role, frolich2007nonparametric} and delivers new semiparametric efficiency bounds for multiple applications (e.g., \cite{heckman1999local} and \cite{mogstad2021causal}).
In turn, Theorem \ref{th:effmain}(ii) provides, to our knowledge, the first characterization of when causal parameters are not root-$n$ estimable in these models.
For our analysis, an important implication of Theorem \ref{th:effmain} is its ability to assess whether our proposed estimators are efficient.
The next corollary accomplishes this task by providing sufficient conditions for the estimators proposed in Sections \ref{subsec:inftype} and \ref{subsec:infout} to be efficient.

\begin{corollary}\label{cor:justid}
Suppose the conditions of Theorem \ref{th:typenorm} (resp. Theorem \ref{th:outnorm}) hold with $\max_{t} \delta_t^\beta \vee \delta_t^\gamma = o(1)$ (resp. $\delta^\beta \vee \delta^\gamma = o(1)$) and the conditions of Theorem \ref{th:effmain}(i) hold with a $\kappa$ satisfying Assumption \ref{ass:proptype}(ii) (resp. Assumption \ref{ass:propout}(ii)).\vspace{-0.1 in}
\begin{packed_enum}
    \item[(i)] If $s=0$ is the only $s\in L^2(P)$ satisfying $\|\Id(s)\|_{\dom,2} = 0$ and $E_P[s(Y,T,Z,X)|Z,X] = 0$, then the estimator of Section \ref{subsec:inftype} (resp. Section \ref{subsec:infout}) attains the efficiency bound.
    
    \item[(ii)] Let Assumption \ref{ass:4out}(ii) hold, $\delta_t(T)\equiv 1\{T = t\}$ and suppose, for any $g\in L^2(P)$ and $t\in \mathbf T$, $\|\Id(g\delta_t)\|_{\dom,2} = 0$ implies $\|g\delta_t\|_{\dom,2} = 0$.  If $s\in L^2(P_{TZX})$ satisfying $\|\Id(s)\|_{\dom,2} = 0$ implies that $s\in L^2(P_{ZX})$, then it follows that the estimator of Section \ref{subsec:inftype} (resp. Section \ref{subsec:infout}) attains the efficiency bound.
\end{packed_enum}
\end{corollary}

Corollary \ref{cor:justid}(i)  provides sufficient conditions for our estimators to be efficient by ensuring $P$ is just identified in the sense of \cite{chen2018overidentification}.
In turn, Corollary \ref{cor:justid}(ii) imposes additional restrictions under which verifying whether our estimators are efficient reduces to a more stringent (hence easier to verify) condition than the one obtained in part (i).
The requirements of Corollary \ref{cor:justid} are easily verified in the examples of Section \ref{subsec:ex} to which our semiparametric efficiency analysis applies; see our discusion of Examples \ref{ex:disc} and \ref{ex:pm} below.
Moreover, we note that Corollary \ref{cor:justid} further implies that our estimators can be used to efficiently estimate parameters that are differentiable functions of multiple $\lambda_{\tr}$ with the structure in \eqref{eq:eff1} \citep{van1991efficiency}.
More generally, however, it is important to note that our estimators may fail to be efficient in applications in which $P$ is overidentified in the sense of \cite{chen2018overidentification}.

\subsubsection{Examples Revisited}


\noindent {\bf Example \ref{ex:disc} (cont.)}  
In this context, Corollary \ref{cor:justid}(ii) can be used to show that the estimators of Section \ref{subsec:inftype} and \ref{subsec:infout} are efficient provided that: (i) For all $t$, the matrix
\begin{equation}\label{ex:disc11}
\left(\begin{array}{ccc} 1\{t^*_1(z_1)=t\} & \ldots & 1\{t^*_1(z_q)=t\} \\ \vdots & \ddots & \vdots \\
1\{t^*_r(z_1)=t\} & \ldots & 1\{t^*_r(z_q)=t\}\end{array}\right)
\end{equation}
has rank $q$; and (ii) Any function $f$ of $(T,Z)$ satisfying the system of equations
\begin{equation}\label{ex:disc12}
 \sum_{i=1}^d \sum_{j=1}^q 1\{t^*_k(z_j) = t_i\}f(t_i,z_j) = 0 \text{ for all } 1\leq k \leq r   
\end{equation}
must be such that $f(t,z) = f(t^\prime,z)$ for any $t\neq t^\prime$ and any $z$.
The second requirement may be verified analytically or numerically through a linear program.
For instance, it is straightforward to analytically verify both requirements in the model of \cite{kline2016evaluating} and hence that our estimators are efficient in that application. \qed

\noindent {\bf Example \ref{ex:pm} (cont.)}  
For this application, Corollary \ref{cor:justid}(ii) can be used to show that our estimators are efficient provided the support of $K_0^*$ and $K_1^*$ contains $[\underline{\bf w},\bar {\bf w}]$ -- i.e.\ provided a marginal change in $W$ always induces some individuals into treatment.
We also note that in Section \ref{subsec:infout} we discussed estimation of parameters such as
\begin{equation}\label{ex:pm11}
E_{\tr}[\rho(Y^*(t))\ell(K_c^*,X)]
\end{equation}
and found our estimators to be root-$n$ consistent when $\ell$ is differentiable in $K_c^*$, but slower than root-$n$ consistent when we set $\ell$ to equal an indicator function.
Theorem \ref{th:effmain} provides an explanation for this difference, as it implies that \eqref{ex:pm11} has a finite efficiency bound when $\ell$ is differentiable, and an infinite one when $\ell$ is an indicator function. \qed

\section{Conclusion}

We proposed and developed a class of potential outcomes models that unifies and extends multiple identification strategies in the literature. 
By leveraging the rich structure of this class of models, we further derived widely applicable identification and estimation results.
We believe that our findings will be valuable to researchers, both in the context of existing models and in the development of novel identification strategies.

\newpage


\begin{center}
    {\LARGE  {\bf Appendix}}
\end{center}

\renewcommand{\thesection}{A.\arabic{section}}
\renewcommand{\theequation}{A.\arabic{equation}}
\renewcommand{\thelemma}{A.\arabic{lemma}}
\renewcommand{\thecorollary}{A.\arabic{corollary}}
\renewcommand{\thetheorem}{A.\arabic{theorem}}
\renewcommand{\theassumption}{A.\arabic{assumption}}
\setcounter{lemma}{0}
\setcounter{theorem}{0}
\setcounter{corollary}{0}
\setcounter{equation}{0}
\setcounter{remark}{0}
\setcounter{section}{0}
\setcounter{assumption}{0}

This Appendix contains the proofs for all the results stated in the paper.
Throughout, we employ the notation $Q_V$ to denote the marginal distribution of a random variable $V$ under $Q$ and $Q_{V|W}$ to denote the conditional distribution of $V$ given $W$ under $Q$.
When employing $\dom^{\rm it}$ (as in Section \ref{sec:idtype}) and $\dom^{\rm io}$ (as in Section \ref{sec:idout}), we implicitly assume the conditional distributions $Q_{Y^*|X}$ and $Q_{Y^*(t)|X}$ exist.
Finally, distributions $Q$ for $(Y^*,T^*,Z,X)$ are assumed to be defined on a product $\sigma$-field generated by $\mathcal F_{Y^*}\times \mathcal F_{T^*} \times \mathcal F_{Z} \times \mathcal F_{X}$, where $\mathcal F_V$ denotes the $\sigma$-field on which $Q_V$ is defined.

\vspace{0.2 in}


\noindent {\Large {\bf A.1 ~ Proofs for Section \ref{sec:id}}}\\


\noindent \emph{Proof of Lemma \ref{th:idom}.} 
We first establish the existence of the dominating measure $\dom \in \Theta_0$. 
To this end, first note that since $\mu$ is separable by Assumption \ref{ass:model}(ii), Lemma 13.14 in \cite{aliprantis:border:2006} implies that $L^1(\mu)$ is separable under $\|\cdot\|_{\mu,1}$.
Next set $D_0 \equiv \{dQ/d\mu : Q \in \Theta_0\}$ and note that Corollary 3.5 in \cite{aliprantis:border:2006}, $D_0\subset L^1(\mu)$, and $L^1(\mu)$ being separable imply $D_0$ is also separable under $\|\cdot\|_{\mu,1}$.
Hence, there exists a countable set $\mathcal D \equiv \{Q_i\}_{i=1}^\infty \subseteq \Theta_0$ such that for any $Q \in \Theta_0$ and $\epsilon > 0$ 
\begin{equation}\label{th:idom2}
\|\frac{dQ}{d\mu} - \frac{dQ_i}{d\mu}\|_{\mu,1} < \epsilon
\end{equation}
for some $Q_i \in \mathcal D$.
Next note that by Assumption \ref{ass:model}(iii), $\mathcal Q$ is a closed convex subset of a Banach space $\mathbf Q$ with norm $\|\cdot\|_{\mathbf Q}$.
For any $2\leq n < \infty$ then define
\begin{equation}\label{th:idom3}
\lambda_{in} = \left\{
\begin{array}{cl}
     1 - \sum_{i=2}^n \lambda_{in} & \text{ if } i = 1\\
     2^{-i}/\max\{1,\|dQ_i/d\mu\|_{\mathbf Q}\}  & \text{ if } 2\leq i \leq n \\
     0 & \text{ if } i > n
\end{array}
\right.
\end{equation}
and note that $\sum_{i=1}^n \lambda_{in} = 1$ and $\lambda_{in} > 0$ for any $1\leq i \leq n$ due to $\sum_{i=1}^\infty 2^{-i} = 1$.
Therefore, since $\mathcal Q$ is convex by Assumption \ref{ass:model}(iii), it follows that 
\begin{equation*}
f_n \equiv \sum_{i=1}^n \lambda_{in} \frac{dQ_i}{d\mu}
\end{equation*}
belongs to $\mathcal Q$ for all $n$.
Moreover, for any $n < m$ the triangle inequality yields that
\begin{align}\label{th:idom5}
\|f_n - f_{m}\|_{\mathbf Q} & \leq |\lambda_{1n} - \lambda_{1m}|\|\frac{dQ_1}{d\mu}\|_{\mathbf Q} + \sum_{i=n+1}^m \lambda_{im}\|\frac{dQ_i}{d\mu}\|_{\mathbf Q} \notag \\
& \leq (\lambda_{1n}-\lambda_{1m})\|\frac{dQ_1}{d\mu}\|_{\mathbf Q} + \sum_{i=n+1}^\infty \frac{1}{2^i},
\end{align}
where we employed that $\lambda_{im}\|dQ_i/d\mu\|_{\mathbf Q} \leq 2^{-i}$ and $\lambda_{1n}$ is decreasing in $n$ by \eqref{th:idom3}.
Furthermore,  since $1/2 < \lambda_{1n}$ by \eqref{th:idom3} and $\sum_{i=1}^\infty 2^{-i} = 1$, it follows from $\lambda_{1n}$ being decreasing in $n$ that the sequence $\{\lambda_{1n}\}_{n=1}^\infty$ has a limit in $\mathbf R$.
Hence, by \eqref{th:idom5} we obtain that the sequence $\{f_n\}_{n=1}^\infty$ is Cauchy in $\mathbf Q$.
By completeness of $\mathbf Q$, there therefore exists a $q_0 \in \mathbf Q$ such that $\|f_n- q_0\|_{\mathbf Q} = o(1)$ and since $\mathcal Q$ is closed in $\mathbf Q$ we obtain that $q_0 \in \mathcal Q \subseteq L^1(\mu)$.
Finally, we define $\dom$ satisfying $\dom \ll \mu$  and $d\dom/d\mu \in \mathcal Q$ by setting
\begin{equation}\label{th:idom6}
\dom(A) \equiv \int_A q_0 d\mu
\end{equation}
for any measurable set $A$.
Since by Assumption \ref{ass:model}(iii) we have $\|\cdot\|_{\mu,1}\lesssim \|\cdot\|_{\mathbf Q}$ we can also conclude that $\|f_n - q_0\|_{\mu,1} = o(1)$.
Therefore, for any bounded $g$ we obtain
\begin{equation}\label{th:idom7}
    \lim_{n\to \infty} |\int g d\dom - \sum_{i=1}^n \lambda_{in}\int g dQ_i| \leq\lim_{n\to \infty} \|g\|_\infty \int |q_0 - \sum_{i=1}^n \lambda_{in}\frac{dQ_i}{d\mu}|d\mu = 0.
\end{equation}
In particular, we note that \eqref{th:idom7} immediately yields $\int d\dom = 1$ and $0\leq \dom(A)\leq 1$ for any measurable set $A$ and hence by \eqref{th:idom6} that $\dom$ is indeed a probability measure.

We next show that $\dom \in \Theta_0$.
To this end note that \eqref{th:idom7} and $Q_i\in \Theta_0$ implying $Q_i$ induces $P$ yields that for any value $t\in \mathbf T$ and (measurable) set $V$ we must have
\begin{align}\label{th:idom8}
\dom (T^\star(Z) = t, (Y^\star(t),Z,X) \in V) & =  \lim_{n\to \infty} \sum_{i=1}^n \lambda_{in} Q_i(T^\star(Z) = t, (Y^\star(t),Z,X) \in V) \notag \\
& = \lim_{n\to \infty} \sum_{i=1}^n \lambda_{in} P(T = t, (Y,Z,X) \in V) \notag \\
& = P(T = t, (Y,Z,X) \in V),
\end{align}
since $\sum_{i=1}^n \lambda_{in} = 1$, which implies $\dom$ also induces $P$.
Next let $f$ and $g$ be arbitrary bounded functions of $(Y^\star,T^\star,X)$ and $(Z,X)$ respectively and note that
\begin{align}\label{th:idom9}
E_{\dom}[g(Z,X)f(Y^\star,T^\star,X)] & = \lim_{n\to\infty}\sum_{i=1}^n \lambda_{in} E_{Q_i}[g(Z,X)E_{Q_i}[f(Y^\star,T^\star,X)|X]] \notag \\
& = \lim_{n\to \infty} \sum_{i=1}^n\lambda_{in} E_{P}[g(Z,X)E_{Q_i}[f(Y^\star,T^\star,X)|X]]\notag \\
&= E_{\dom}[g(Z,X)(\lim_{n\to\infty} \sum_{i=1}^n \lambda_{in} E_{Q_i}[f(Y^\star,T^\star,X)|X])],
\end{align}
where the first equality follows from  \eqref{th:idom7} and  $(Y^\star,T^\star)\indep Z|X$ under $Q_i$ and the second equality from $Q_i$ inducing $P$ due to $Q_i \in \Theta_0$.
In turn, the third equality in \eqref{th:idom9} follows from the dominated convergence theorem and $\dom $ inducing $P$ as shown in \eqref{th:idom8}.
Since \eqref{th:idom9} holds for any bounded  $g$, Definition 10.1.1 in \cite{bogachev2:2007} we obtain
\begin{align}\label{th:idom10}
E_{\dom}[f(Y^\star,T^\star,X)|X,Z] & = \lim_{n\to\infty} \sum_{i=1}^n\lambda_{in} E_{Q_i}[f(Y^\star,T^\star,X)|X] \notag \\
& = E_{\dom}[f(Y^\star,T^\star,X)|X],
\end{align}
where the second equality can be deduced by applying the equalities in \eqref{th:idom9} evaluated at functions $g$ of $X$ only.
We have so far shown that result \eqref{th:idom10} holds for any arbitrary bounded $f$.
To extend the result to any $f \in L^1(\dom)$ let $f_M \equiv f1\{|f|\leq M\}$ and note that result \eqref{th:idom10} and Proposition 10.1.7 in \cite{bogachev2:2007} imply that
\begin{multline}\label{th:idom10p5}
E_{\dom}[f(Y^\star,T^\star,X)|X,Z] = \lim_{M\uparrow \infty} E_{\dom}[f_M(Y^\star,T^\star,X)|X,Z]\\ = \lim_{M\uparrow \infty} E_{\dom}[f_M(Y^\star,T^\star,X)|X]  = E_{\dom}[f(Y^\star,T^\star,X)|X].
\end{multline}
Since \eqref{th:idom10p5} holds for any integrable $f$, we can conclude that $(Y^\star,T^\star)\indep Z|X$ under $\dom$ and therefore, by the preceding results, that $\dom \in \Theta_0$.
Next, fix an arbitrary $Q\in \Theta_0$ and set $A$ with $Q(A) > 0$ and note that there exists a $Q_k\in \mathcal D$ satisfying
\begin{equation*}
\|\frac{dQ}{d\mu}-\frac{dQ_k}{d\mu}\|_{\mu,1} < \frac{Q(A)}{2}
\end{equation*}
by result \eqref{th:idom2}.
Hence, the triangle and Jensen's inequalities allow us to conclude that
\begin{equation}\label{th:idom11}
    Q_k(A) \geq Q(A) - |\int_A(\frac{dQ}{d\mu} - \frac{dQ_k}{d\mu})d\mu| \geq Q(A) - \|\frac{dQ}{d\mu}-\frac{dQ_k}{d\mu}\|_{\mu,1} > \frac{Q(A)}{2} 
\end{equation}
and thus that $Q_k(A) > 0$ as well.
Since definition \eqref{th:idom3} implies that the sequence $\{\lambda_{kn}\}_{n=1}^\infty$ is bounded away from zero for $n$ sufficiently large, we can combine results \eqref{th:idom7} and \eqref{th:idom11} to obtain that $\dom(A) \geq \liminf_{n \to \infty} \lambda_{kn} Q_k(A) > 0$.
In particular, since $A$ was arbitrary we can conclude that $Q \ll \dom$ as desired. 

In order to establish $\Theta_0$ is convex, let $Q_1,Q_2 \in \Theta_0$, $\gamma \in [0,1]$, and define $Q_\gamma \equiv \gamma Q_1 + (1-\gamma)Q_2$. Then note: (i) $dQ_\gamma/d\mu = \gamma dQ_1/d\mu + (1-\gamma)dQ_2/d\mu\in \restr$ by Assumption \ref{ass:model}(iii); (ii) $Q_\gamma$ induces $P$ by the arguments in \eqref{th:idom8} applied with $Q_\gamma$ in place of $\dom$; and (iii) $(Y^\star,T^\star)\indep Z|X$ under $Q_\gamma$ by the arguments in \eqref{th:idom9} and \eqref{th:idom10} applied with $Q_\gamma$ in place of $\dom$. It follows that $Q_\gamma \in \Theta_0$ and therefore that $\Theta_0$ is convex. \qed

\noindent \emph{Proof of Lemma \ref{lm:warmup}.}
First note that $\dom(\Id(\kappa) = \ell) = 1$ and Lemma \ref{th:idom}  together imply that $Q(\Id(\kappa) = \ell) = 1$ for all $Q\in \Theta_0$. 
By Corollary \ref{cor:cond} we can thus conclude
\begin{equation}\label{lm:warmup1}
f(Y^\star,T^\star,X) = E_Q[\kappa(Y,T,Z,X)|Y^\star,T^\star,X]
\end{equation}
$Q$-almost surely for any $Q\in \Theta_0$.
Result \eqref{lm:warmup1} and $\kappa \in L^1(P)$ therefore yields that $f \in L^1(Q)$ for all $Q\in \Theta_0$ as claimed.
Hence, for any $Q\in \Theta_0$ we can conclude
\begin{equation*}
E_{Q}[f(Y^\star,T^\star,X)] = E_{Q}[\kappa(Y,T,Z,X)] = E_P[\kappa(Y,T,Z,X)]
\end{equation*}
where the first equality follows from \eqref{lm:warmup1} and the law of iterated expectations, and the second equality follows from $Q\in \Theta_0$. \qed

\noindent \emph{Proof of Corollary \ref{cor:warmup}.}
By Lemma \ref{lm:muPswitch}, we have $\mu(\Id(\kappa) = \ell) = 1$ and hence also that $\dom(\Id(\kappa) = \ell) = 1$ due to $\dom \ll \mu$. Thus, the result is immediate from Lemma \ref{lm:warmup}. \qed

\noindent \emph{Proof of Theorem \ref{th:genmain}.} We first show that $\lambda_{\tr}$ being identified implies that $\Lambda$ belongs to the $\tau$-closure of $\mathcal R$.
To this end, we let ${\mathcal L}^\prime$ denote the linear span of $\{\mathcal R \cup \Lambda\}$ and note that every $L\in \mathcal L$ can be identified with a linear functional on $\mathcal L^\prime$ through the relation $L^\prime \mapsto L^\prime(L)$.
We endow $\mathcal L^\prime$ with the weak topology generated by $\mathcal L$, which we denote by $\sigma(\mathcal L^\prime,\mathcal L)$, and observe $(\mathcal L^\prime,\sigma(\mathcal L^\prime,\mathcal L))$ is a topological vector space and its topological dual is $\mathcal L$; see, e.g., Example 1.3.23 in \cite{bogachev2017topological}.
Moreover, since $\sigma(\mathcal L^\prime,\mathcal L)$ is generated by the family of seminorms $\{|L |\}_{L\in \mathcal L}$, Theorem 5.73 in \cite{aliprantis:border:2006} implies that $(\mathcal L^\prime,\sigma(\mathcal L^\prime,\mathcal L))$ is additionally locally convex.
We also note that by Lemma 2.53 in \cite{aliprantis:border:2006}, the $\tau$ topology on $\{\mathcal R \cup \Lambda\}$ coincides with the relative topology on $\{\mathcal R \cup \Lambda\}$ that is induced by the topology $\sigma(\mathcal L^\prime,\mathcal L)$ on $\mathcal L^\prime$.
Therefore, $\Lambda$ belongs to the $\tau$-closure of $\mathcal R$ (in $\{\mathcal R \cup \Lambda\}$) if and only if $\Lambda$ belongs to the $\sigma(\mathcal L^\prime,\mathcal L)$-closure of $\mathcal R$ (in $\mathcal L^\prime$); see, e.g., Theorem 17.4 in \cite{munkres2000topology}.  
Letting $\bar {\mathcal R}$ denote the $\sigma(\mathcal L^\prime,\mathcal L)$-closure of $\mathcal R$ in $\mathcal L^\prime$, it then follows that in order to show that $\Lambda$ belongs to the $\tau$-closure of $\mathcal R$ it suffices to establish that $\Lambda \in \bar{\mathcal R}$.

We proceed by contradiction and suppose that $\Lambda \notin \bar {\mathcal R}$. 
Since, as argued, $(\mathcal L^\prime,\sigma(\mathcal L^\prime,\mathcal L))$ is a locally convex topological vector space and $\mathcal L$ is its topological dual, Corollary 5.80 in \cite{aliprantis:border:2006} implies there then is an $L_0\in \mathcal L$ satisfying
\begin{equation}\label{th:genmaineq1}
L_0(\Lambda) \neq 0 \hspace{0.5 in} L_0(L^\prime) = 0 \text{ for all } L^\prime \in \bar {\mathcal R}.
\end{equation}
Moreover, by definition of $\mathcal L$ there is a finite collection $\{(s_j,Q_j)\}_{j=1}^J$ with $s_j \in \mathcal S_{Q_j}$ and $Q_j \in \Theta_0$ for all $1\leq j \leq J$ and such that for all $L^\prime \in \mathcal L^\prime$ we have $L_0(L^\prime) = L^\prime(\sum_{j=1}^J \langle \cdot, s_j\rangle_{Q_j})$.
Hence, by definition of $\mathcal R$ and $\Id$ we obtain for any $f\in L^1(P)$ that
\begin{align}\label{th:genmaineq3}
0 & = \sum_{j=1}^J E_{Q_j}[(\sum_{t\in \mathbf T} E_{P_{Z|X}}[f(Y^\star(t),t,Z,X)1\{T^\star(Z)=t\}])s_j(Y^\star,T^\star,X)]\notag \\
& =\sum_{j=1}^J E_{Q_j}[f(Y,T,Z,X) E_{Q_j}[s_j(Y^\star,T^\star,X)|Y,T,Z,X]] \notag \\
& = E_P[f(Y,T,Z,X)(\sum_{j=1}^J E_{Q_j}[s_j(Y^\star,T^\star,X)|Y,T,Z,X])],
\end{align}
where the first equality follows from $L_0(L^\prime) = 0$ for all $L^\prime \in \mathcal R$, the second from Corollary \ref{cor:cond} and the law of iterated expectations, and the third from $Q_j \in \Theta_0$ for all $1\leq j \leq J$.
Thus, since \eqref{th:genmaineq3} was shown to hold for any $f\in L^1(P)$ we can conclude that
\begin{equation*}
P(\sum_{j=1}^J E_{Q_j}[s_j(Y^\star,T^\star,X)|Y,T,Z,X] = 0) = 1.    
\end{equation*}
Hence, by Lemma \ref{lm:Pert} there are $\tilde Q, Q^{\rm a} \in \Theta_0$ and $\eta > 0$ such that for all (measurable) $A$ 
\begin{equation*}
\tilde Q(A) = Q^{\rm a}(A) + \eta \sum_{j=1}^J E_{Q_j}[s_j(Y^\star,T^\star,X)1\{(Y^\star,T^\star,Z,X)\in A\}].
\end{equation*}
Letting ${\bf 1}$ denote the function in $L^\infty(\dom)$ that takes a constant value of one, we then obtain by definition of $\lambda_Q$, $L_0(\Lambda) = \Lambda(L_0)$ and the definition of $\Lambda$ that
\begin{equation*}
    \lambda_{\tilde Q} = \lim_{k\to \infty} \langle \ell_k ,{\bf 1}\rangle_{\tilde Q} = \lim_{k\to \infty} \{\langle \ell_k ,{\bf 1}\rangle_{Q^{\rm a}} + \eta \sum_{j=1}^J \langle \ell_k, s_j\rangle_{Q_j}\}
    = \lambda_{Q^{\rm a}} + \eta L_0(\Lambda).
\end{equation*}
However, $\eta > 0$ and $L_0(\Lambda)\neq 0$ by \eqref{th:genmaineq1} together imply that $\lambda_{\tilde Q} \neq \lambda_{Q^{\rm a}}$.
Thus, since $\tilde Q,Q^{\rm a}\in \Theta_0$ we obtain that $\lambda_Q$ is not identified reaching a contradiction.
We therefore conclude that if $\lambda_Q$ is identified, then $\Lambda$ must belong to the $\tau$-closure of $\mathcal R$.

For the converse direction, we now suppose that $\Lambda$ belongs to the $\tau$-closure of $\mathcal R$.
Since $\mathcal L$ is identified (because it only depends on $\Theta_0$), $\Lambda$ is identified (because $\{\ell_j\}_{j=1}^\infty$ is identified), and $\mathcal R$ is identified (because $\Id:L^1(P)\to L^1(\dom)$ is identified), Theorem 2.4 in \cite{aliprantis:border:2006} implies there is an identified net $\{L^\prime_\alpha\}_{\alpha \in \mathcal A}$ with
\begin{equation}\label{th:genmaineq7}
\lim_{\alpha} L^\prime_\alpha(L) = \Lambda(L) \text{ for all } L\in \mathcal L
\end{equation}
and $L_\alpha^\prime \in \mathcal R$ for all $\alpha \in \mathcal A$.
Therefore, for any $Q_1,Q_2\in \Theta_0$ we can then conclude that
\begin{equation*}
\lambda_{Q_1} =  \Lambda(\langle \cdot,{\bf 1}\rangle_{Q_1}) = \lim_{\alpha} L_\alpha^\prime(\langle \cdot,{\bf 1}\rangle_{Q_1}) = \lim_\alpha L_\alpha^\prime(\langle \cdot,{\bf 1}\rangle_{Q_2}) = \Lambda(\langle \cdot, {\bf 1}\rangle_{Q_2}) = \lambda_{Q_2},
\end{equation*}
where the first and last equalities follow by definition of $\Lambda$, the second and fourth equalities by result \eqref{th:genmaineq7}, and the third equality by Lemma \ref{lm:warmup} and $L^\prime_\alpha \in \mathcal R$.
Thus, we conclude that $\lambda_Q$ is constant in $Q\in \Theta_0$ and is therefore identified. \qed

\noindent \emph{Proof of Corollary \ref{cor:int}.} First note that Assumptions \ref{ass:setup} and \ref{ass:model} were directly imposed. 
Moreover, Assumption \ref{ass:genpar}(i) is satisfied since $\ell \in L^1(\mu)$, $dQ/d\mu \in L^\infty(\mu)$ for all $Q\in \Theta_0$ by Assumption \ref{ass:model}(iii), and Holder's inequality imply for any $Q\in \Theta_0$ that
\begin{equation}\label{eq:corint1}
    E_Q[|\ell(Y^\star,T^\star,X)|] = \int |\ell |\frac{dQ}{d\mu} d\mu \leq \|\ell\|_{\mu,1} \|\frac{dQ}{d\mu}\|_{\mu,\infty} < \infty.
\end{equation}
Also note that Assumption \ref{ass:genpar}(ii) is immediate since here the sequence $\{\ell_j\}$ is constant, while Assumption \ref{ass:genpar}(iii) trivially holds due to $\restr = \mathbf Q$.
Thus, Theorem \ref{th:genmain} implies that $\lambda_Q$ is identified if and only if $\Lambda$ belongs to the $\tau$-closure of $\mathcal R$.

To establish part (i), note that since $\ell \in L^1(Q)$ for all $Q\in \Theta_0$ and $s d\dom/d\mu \in L^\infty(\mu)$ for any $s\in L^\infty(\dom_{Y^\star T^\star X})$ (because $\dom \ll \mu$ and $d\dom/d\mu \in L^\infty(\mu)$), it follows that in this application $\mathcal S_{\dom} = L^\infty(\dom_{Y^\star T^\star X})$.
Therefore, Lemma \ref{lm:seqclos} and the definition of $\mathcal R$ imply that there is a sequence $\{\kappa_j\} \subseteq L^1(P)$ satisfying  $\|\ell - \Id(\kappa_j)\|_{\dom,1} = o(1)$.
Since $\mu \ll \dom$ and $d\mu/d\dom$ is bounded, we can conclude $\{\kappa_j\}$ also satisfies $\|\ell - \Id(\kappa_j)\|_{\mu,1} = o(1)$.
For the converse, note that if there is a sequence $\{\kappa_j\} \subset L^1(P)$ satisfying $\|\ell - \Id(\kappa_j)\|_{\mu,1} = o(1)$, then $dQ/d\mu \in L^\infty(\mu)$ for all $Q\in \Theta_0$ and Holder's inequality yields
\begin{equation}\label{eq:corint4}
\lim_{j\to \infty} \|\ell - \Id(\kappa_j)\|_{Q,1} \leq \|\frac{dQ}{d\mu}\|_{\mu,\infty} \times \lim_{j\to \infty} \|\ell - \Id(\kappa_j)\|_{\mu,1} = 0.
\end{equation}
Therefore, for any $Q_1,Q_2\in \Theta_0$, Lemma \ref{lm:warmup} and result \eqref{eq:corint4} together establish that
\begin{equation*}
\lambda_{Q_1} = \lim_{j\to \infty} \int \Id(\kappa_j) dQ_1 = \lim_{j\to \infty} \int \kappa_j dP = \lim_{j\to \infty} \int \Id(\kappa_j) dQ_2 = \lambda_{Q_2},
\end{equation*}
which establishes $\lambda_{\tr}$ is identified and $\lambda_{\tr} = \lim_{j\to \infty} E_P[\kappa_j(Y,T,Z,X)]$. 
In turn part (ii) of the Corollary follows from part (i) and Lemma \ref{lm:muPswitch}. \qed

\noindent \emph{Proof of Theorem \ref{th:typemain}.} By Theorem \ref{th:genmain}, $\lambda_{\tr}$ is identified if and only if $\Lambda$ belongs to the $\tau$-closure of $\mathcal R$.
Since $\Rf \subseteq \mathcal R$, it immediately follows that if $\Lambda$ is in the $\tau$-closure of $\Rf$, then $\lambda_{\tr}$ is identified. 
Thus, to establish the theorem it suffices to show that if $\Lambda$ is in the $\tau$-closure of $\mathcal R$, then it must also belong to the $\tau$-closure of $\Rf$.
To this end, note that if $\Lambda$ belongs to the $\tau$-closure of $\mathcal R$, then the definition of $\mathcal R$ and Theorem 2.14 in \cite{aliprantis:border:2006} imply that there exists a net $\{f_\alpha\}_{\alpha \in \mathcal A} \subseteq L^1(P)$ satisfying
\begin{equation}\label{th:typemain1}
\lim_\alpha  \langle s, \Id(f_\alpha)\rangle_Q = \Lambda(\langle s,\cdot\rangle_Q)    
\end{equation}
for all $s\in \mathcal S_Q$ and $Q\in \Theta_0$. 
Next, set $g_\alpha(t,Z,X) \equiv E_{\dom_{Y^\star|X}}[f_\alpha(Y^\star(t),t,Z,X)]$ for any $t\in \{t_1,\ldots, t_d\}$.
By Jensen's inequality, $\dom \in \Theta_0$, and the definition of $\dom^{\rm it}$ we then obtain
\begin{align}\label{th:typemain3}
E_P[|g_\alpha(T,Z,X)|] & = E_{\dom_{T^\star Z X}}[\sum_{t\in \mathbf T} 1\{T^\star(Z) = t\}|E_{\dom_{Y^\star|X}}[f_\alpha(Y^\star(t),t,Z,X)]|] \notag\\
 & \leq E_{\dom^{\rm it}}[\sum_{t\in \mathbf T} 1\{T^\star(Z) =t\}|f_\alpha(Y^\star(t),t,Z,X)|] \notag \\
 & = E_{\dom^{\rm it}_{Y^* T^* X}}[\sum_{t\in \mathbf T} E_{\dom_{Z|X}}[1\{T^\star(Z) = t\}|f_\alpha(Y^\star(t),t,Z,X)|]],
\end{align}
where the final equality follows from Lemma \ref{lm:cond} and $\dom^{\rm it}_{ZX} = \dom_{ZX}$.
Letting ${\bf 1}$ denote the function of $(Y^\star,T^\star,X)$ taking a constant value of $1$, note that ${\bf 1}\in \mathcal S_{\dom}$ and Assumption \ref{ass:4types}(ii) imply $d\dom^{\rm it}_{Y^\star T^\star X}/d\dom_{Y^\star T^\star X} \in \mathcal S_{\dom}$.
In particular, since $\mathcal S_{\dom} \subseteq L^\infty(\dom_{Y^\star T^\star X})$, we may conclude that $d\dom^{\rm it}_{Y^\star T^\star X}/d\dom_{Y^\star T^\star X}$ is bounded, which together with \eqref{th:typemain3} yields
\begin{align}\label{th:typemain4}
E_P[|g_\alpha(T,Z,X)|]& \lesssim E_{\dom_{Y^* T^* X}}[\sum_{t\in \mathbf T} E_{\dom_{Z|X}}[1\{T^\star(Z) = t\}|f_\alpha(Y^\star(t),t,Z,X)|]] \notag \\ & = E_P[|f_\alpha(Y,T,Z,X)|], 
\end{align}
where the equality follows from Corollary \ref{cor:cond} and $\dom \in \Theta_0$ implying $\dom_{Z|X} = P_{Z|X}$.
Thus, since $f_\alpha \in L^1(P)$, result \eqref{th:typemain4} implies that $g_\alpha\in L^1(P_{TZX})$.

Next select any $s_0 \in \mathcal S_{\dom} \cap L^\infty(\dom_{T^\star X})$ and note that the definition of $\Id$ yields that
\begin{align}\label{th:typemain5}
    \langle s_0,\Id(g_\alpha)\rangle_{\dom} & = E_{\dom_{T^\star X}}[E_{P_{Z|X}}[E_{\dom_{Y^\star|X}}[\sum_{t\in \mathbf T} 1\{T^\star (Z) = t\} f_\alpha(Y^\star(t),t,Z,X)s_0(T^\star,X)]]] \notag \\
    & = E_{\dom^{\rm it}}[\sum_{t\in \mathbf T} 1\{T^\star (Z) = t\} f_\alpha(Y^\star(t),t,Z,X)s_0(T^\star,X)] \notag \\
    & = E_{\dom^{\rm it}_{Y^* T^* X}}[\sum_{t\in \mathbf T} E_{\dom^{\rm it}_{Z|X}}[1\{T^\star (Z) = t\} f_\alpha(Y^\star(t),t,Z,X)]s_0(T^\star,X)],
\end{align}
where in the second equality we employed Lemma \ref{lm:cond}, $P_{Z|X} = \dom_{Z|X}$ due to $\dom \in \Theta_0$, and the definition of $\dom^{\rm it}$, while the final equality follows from Lemma \ref{lm:cond} and $(Y^\star,T^\star)\indep Z|X$ under $\dom^{\rm it}$.
Further note that because $s_0$ and $\ell_j$ only depend on $(T^\star,X)$ we have
\begin{multline}\label{th:typemain6}
\Lambda(\langle \cdot,s_0\rangle_{\dom}) = \lim_{j \to \infty}\langle \ell_j, s_0\rangle_{\dom_{T^\star X}} = \lim_{j \to \infty}\langle \ell_j, s_0\rangle_{\dom^{\rm i}_{T^\star X}} = \lim_{j\to \infty}\langle \ell_j, s_0\frac{d\dom^{\rm it}_{Y^\star T^\star X}}{d\dom_{Y^\star T^\star X}}\rangle_{\dom}\\
= \lim_{\alpha} \langle \Id(f_\alpha),s_0\frac{d\dom^{\rm it}_{Y^\star T^\star X}}{d\dom_{Y^\star T^\star X}}\rangle_{\dom}
= \lim_{\alpha} \langle \Id(f_\alpha),s_0\rangle_{\dom^{\rm it}} = \lim_\alpha \langle \Id(g_\alpha),s_0\rangle_{\dom},
\end{multline}
where the second equality follows from $\dom_{T^\star X} = \dom^{\rm it}_{T^\star X}$, the fourth from result \eqref{th:typemain1}, $\dom \in \Theta_0$, and $s_0(d\dom^{\rm it}_{Y^\star T^\star X}/d\dom_{Y^\star T^\star X})\in \mathcal S_{\dom}$ by Assumption \ref{ass:4types}(ii), and the sixth from result \eqref{th:typemain5}, the definition of $\Id$, and $\dom_{Z|X}^{\rm it} = \dom_{Z|X} = P_{Z|X}$.
To conclude, let $\Pi_Q(s)(T^\star,X) \equiv E_Q[s(Y^\star,T^\star,X)|T^\star,X]$ and note that for any $s\in \mathcal S_Q$ and $Q\in \Theta_0$ we have
\begin{multline}\label{th:typemain7}
\Lambda(\langle \cdot, s\rangle_Q) = \lim_{j\to \infty} \langle \ell_j,\Pi_Q(s)\rangle_Q = \lim_{j\to \infty} \langle \ell_j,\Pi_Q(s)\frac{dQ_{T^\star X}}{d\dom_{T^\star X}}\rangle_{\dom} \\
= \lim_\alpha \langle \Id(g_\alpha),\Pi_Q(s)\frac{dQ_{T^\star X}}{d\dom_{T^\star X}}\rangle_{\dom} = \lim_\alpha \langle \Id(g_\alpha),s\rangle_Q,
\end{multline}
where the first equality follows from $\ell_j$ depending only on $(T^\star,X)$, the third equality follows from result \eqref{th:typemain6} and $\Pi_Q(s)dQ_{T^\star X}/d\dom_{T^\star X} \in \mathcal S_{\dom}$ by Assumptions \ref{ass:4types}(iii) and \ref{ass:4types}(iv), while the final equality follows from the law of iterated expectations.
Thus, result \eqref{th:typemain7} and Theorem 2.14 in \cite{aliprantis:border:2006} implies that $\Lambda$ belongs to the $\tau$-closure of $\Rf$, which establishes the claim of the theorem. \qed

\noindent \emph{Proof of Corollary \ref{cor:discint}.}
The fact that existence of a $\kappa \in L^1(P_{TZX})$ satisfying $\dom(\Id(\kappa)=\ell) = 1$ implies $\lambda_{\tr}$ is identified follows from Lemma \ref{lm:warmup}.
To establish the converse, we verify the conditions for Theorem \ref{th:typemain}.
To this end, note Assumptions \ref{ass:setup} and \ref{ass:model} were directly assumed, while Assumption \ref{ass:genpar}(iii) is immediate from $\restr = \mathbf Q = L^1(\mu)$.
Define
\begin{equation*}
\mathbf T_0^\star \equiv \{t^\star \in \mathbf T^\star : \dom(T^\star = t^\star) > 0\},
\end{equation*}
and note that $Q(T^\star \in \mathbf T_0^\star) = 1$ for any $Q\in \Theta_0$ due to $Q\ll \dom$.
Moreover, since any $Q\in \Theta_0$ must satisfy $Q_X = \dom_X = P_X$, it follows by direct calculation that 
\begin{equation}\label{cor:discint2}
\frac{dQ_{T^\star X}}{d\dom_{T^\star X}}(T^\star,X) = \sum_{t^\star \in \mathbf T^\star_0}1\{T^\star = t^\star\}\frac{Q(T^\star = t^\star|X)}{\dom(T^\star = t^\star|X)}    
\end{equation}
for any $Q\in \Theta_0$.
In particular, since $\dom (T^\star = t^\star|X) \geq \varepsilon$ a.s.\ result \eqref{cor:discint2} implies that
\begin{equation}\label{cor:discint3}
\|\frac{d Q_{T^\star X}}{d\dom_{T^\star X}}\|_{\dom,\infty} \leq \frac{1}{\varepsilon}.    
\end{equation}
Hence, $\ell\in L^1(\dom_{T^\star X})$ and \eqref{cor:discint3} imply $\ell \in L^1(Q)$ for any $Q\in \Theta_0$, verifying Assumptions \ref{ass:genpar}(i)(ii).
Further note that in this application  $\mathcal S_Q = L^\infty(Q_{Y^\star T^\star X})$ for any $Q\in \Theta_0$ and therefore Assumptions \ref{ass:4types}(i)(ii) hold (because we assumed $d\dom^{\rm it}/d\dom$ is bounded), Assumption \ref{ass:4types}(iii) follows from \eqref{cor:discint3}, and Assumption \ref{ass:4types}(iv) holds by Jensen's inequality.
Thus, all the conditions of Theorem \ref{th:typemain} are satisfied and we can conclude that if $\lambda_{\tr}$ is identified, then $\Lambda$ belongs to the $\tau$-closure of $\Rf$.
By applying Lemma \ref{lm:seqclos} we can then conclude that there exists a sequence $\{\kappa_j\}\subseteq L^1(P_{TZX})$ satisfying
\begin{equation}\label{eq:discint5}
\lim_{j\to \infty}\|\ell - \Id(\kappa_j)\|_{\dom,1} = 0.
\end{equation}

Next let $\mathbf S_0 \equiv \{(t,z) \in \mathbf T\times \mathbf Z : P(T=t,Z=z) > 0\}$ and note that $\dom \in \Theta_0$ implying $\dom$ is observationally equivalent to $\tr$ and $(Y^\star,T^\star)\indep Z | X$ under $\dom$ yield 
\begin{multline}\label{eq:discint6}
P(T = t,Z=z|X) = \dom(T^\star(z) = t,Z = z|X) \\
= \sum_{t^\star \in \mathbf T_0^\star : t^\star(z) = t} \dom(T^\star = t^\star|X)P(Z = z|X) \geq \varepsilon^2
\end{multline}
for any $(t,z)\in \mathbf S_0$, and where in the final inequality we employed that $\dom(T^*=t^*|X) \geq \varepsilon$ for any $t^* \in \mathbf T^*_0$ and $P(Z=z|X) \geq \varepsilon$ for any $z\in \mathbf Z$ by hypothesis.
Hence, for any event $E$ with $P(X\in E) > 0$ and $(t,z)\in \mathbf S_0$, Bayes' rule and result \eqref{eq:discint6} yield
\begin{equation*}
P(X \in E) =\frac{ P(X\in E|T = t, Z =z)P(T=t,Z=z)}{P(T=t,Z=z|X\in E)} \leq \frac{P(X\in E|T = t, Z =z)}{\varepsilon^2}.    
\end{equation*}
Letting $P_{X|t,z}$ denote the distribution of $X$ conditional on $(T,Z)=(t,z)$ for any $(t,z)\in \mathbf S_0$, it therefore follows that $P_X\ll P_{X|t,z}$ and $dP_X/dP_{X|t,z} \leq \varepsilon^{-2}$ almost surely under $P_{X|t,z}$.
In particular, we can conclude for any $(t,z)\in \mathbf S_0$ and $1 \leq j <\infty$ that
\begin{equation}\label{eq:discint8}
E_{P_X}[|\kappa_j(t,z,X)|] \leq \frac{1}{\varepsilon^2}E_P[|\kappa_j(t,z,X)||T=t,Z=z] \leq \frac{1}{\varepsilon^4}E_P[|\kappa_j(T,Z,X)|]
\end{equation}
where the final inequality follows by noting \eqref{eq:discint6} implies $P(T =t,Z=z) \geq \varepsilon^2$.
Thus, $\kappa_j \in L^1(P_{TZX})$ and result \eqref{eq:discint8} together imply that $\kappa_j(t,z,\cdot)\in L^1(P_X)$ for any $(t,z)\in \mathbf S_0$.
Next set $\tilde \kappa_j(t,z,X)\equiv \kappa_j(t,z,X)P(Z=z|X)$ and note that
\begin{align}\label{eq:discint9}
\|\ell - \Id(\kappa_j)\|_{\dom,1} & =   E_{\dom}[|\ell(T^\star,X)- \sum_{(t,z)\in \mathbf S_0}1\{T^\star(z)  = t\}\kappa_j(t,z,X)P(Z=z|X)|] \notag \\ & = E_{\dom}[|\ell(T^\star,X)- \sum_{(t,z)\in \mathbf S_0}1\{T^\star(z)  = t\}\tilde \kappa_j(t,z,X)|].
\end{align}
Since $\tilde \kappa_j(t,z,\cdot)\in L^1(P_X)$ for all $(t,z)\in \mathbf S_0$, results \eqref{eq:discint5}, \eqref{eq:discint9}, and Lemma \ref{lm:closedR} imply there are functions $\{f_0(t,z,X)\}_{(t,z)\in \mathbf S_0}$ satisfying $f_0(t,z,\cdot)\in L^1(P_X)$  and 
\begin{equation*}
E_{\dom}[|\ell(T^\star,X) - \sum_{(t,z)\in \mathbf S_0}1\{T^\star(z) = t\} f_0(t,z,X)|]=0.
\end{equation*}
Finally, set $\kappa(t,z,X) \equiv f_0(t,z,X)/P(Z=z|X)$ and note that $\kappa\in L^1(P_{TZX})$ because $P(Z=z|X) \geq\varepsilon>0$ a.s.\ and $f_0(t,z,\cdot)\in L^1(P_X)$ for any $(t,z)\in \mathbf S_0$.
We then obtain
\begin{multline*}
E_{\dom}[|\ell(T^\star,X) - \sum_{t\in\mathbf T} E_{P_{Z|X}}[1\{T^\star(Z)=t\}\kappa(t,Z,X)]|] \\
= E_{\dom}[|\ell(T^\star,X) - \sum_{(t,z)\in \mathbf S_0} 1\{T^\star(z)=t\}f_0(t,z,X)|]  = 0,
\end{multline*}
yielding that identification of $\lambda_Q$ implies the existence of the desired $\kappa$. 
Hence, we have shown that $\lambda_{\tr}$ is identified if and only if $\dom(\ell = \Id(\kappa))=1$ for some $\kappa\in L^1(P_{TZX})$, which establishes part (i) of the corollary. 
Part (ii) of the corollary is immediate from part (i), Lemma \ref{lm:muPswitch}, $\mu \ll \dom$ by assumption, and $\dom \ll \mu$ since $\dom \in \Theta_0$. \qed

\noindent \emph{Proof of Theorem \ref{th:outmain}.} We first show that if $\Lambda$ belongs to the $\tau$-closure of $\Rs$, then $\lambda_{\tr}$ is identified.
To this end, note that $\mathcal L$ is identified (because $\Theta_0$ is identified), $\Lambda$ is identified (because $\{\ell_j\}$ is identified), and $\Rs$ is identified (because $\Id:L^1(P)\to L^1(\dom)$ is identified).
Hence, Theorem 2.14 in \cite{aliprantis:border:2006} and $\Lambda$ being in the $\tau$-closure of $\Rs$ imply there is an identified net $\{L_\alpha^\prime\}_{\alpha \in \mathcal A}$ satisfying
\begin{equation}\label{th:outmain1}
\lim_\alpha L^\prime_\alpha(L) = \Lambda(L) \text{ for all } L\in \mathcal L.
\end{equation}
Next, let $\Pi_Q(s)(T^\star,X) \equiv E_Q[s(Y^\star,T^\star,X)|T^\star,X]$ for any $s \in L^1(Q)$, and note that Assumption \ref{ass:4outbasic}(i) and the law of iterated expectations imply for any $Q\in \Theta_0$ that
\begin{equation}\label{th:outmain2}
    \lambda_Q = \Lambda(\langle \cdot, \Pi_Q(\rho) \rangle_Q) = \lim_{\alpha} L_\alpha^\prime(\langle \cdot,\Pi_Q(\rho)\rangle_Q),
\end{equation}
where the second equality follows from \eqref{th:outmain1} and $\Pi_Q(\rho)\in \mathcal S_Q$ by Assumption \ref{ass:4out}(i).
Also note that, by definition of $\Rs$, there exists a net $\{f_\alpha\}_{\alpha \in \mathcal A} \subseteq L^1(P_{tZX})$ satisfying $L_\alpha^\prime(\langle \cdot, s\rangle_Q) = \langle \Id(f_\alpha),s\rangle_Q$ for any $Q\in \Theta_0$ and $s\in \mathcal S_Q$.
Noting that $f_\alpha(T,Z,X) = g_\alpha(Z,X)1\{T=t\}$ for some function $g_\alpha$, we then obtain that
\begin{multline}\label{th:outmain3}
L_\alpha^\prime(\langle \cdot, \Pi_Q(\rho)\rangle_Q) = E_{Q}[E_{Q}[\rho(Y^\star(t))|T^\star,X]E_{P_{Z|X}}[g_\alpha(Z,X)1\{T^\star(Z)=t\}]] \\
= E_{Q}[E_{P_{Z|X}}[\rho(Y^\star(t))g_\alpha(Z,X)1\{T^\star(Z)=t\}]] = E_P[\rho(Y)g_\alpha(Z,X)1\{T=t\}],
\end{multline}
where the final equality follows from Corollary \ref{cor:cond}, the law of iterated expectations, and $Q\in \Theta_0$.
Since \eqref{th:outmain2} and \eqref{th:outmain3} hold for any $Q\in \Theta_0$, it follows that $\lambda_{\tr}$ is identified.

We next establish that if $\lambda_{\tr}$ is identified, then $\Lambda$ must belong to the $\tau$-closure of $\Rs$.
To this end, we first define the spaces $\mathcal S_{Q,\rho}$ and $\mathcal L_\rho$ to be given by 
\begin{align*}
\mathcal S_{Q,\rho} &\equiv \{s \in L^\infty(Q_{Y^\star T^\star X}) : |\lim_{j\to \infty} \langle s,\rho \ell_j\rangle_Q|<\infty \text{ and } s\frac{dQ}{d\mu} \in \mathbf Q\} \notag \\
\mathcal L_\rho & \equiv \text{span}\{L :\bigcap_{Q\in \Theta_0} L^1(Q)\to \mathbf R \text{ s.t. } L = \langle \cdot,s\rangle_Q \text{ for some } s\in \mathcal S_{Q,\rho} \text{ and } Q\in \Theta_0\},
\end{align*}
let $\Lambda_\rho(L) \equiv \lim_{j\to \infty} L(\ell_j\rho)$ for any $L\in \mathcal L_\rho$, and $\tau_{\rho}$ denote the weak topology on $\{\mathcal R\cup \Lambda_{\rho}\}$ that is generated by $\mathcal L_{\rho}$ -- i.e.\ $\mathcal S_{Q,\rho},$ $\mathcal L_\rho,$ $\Lambda_\rho,$ and $\tau_{\rho}$ correspond to our definitions for $\mathcal L, \mathcal S_Q, \Lambda,$ and $\tau$ applied with $\{\ell_j\rho\}$ in place of $\{\ell_j\}$.
By Theorem \ref{th:genmain} and $\lambda_{\tr}$ being identified, it then follows that $\Lambda_\rho$ belongs to the $\tau_{\rho}$-closure of $\mathcal R$.
Hence, Theorem 2.14 in \cite{aliprantis:border:2006} implies there is a net $\{L_\alpha^\prime\}_{\alpha \in \mathcal A}\subseteq \mathcal R$ satisfying
\begin{equation}\label{th:outmain5}
\lim_{\alpha} L^\prime_\alpha(L) = \Lambda_\rho(L) \text{ for all } L \in \mathcal L_\rho   .
\end{equation}
Next note that $Y^\star(t)$ being independent of $T^\star$ conditionally on $X$ under $\dom^{\rm io}$, the marginal distribution of $(Y^\star(t),X)$ being the same under $\dom$ and $\dom^{\rm io}$, the law of iterated expectations, and Assumptions \ref{ass:4out}(ii)(iii) allow us to conclude that
\begin{equation}\label{th:outmain6}
\dom^{\rm io}(E_{\dom^{\rm io}}[\rho(Y^\star(t)) \phi_{\dom,\rho}(Y^\star(t),X)|T^\star,X] = 1) = 1.
\end{equation}
Fixing an arbitrary $s\in \mathcal S_{\dom}$, then note that the law of iterated expectations, the marginal distribution of $(T^\star,X)$ being the same under $\dom$ and $\dom^{\rm io}$ and result \eqref{th:outmain6} yield
\begin{multline}\label{th:outmain7}
\lim_{j\to \infty} \langle \ell_j,s\rangle_{\dom} 
= \lim_{j\to \infty} \langle \ell_j,\Pi_{\dom}(s)\rangle_{\dom} 
= \lim_{j\to \infty} \langle \ell_j,\Pi_{\dom}(s)\rangle_{\dom^{\rm io}} \\ 
= \lim_{j\to \infty} \langle \ell_j \rho, \phi_{\dom,\rho} \Pi_{\dom}(s)\rangle_{\dom^{\rm io}} 
= \lim_{j\to \infty}\langle \ell_j \rho, \phi_{\dom,\rho} \Pi_{\dom}(s) \frac{d\dom^{\rm io}_{Y^\star T^\star X}}{d\dom_{Y^\star T^\star X}} \rangle_{\dom},
\end{multline}
where the final equality follows from Assumption \ref{ass:4out}(ii).
Since the limit in \eqref{th:outmain7} exists due to $s\in \mathcal S_{\dom}$, Assumptions \ref{ass:4out}(i)(iii) imply $\phi_{\dom,\rho}\Pi_{\dom}(s)(d\dom^{\rm io}_{Y^\star T^\star X}/d\dom_{Y^\star T^\star X})\in \mathcal S_{\dom,\rho}$.
Next note that by definition of $\mathcal R$, there exists a net $\{v_\alpha\}_{\alpha \in \mathcal A} \subseteq L^1(P)$ such that $L^\prime_\alpha(\langle \cdot,s\rangle_Q) = \langle \Id(v_\alpha),s\rangle_Q$ for any $Q\in \Theta_0$ and $s\in \mathcal S_{Q,\rho}$.
In particular, we have 
\begin{equation}\label{th:outmain8}
  \lim_{j\to \infty} \langle \ell_j,s\rangle_{\dom} = \lim_{\alpha} \langle \Id(v_\alpha), \phi_{\dom,\rho} \Pi_{\dom}(s) \frac{d\dom^{\rm io}_{Y^\star T^\star X}}{d\dom_{Y^\star T^\star X}}\rangle_{\dom} = \lim_\alpha \langle \Id(v_\alpha), \phi_{\dom,\rho} \Pi_{\dom}(s)\rangle_{\dom^{\rm io}} 
\end{equation}
due to \eqref{th:outmain5} and \eqref{th:outmain7}.
Next set $f_\alpha(T,Z,X) \equiv 1\{T=t\}g_\alpha(Z,X)$ with $g_\alpha$ given by
\begin{equation*}
g_\alpha(Z,X) \equiv E_{\dom_{Y^\star(t)|X}}[v_\alpha(Y^\star(t),t,Z,X)\phi_{\dom,\rho}(Y^\star(t),X)],
\end{equation*}
and where in the expectation $Z$ and $X$ are kept constant.
Also note that $\dom \in \Theta_0$, Jensen's inequality,  and $\phi_{\dom,\rho} \in L^\infty(\dom)$ by Assumption \ref{ass:4out}(iii) imply that
\begin{align}\label{th:outmain10}
E_P[|f_\alpha(T,Z,X)|] &  \lesssim E_{\dom_{T^\star Z X}}[1\{T^\star(Z) = t\}E_{\dom_{Y^\star(t)|X}}[|v_\alpha(Y^\star(t),t,Z,X)|] \notag  \\
& = E_{\dom^{\rm io}}[1\{T^\star(Z)=t\}|v_\alpha(Y^\star(t),t,Z,X)|] \notag \\
& \lesssim  E_{\dom}[1\{T^\star(Z)=t\}|v_\alpha(Y^\star(t),t,Z,X)|] \notag \\
& = E_P[|1\{T=t\}v_\alpha(Y,T,Z,X)|],
\end{align}
where the first equality holds by definition of $\dom^{\rm io}$; the second inequality follows from $d\dom^{\rm io}/d\dom \in L^\infty(\dom)$ by Assumption \ref{ass:4out}(ii); and the final equality holds because $\dom \in \Theta_0$.
In particular, since $v_\alpha \in L^1(P)$, result \eqref{th:outmain10} implies $f_\alpha \in L^1(P_{tZX})$ and therefore that $\Id(f_\alpha) \in \Rs$.
Finally, we observe that the law of iterated expectations, $\dom \in \Theta_0$, Corollary \ref{cor:cond}, and $\dom^{\rm io}_{T^\star ZX} = \dom_{T^\star Z X}$ allow us to conclude for any $s\in \mathcal S_{\dom}$ that
\begin{align}\label{th:outmain11}
\langle \Upsilon (&f_\alpha),  s\rangle_{\dom} \notag \\
& = E_{\dom^{\rm io}}[g_\alpha(Z,X)1\{T^\star(Z) = t\}E_{\dom}[s(Y^\star,T^\star,X)|T^\star,X]] \notag \\
& = E_{\dom^{\rm io}}[v_\alpha(Y^\star(t),t,Z,X)1\{T^\star(Z) =t \}\phi_{\dom,\rho}(Y^\star(t),X)E_{\dom}[s(Y^\star,T^\star,X)|T^\star,X]] \notag \\
& = \sum_{\tilde t \in \mathbf T}E_{\dom^{\rm io}}[v_\alpha(Y^\star(\tilde t),\tilde t,Z,X)1\{T^\star(Z) =\tilde t \}\phi_{\dom,\rho}(Y^\star(t),X)E_{\dom}[s(Y^\star,T^\star,X)|T^\star,X]] \notag \\
& = \langle \Id(v_\alpha), \phi_{\dom,\rho} \Pi_{\dom}(s)\rangle_{\dom^{\rm io}},
\end{align}
where the second equality follow from the definition of $\dom^{\rm io}$ and $g_\alpha$; the third equality from $(Y^\star(\tilde t),T^\star,Z)$ being independent of $Y^\star(t)$ conditionally on $X$ under $\dom^{\rm io}$ whenever $\tilde t \neq t$, $E_{\dom^{\rm io}}[\phi_{\dom,\rho}(Y^\star(t),X)|X] = 0$ by definition of $\phi_{\dom,\rho}$ and $\dom^{\rm io}_{Y^\star(t) X} = \dom_{Y^\star(t)X}$; and the final equality holds by Lemma \ref{lm:cond} and $\dom^{\rm io}_{ZX} = \dom_{ZX} = P_{ZX}$. Thus, combining results \eqref{th:outmain8} with \eqref{th:outmain11} allows us to conclude that for any $s\in \mathcal S_{\dom}$ we have
\begin{equation}\label{th:outmain12}
\lim_{j\to \infty} \langle \ell_j,s\rangle_{\dom} = \lim_\alpha \langle \Id(f_\alpha),s\rangle_{\dom}.
\end{equation}
To conclude, note that for any $Q\in \Theta_0$ and $s\in \mathcal S_Q$, the law of iterated expectations, $Q \ll \dom$, $\Pi_Q(s) dQ_{T^\star X}/d\dom_{T^\star X}\in \mathcal S_{\dom}$ by Assumptions \ref{ass:4out}(i)(iv), and result \eqref{th:outmain12} yield
\begin{multline*}
\Lambda(\langle \cdot, s\rangle_Q) = \lim_{j\to \infty} \langle \ell_j,\Pi_Q(s)\rangle_Q = \lim_{j\to \infty} \langle \ell_j,\Pi_Q(s)\frac{dQ_{T^\star X}}{d\dom_{T^\star X}}\rangle_{\dom} \\
= \lim_\alpha \langle \Id(f_\alpha),\Pi_Q(s)\frac{dQ_{T^\star X}}{d\dom_{T^\star X}}\rangle_{\dom} = \lim_\alpha \langle \Id(f_\alpha),s\rangle_Q.
\end{multline*}
Hence, since $\Upsilon(f_\alpha)\in \Rs$, we conclude that $\Lambda$ belongs to the $\tau$-closure of $\Rs$. \qed

\noindent \emph{Proof of Corollary \ref{cor:discintout}.} The proof is similar to that of Corollary \ref{cor:discint} and  we therefore omit some of the details.
We first verify that the assumptions of Theorem \ref{th:outmain} are satisfied.
To this end, note that Assumptions \ref{ass:setup} and \ref{ass:model} were directly imposed, while $\restr = \mathbf Q = L^1(\mu)$ implies Assumption \ref{ass:genpar}(iii) holds.
Letting $\mathbf T_0^\star \equiv \{t^\star \in \mathbf T^\star : \dom(T^\star = t^\star) > 0\}$, it can then be shown that $\dom(T^\star = t^\star|X) \geq \varepsilon > 0$ a.s.\ and $Q_X = P_X$ for any $Q\in \Theta_0$ yield
\begin{equation}\label{cor:discintout1}
\|\frac{dQ_{T^\star X}}{d\dom_{T^\star X}}\|_{\dom,\infty} \leq \frac{1}{\varepsilon}    .
\end{equation}
In particular, $\ell \in L^1(\dom_{T^\star X})$ and result \eqref{cor:discintout1} imply that Assumptions \ref{ass:4outbasic}(i)(ii) also hold. 
Moreover, since in this application $\mathcal S_Q = L^\infty(Q_{Y^\star T^\star X})$ for any $Q\in \Theta_0$, Assumptions \ref{ass:4out}(i) and (iv) hold by Jensen's inequality and result \eqref{cor:discintout1} respectively.
Similarly, we note that Assumption \ref{ass:4out}(ii) was directly imposed, while Assumption \ref{ass:4out}(iii) is satisfied since we assumed $\rho \in L^\infty(\dom)$ and $\text{Var}_{\dom}\{\rho(Y^\star(t_0))|X\} \geq \varepsilon > 0$ a.s.\ under $\dom$.
Thus, the conditions of Theorem \ref{th:outmain} hold.

Next note that if (i) holds, then Theorem \ref{th:outmain} implies $\Lambda$ belongs to the $\tau$-closure of $\Rs$.
By Lemma \ref{lm:seqclos}, there therefore exists a sequence $\{\kappa_j\}\in L^1(P_{ZX t})$ satisfying
\begin{equation}\label{cor:discintout2}
\lim_{j\to \infty} \|\ell - \Id(\kappa_j)\|_{\dom,1} = 0.
\end{equation}
Letting $\mathbf Z_0 \equiv \{z \in \mathbf Z : P(T = t, Z = z) > 0\}$ and noting that $\kappa_j(T,Z,X) = 1\{T = t\}g_j(Z,X)$ for some function $g_j$ by definition of $L^1(P_{tZX})$, it then follows from the same arguments employed in Corollary \ref{cor:discint} that $g_j(z,\cdot)\in L^1(P_X)$ for any $z\in \mathbf Z_0$.
Next set $\tilde g_j(z,X) \equiv g_j(z,X)P(Z=z|X)$ and observe that by definition of $\Id$ we have
\begin{align}\label{cor:discintout3}
\|\ell- \Id(\kappa_j)\|_{\dom,1} & = E_{\dom}[|\ell(T^\star,X) - \sum_{z\in \mathbf Z_0}1\{T^\star(z) = t\}g_j(z,X)P(Z=z|X)|] \notag \\
& = E_{\dom}[|\ell(T^\star,X) - \sum_{z\in \mathbf Z_0}1\{T^\star(z) = t\}\tilde g_j(z,X)|].
\end{align}
Combining results \eqref{cor:discintout2} and \eqref{cor:discintout3} with Lemma \ref{lm:closedR} then implies that there are functions $\{f_0(z,X)\}_{z\in \mathbf Z_0}$ satisfying $f_0(z,\cdot)\in L^1(P_X)$ for all $z\in \mathbf Z_0$ and
\begin{equation}\label{cor:discintout4}
E_{\dom}[|\ell(T^\star, X) - \sum_{z\in \mathbf Z_0} 1\{T^\star(z) = t\} f_0(z,X)|] = 0.
\end{equation}
Hence, setting $\kappa(T,Z,X) \equiv 1\{T = t\}\sum_{z\in \mathbf Z_0}1\{Z=z\}f_0(Z,X)/P(Z=z|X)$ we obtain from $P(Z=z|X)\geq \varepsilon > 0$ a.s.\ that $\kappa \in L^1(P_{tZX})$ and from \eqref{cor:discintout4} that $\dom(\Id(\kappa) = \ell)=1$.
Since $\dom(\Id(\kappa) = \ell) = 1$ implies  $\dom(\Id(f\kappa) = f\ell) = 1$ for any bounded $f$, Lemma \ref{lm:warmup} allows us to conclude that (i) implies (iii).
Thus, because (iii) trivially implies (ii) and (ii) trivially implies (i), the claim of the corollary follows. \qed

\begin{lemma}\label{lm:cond}
Let $Q$ be a distribution for $(Y^\star,T^\star,Z,X)$ satisfying $(Y^\star,T^\star)\indep Z|X$ under $Q$. Then it follows that for any $f\in L^1(Q)$ we have:
$$E_Q[f(Y^\star,T^\star,Z,X)|Y^\star,T^\star,X] = E_{Q_{Z|X}}[f(Y^\star,T^\star,Z,X)].$$
\end{lemma}

\noindent \emph{Proof.}
Let $\mathcal G$ denote the $\sigma$-field on which $Q$ is defined, which recall we set to equal the $\sigma$-field generated by $(\mathcal G_{Y^\star}\times \mathcal G_{T^\star}\times \mathcal G_{Z}\times \mathcal G_{X})$ where $\mathcal G_V$ denotes the $\sigma$-field on which the marginal distribution $Q_V$ is defined.
We further define the class of sets
\begin{equation*}
\mathcal A \equiv \{A \in \mathcal G : E_Q[1\{(Y^\star,T^\star, Z,X) \in A\}|Y^\star,T^\star,X] = E_{Q_{Z|X}}[1\{(Y^\star,T^\star, Z, X) \in A\}]\}
\end{equation*}
and note that $\mathbf Y^\star\times \mathbf T^\star \times \mathbf Z \times \mathbf X \in \mathcal A$.
Also observe that if $A_1,A_2 \in \mathcal A$ and $A_1\subseteq A_2$ then
\begin{align}\label{lm:cond2}
E_Q[1\{(Y^\star,T^\star,& Z,X) \in A_2\setminus A_1\}|Y^\star,T^\star, X] \notag \\ &  = E_Q[1\{(Y^\star,T^\star,Z,X)\in A_2\} -1\{(Y^\star,T^\star,Z,X)\in A_1\}|Y^\star,T^\star, X]  \notag  \\
& = E_{Q_{Z|X}}[1\{(Y^\star,T^\star,Z,X)\in A_2\}] - E_{Q_{Z|X}}[1\{(Y^\star,T^\star,Z,X)\in A_1\}] \notag \\
& = E_{Q_{Z|X}}[1\{(Y^\star,T^\star,Z,X)\in A_2\setminus A_1\}],
\end{align}
where the first and third equalities follow from $A_1 \subseteq A_2$ while the second equality follows from $A_1,A_2 \in \mathcal A$. 
In particular, result \eqref{lm:cond2} implies that $A_2\setminus A_1\in \mathcal A$.
Next, let $\{A_i\}_{i=1}^\infty\subset \mathcal A$ be a sequence of pairwise disjoint sets and note that
\begin{align}\label{lm:cond3}
    E_Q[1\{(Y^\star,T^\star, Z,X)& \in \bigcup_{i=1}^\infty A_i\}|Y^\star,T^\star,X] \notag \\
    & = \lim_{n\to \infty} E_Q[\sum_{i=1}^n1\{(Y^\star,T^\star, Z,X)\in A_i\}|Y^\star,T^\star,X] \notag \\
    & = \lim_{n\to \infty} E_{Q_{Z|X}}[\sum_{i=1}^n1\{(Y^\star,T^\star, Z,X)\in  A_i\}] \notag \\
    & = E_{Q_{Z|X}}[1\{(Y^\star,T^\star, Z,X)\in \bigcup_{i=1}^\infty A_i\}],
\end{align}
where the first and third equalities follow from Theorem 10.1.5(4) in \cite{bogachev2:2007} and $\{A_i\}_{i=1}^\infty$ being disjoint, while the second holds due to $A_i\in \mathcal A$ for all $i$.
In particular, \eqref{lm:cond3} implies $\bigcup_{i=1}^\infty A_i \in \mathcal A$, and we can therefore conclude that $\mathcal A$ is a $\lambda$-system.

Let $A_{Y^\star}\in \mathcal G_{Y^\star}$, $A_{T^\star}\in \mathcal G_{T^\star}$, $A_{Z}\in \mathcal G_{Z}$, and $A_{X}\in \mathcal G_{X}$ be arbitrary, and then observe
\begin{align}\label{lm:cond4}
E_Q[1\{(Y^\star,T^\star,&Z,X) \in  A_{Y^\star}\times A_{T^\star}\times A_Z\times A_X\}|Y^\star, T^\star,X] \notag \\ 
& = 1\{Y^\star \in A_{Y^\star}\}1\{T^\star \in A_{T^\star}\}1\{X\in A_X\}E_{Q_{Z|X}}[1\{Z\in A_Z\}] \notag \\
& = E_{Q_{Z|X}}[1\{(Y^\star,T^\star,Z,X) \in A_{Y^\star}\times A_{T^\star}\times A_Z\times A_X\}]
\end{align}
where the first equality follows from $(Y^\star,T^\star) \indep Z|X$ under $Q$ and the second by direct manipulation.
Result \eqref{lm:cond4} implies $A_{Y^\star}\times A_{T^\star}\times A_Z\times A_X\in \mathcal A$ and, since $A_{Y^\star},A_{T^\star}, A_Z,A_X$ were arbitrary, that $(\mathcal G_{Y^\star}\times \mathcal G_{T^\star}\times \mathcal G_Z\times \mathcal G_X) \subseteq \mathcal A$.
Since $(\mathcal G_Z\times \mathcal G_X\times \mathcal G_{T^\star}\times \mathcal G_{Y^\star})$ is a $\pi$-system and $\mathcal G$ equals the $\sigma$-field generated by $(\mathcal G_{Y^\star}\times \mathcal G_{T^\star}\times \mathcal G_Z\times \mathcal G_X)$, the $\pi-\lambda$ theorem (see, e.g., Theorem 2.38 in \cite{pollard2002user}) then implies $\mathcal A = \mathcal G$.

To conclude, let $f\in L^1(Q)$ be arbitrary and $\{f_n\}$ be a sequence of simple functions satisfying $|f_n|\leq |f|$ and $f_n(Y^\star,T^\star,Z,X) \to f(Y^\star,T^\star,Z,X)$ on a set with $Q$-probability one.
By Proposition 10.1.7 in \cite{bogachev2:2007} we can then conclude that
\begin{multline*}
E_Q[f(Y^\star,T^\star,Z,X)|Y^\star,T^\star,X] = \lim_{n\to \infty} E_Q[f_n(Y^\star,T^\star,Z,X)|Y^\star,T^\star,X] \\
= \lim_{n\to \infty} E_{Q_{Z|X}}[f_n(Y^\star,T^\star,Z,X)] = E_{Q_{Z|X}}[f(Y^\star,T^\star,Z,X)],
\end{multline*}
where the second equality holds due to $f_n$ being a simple function and $\mathcal A = \mathcal G$. \qed

\begin{corollary}\label{cor:cond}
If Assumption \ref{ass:setup} holds, $Q\in \Theta_0$, and $f\in L^1(P)$, then it follows
$$E_Q[f(Y,T,Z,X)|Y^\star,T^\star,X] = \sum_{t\in \mathbf T} E_{P_{Z|X}}[f(Y^\star(t),t,Z,X)1\{T^\star(Z) =t\}].$$
\end{corollary}

\noindent \emph{Proof.} The claim is immediate from Lemma \ref{lm:cond} and noting that: (i) $Y = Y^\star(T)$ and $T = T^\star(Z)$ by Assumption \ref{ass:setup} imply $f(Y,T,Z,X) = \sum_{t\in \mathbf T} f(Y^\star(t),t,Z,X)1\{T^\star(Z) = t\}$, and (ii) $Q_{Z|X} = P_{Z|X}$ due to $Q\in \Theta_0$ by hypothesis. \qed

\begin{lemma}\label{lm:muPswitch}
Let Assumptions \ref{ass:setup} and \ref{ass:model} hold, and suppose that $\mu(\pi(Z,X) > \delta) = 1$ for some $\delta > 0$. If $\nu \in L^1(P)$, then it follows $\kappa = \nu/\pi$ satisfies $\kappa \in L^1(P)$ and 
$$E_{\mu_{Z|X}} [\nu (Y^*(t),t,Z,X)1\{T^\star(Z) = t\}] = E_{P_{Z|X}}[\kappa(Y^*(t),t,Z,X)1\{T^*(Z) = t\}].$$
\end{lemma}

\noindent {\emph Proof.} First note that since $\mu(\pi(Z,X) > \delta) = 1$ and $\dom \in \Theta_0$ must satisfy $\dom \ll \mu$, it follows that $\dom(\pi(Z,X) > \delta) = 1$. 
Moreover, $\dom_{ZX} = P_{ZX}$ due to $\dom \in \Theta_0$ and therefore $P(\pi(Z,X) > \delta) = 1$ as well. 
Hence, we can conclude that $1/\pi \in L^\infty(P)$, which together with $\nu \in L^1(P)$ yields that $\kappa = \nu/\pi \in L^1(P)$.
Next note that since $\mu(\pi(Z,X) > 0) = 1$, it follows that on a set with $\mu$-probability one we must have
\begin{align*}
E_{\mu_{Z|X}} [\nu & (Y^*(t),t,Z,X)1\{T^\star(Z) = t\}]\\
&  = E_{\mu_{Z|X}}[\nu(Y^*(t),t,Z,X)1\{T^\star(Z) = t\}1\{\pi(Z,X) > 0\}] \\ 
& = E_{P_{Z|X}}[\kappa(Y^*(t),t,Z,X)1\{T^*(Z) = t\}],
\end{align*}
where the final equality follows from the definitions of $\kappa$ and $\pi$. \qed

\begin{lemma}\label{lm:Pert}
Let Assumptions \ref{ass:setup}, \ref{ass:model}, and \ref{ass:genpar}(iii) hold, and suppose that a finite collection $\{s_j,Q_j\}_{j=1}^J$ with $s_j \in \mathcal S_{Q_j}$ and $Q_j \in \Theta_0$ for all $1\leq j \leq J$ satisfies 
\begin{equation}\label{lm:Pertdisp}
P(\sum_{j=1}^J E_{Q_j}[s_j(Y^\star,T^\star,X)|Y,T,Z,X] = 0) = 1.   
\end{equation}
Then, there exist a $Q^{\rm a}\in \Theta_0$ and a constant $\eta > 0$ such that the measure $\tilde Q$ given by
$$\tilde Q(A) \equiv Q^{\rm a}(A) + \sum_{j=1}^J \eta E_{Q_j}[s_j(Y^\star,T^\star,X)1\{(Y^\star,T^\star,Z,X)\in A\}]$$
(for any $A$ in the domain of $\dom$) belongs to the identified set $\Theta_0$.
\end{lemma}

\noindent \emph{Proof.} First note that by Assumption \ref{ass:genpar}(iii) there exists a measure $Q^{\rm i}\in \Theta_0$ such that $dQ^{\rm i}/d\mu$ belongs to the interior of $\restr$ in $\mathbf Q$.
Therefore setting $Q^{\rm a}$ to equal
\begin{equation}\label{lm:Pert1}
    Q^{\rm a} = \lambda Q^{\rm i} + \frac{1-\lambda}{J} \sum_{j=1}^J Q_j
\end{equation}
we can conclude that $dQ^{\rm a}/d\mu$ also belongs to the interior of $\restr$ in $\mathbf Q$ provided $\lambda \in (0,1)$ is chosen sufficiently large, and moreover that $Q^{\rm a} \in \Theta_0$ due to $\Theta_0$ being convex and $Q_j \in \Theta_0$ for all $1\leq j \leq J$.
Next note that  $\tilde Q(A)$ is well defined for any $A$ in the domain of $\dom$ and that $\tilde Q$ is countably additive by the dominated convergence theorem.
Moreover, observe that $\|s_j\|_{Q_j,\infty} < \infty$ for all $1 \leq j \leq J$ since $s_j\in \mathcal S_{Q_j} \subseteq L^\infty(Q_j)$.
Hence, by the choice of $Q^{\rm a}$ in \eqref{lm:Pert1} we obtain for any $A$ in the domain of $\dom$ that
\begin{equation}\label{lm:Pert2}
\tilde Q(A) \geq \sum_{j=1}^J \{\frac{1-\lambda}{J} Q_j(A) - \eta \|s_j\|_{Q_j,\infty} \int_A dQ_j\} = \sum_{j=1}^JQ_j(A)(\frac{1-\lambda}{J}-\eta \|s_j\|_{Q_j,\infty}).
\end{equation}
Thus, result \eqref{lm:Pert2} implies $\tilde Q$ is a positive measure provided we set $\eta>0$ to satisfy $\eta < (1-\lambda)/(J\max_j \|s\|_{\dom,\infty})$ (which is possible because $\lambda <1$).
Further observe
\begin{equation}\label{lm:Pert3}
\tilde Q(\mathbf Y^\star \times \mathbf T^\star\times \mathbf Z\times \mathbf X) = Q^{\rm a}(\mathbf Y^\star \times \mathbf T^\star\times \mathbf Z\times \mathbf X) + \eta \sum_{j=1}^J E_{Q_j}[s_j(Y^\star,T^\star,X)] = 1,
\end{equation}
where the final equality follows from $Q^{\rm a}$ being a probability measure, condition \eqref{lm:Pertdisp}, the law of iterated expectations, and $Q_j$ being observationally equivalent to $\tr$ due to $Q_j \in \Theta_0$ for all $1\leq j \leq J$.
Given the already verified positivity of $\tilde Q$ (for $\eta$ sufficiently small), result \eqref{lm:Pert3} implies that $\tilde Q$ is indeed a probability measure.
Also note that since $Q_j \ll \mu$ due to $Q_j \in \Theta_0$ for all $1\leq j \leq J$, it follows $\tilde Q \ll \mu$ and $d\tilde Q/d\mu = dQ^{\rm a}/d\mu + \eta \sum_j s_j dQ_j/d\mu$.
Thus, since $dQ^{\rm a}/d\mu$ belongs to the interior of $\restr$ in $\mathbf Q$ and $s_j dQ_j/d\mu \in \mathbf Q$ by definition of $\mathcal S_{Q_j}$, it follows that $d\tilde Q/d\mu \in \restr$ for $\eta >0$  small.

We next show that $\tilde Q$ is observationally equivalent to $\tr$.
To this end, note that Assumption \ref{ass:setup}(ii) and $Q^{\rm a} \in \Theta_0$ imply for any $t \in \mathbf T$ and (measurable) set $V$ that
\begin{equation}\label{lm:Pert4}
P(T=t, (Y,Z,X)\in V) = Q^{\rm a}(T^\star(Z) = t, (Y^\star(t),Z,X)\in V).
\end{equation} 
However, for any $t \in \mathbf T$ and (measurable) set $V$, Assumption \ref{ass:setup}(ii) also yields that
\begin{multline}\label{lm:Pert5}
 \sum_{j=1}^J E_{Q_j}[s_j(Y^\star,T^\star,X)1\{T^\star(Z) = t, (Y^\star(t),Z,X)\in V)\}] \\
 = \sum_{j=1}^J E_{Q_j}[E_{Q_j}[s_j(Y^\star,T^\star,X)|Y,T,Z,X]1\{T=t,(Y,Z,X)\in V\}] = 0,
\end{multline}
where the final equality follows from condition \eqref{lm:Pertdisp} and $Q_j$ being observationally equivalent to $\tr$ due to $Q_j \in \Theta_0$ for all $1\leq j \leq J$.
Hence, \eqref{lm:Pert4}, \eqref{lm:Pert5}, and the definition of $\tilde Q$ imply that $\tilde Q$ is indeed observationally equivalent to $\tr$. 

To conclude the proof, it only remains to show that $(Y^\star,T^\star)\indep Z|X$ under $\tilde Q$.
To this end, select an $f\in L^1(P_{ZX})$ and let $g$ be any bounded function of $(Y^\star,T^\star,X)$.
Then note that since $(Y^\star,T^\star)\indep Z|X$ under $Q^{\rm a}$ and all $Q_j$ (due to $Q^{\rm a},Q_j\in \Theta_0$) we obtain
\begin{multline}\label{lm:Pert6}
E_{\tilde Q} [g(Y^\star,T^\star,X)f(Z,X)]  = E_{Q^{\rm a}}[g(Y^\star,T^\star,X)E_{Q^{\rm a}}[f(Z,X)|X]]\\+ \eta \sum_{j=1}^J E_{Q_j}[g(Y^\star,T^\star,X)s_j(Y^\star,T^\star,X)E_{Q_j}[f(Z,X)|X]].
\end{multline}
However, since we showed $\tilde Q$ is observationally equivalent to $\tr$ and $Q^{\rm a}$ and $Q_j$ are observationally equivalent to $\tr$ (due to $Q^{\rm a},Q_j\in \Theta_0$) it also follows that $E_{Q^{\rm a}}[f(Z,X)|X] = E_{Q_j}[f(Z,X)|X] = E_{\tilde Q}[f(Z,X)|X]$. 
Combining this observation with \eqref{lm:Pert6} then yields
\begin{equation}\label{lm:Pert7}
E_{\tilde Q} [g(Y^\star,T^\star,X)f(Z,X)] = E_{\tilde Q}[g(Y^\star,T^\star,X)E_{\tilde Q}[f(Z,X)|X]].   
\end{equation}
Since \eqref{lm:Pert7} holds for any bounded  $g$ it follows $E_{\tilde Q}[f(Z,X)|Y^\star,T^\star,X] = E_{\tilde Q}[f(Z,X)|X]$; see, e.g., Definition 10.1.1 in \cite{bogachev2:2007}.
Thus, since $f\in L^1(P_{ZX})$ was also arbitrary, we conclude $(Y^\star,T^\star)\indep Z|X$ under $\tilde Q$ and therefore that $\tilde Q \in \Theta_0$. \qed

\begin{lemma}\label{lm:seqclos}
Let Assumptions \ref{ass:setup}, \ref{ass:model} hold, $\Lambda:\mathcal L \to \mathbf R$ satisfy $\Lambda(\langle \cdot,s\rangle_Q) = \langle \ell,s\rangle_Q$ for some $\ell \in \bigcap_{Q\in \Theta_0} L^1(Q)$, $\mathcal S_{\dom} = L^\infty(\dom_{Y^\star T^\star X})$, and $\mathcal C\subseteq \mathcal R$ be convex. 
If $\Lambda$ belongs to the $\tau$-closure of $\mathcal C$, then there is a sequence $\{f_n\}_{n=1}^\infty \subseteq L^1(\dom_{Y^\star T^\star X})$ such that $\lim_{n\to \infty} \|\ell - f_n\|_{\dom,1} = 0$ and $L_n:\mathcal L \to \mathbf R$ given by $L_n(\langle \cdot,s\rangle_Q)\equiv \langle f_n,s\rangle_Q$ satisfies $L_n \in \mathcal C$ for all $n$.
\end{lemma}

\noindent \emph{Proof.} 
First note that for any $L\in \mathcal C$, it follows from $\mathcal C \subseteq \mathcal R$ that there is a $\psi(L)\in \bigcap_{Q\in \Theta_0}L^1(Q)$ such that $L(\langle \cdot, s\rangle_Q) = \langle \psi(L),s\rangle_Q$ for all $s\in \mathcal S_Q$ and $Q\in \Theta_0$.
Next define $C \equiv \{f \in \bigcap_{Q\in \Theta_0}L^1(Q) : f = \psi(L) \text{ for some }L\in \mathcal C\}$ and note that by Theorem 2.14 in \cite{aliprantis:border:2006} and $\Lambda$ belonging to the $\tau$-closure of $\mathcal C$, there must exist a net $\{c_\alpha\}_{\alpha\in \mathcal A}\subseteq C$ such that for all $s\in \mathcal S_{\dom} = L^\infty(\dom_{Y^\star T^\star X})$ we have 
\begin{equation}\label{lm:seqclos1}
\lim_{\alpha} \langle s, c_\alpha\rangle_{\dom} = \langle s, \ell\rangle_{\dom}.
\end{equation}
A second application of Theorem 2.14 in \cite{aliprantis:border:2006} and result \eqref{lm:seqclos1} imply $\ell\in L^1(\dom_{Y^\star T^\star X})$ belongs to the closure of $C \subseteq L^1(\dom_{Y^\star T^\star X})$ under the weak topology generated by $\mathcal S_{\dom} =L^\infty(\dom_{Y^\star T^\star X})$. 
However, since $L^\infty(\dom_{Y^\star T^\star X})$ is the norm dual of $L^1(\dom_{Y^\star T^\star X})$ and $C\subseteq L^1(\dom_{Y^\star T^\star X})$ is convex (by convexity of $\mathcal C$), it follows that $\ell$ belongs to the closure of $C$ under $\|\cdot\|_{\dom,1}$ (see, e.g., Theorem 3.12 in \cite{rudin1991functional}).
Therefore, by Theorem 2.40(1) in \cite{aliprantis:border:2006} we can conclude that there is a sequence $\{f_n\}\subseteq C$ satisfying $\|\ell - f_n\|_{\dom,1} = o(1)$ as claimed. \qed

\begin{lemma}\label{lm:closedR}
Let Assumptions \ref{ass:setup} and \ref{ass:model} hold, $\mathbf T^\star$ be finite, $\{C_i\}_{i=1}^r$ be a finite collection of subsets of $\mathbf T^\star$, and define $A : \bigotimes_{i=1}^r L^1(P_X)\to L^1(\dom_{T^\star X})$ according to
$$A(f)(T^\star, X) = \sum_{i = 1}^r 1\{T^\star \in C_i\}f_i(X)$$
for any $f = \{f_i\}_{i=1}^r \in \bigotimes_{i=1}^r L^1(P_X)$. 
If $\dom(T^\star = t^\star|X) \geq \varepsilon > 0$ a.s.\ for any $t^\star\in \mathbf T^\star$ satisfying $\dom(T^\star = t^\star) > 0$, then it follows that the range of $A$ is closed.
\end{lemma}

\noindent \emph{Proof.} First let $\mathbf T_0^\star \equiv \{t^\star\in \mathbf T^\star : \dom(T^\star = t^\star) > 0\}$ denote the support of $T^\star$ under $\dom$ and enumerate $\mathbf T_0^\star$ by $\mathbf T_0^\star = \{t^\star_1,\ldots, t^\star_{d^\star}\}$. 
Further interpret any $\{f_i\}_{i=1}^r \in \bigotimes_{i=1}^r L^1(P_X)$ as a column vector $f(X) \equiv (f_1(X),\ldots, f_r(X))^\prime$ and define a $d^\star \times r$ matrix $\Omega$ according to
\begin{equation*}
\Omega \equiv \left(\begin{array}{c} \omega_1^\prime \\ \vdots \\ \omega_{d^\star}^\prime\end{array}\right) \hspace{0.4 in} \omega_j \equiv \left(\begin{array}{c} 1\{t^\star_j\in C_1\} \\ \vdots \\ 1\{t^\star_j \in C_r\}\end{array}\right).
\end{equation*}
Letting $\|a\|_1 \equiv \sum_{i=1}^d |a_i|$ for any $a\equiv (a_1,\ldots,a_d)^\prime \in \mathbf R^d$, then note that $\dom (T^\star = t^\star|X) \geq \varepsilon > 0$ for all $t^\star \in \mathbf T_0^\star$ and $\dom_X = P_X$ due to $\dom\in \Theta_0$ allow us to conclude
\begin{multline}\label{lm:closedR2}
E_{\dom}[|A(f)(T^\star,X)|] = E_{\dom}[\sum_{j=1}^{d^\star} 1\{T^\star = t^\star_j\}|\sum_{i=1}^r 1\{t^\star_j \in C_i\}f_i(X)|] \\ = E_{P_X}[\sum_{j=1}^{d^\star} \dom(T^\star = t_j^\star|X)|\omega_j^\prime f(X)|]\geq  \varepsilon \times E_{P_X}[\|\Omega f(X)\|_1].
\end{multline}
Let $\Omega^\dagger$ denote the Moore-Penrose pseudoinverse of $\Omega$ and note that since $\Omega \Omega^\dagger \Omega = \Omega$ by Proposition 6.11.1(6) in \cite{luenberger:1969} and $\Omega : \text{range}\{\Omega^\dagger\} \to \mathbf R^{d^\star}$ is an invertible map (see Chapter 6.11 in \cite{luenberger:1969}), it follows there is an $\eta > 0$ satisfying
\begin{equation}\label{lm:closedR3}
E_{P_X}[\|\Omega f(X)\|_1] \geq \eta E_{P_X}[\|\Omega^\dagger \Omega f(X)\|_1].
\end{equation}
Next suppose there is a sequence $\{f_n\} \in \bigotimes_{i=1}^r L^1(P_X)$ and an $\ell \in L^1(\dom_{T^* X})$ such that $\|A(f_n) - \ell\|_{\dom,1} = o(1)$.
Combining \eqref{lm:closedR2} and \eqref{lm:closedR3} implies the sequence $\{\Omega^\dagger \Omega f_n\}$ is Cauchy in $\bigotimes_{i=1}^r L^1(P_X)$ under the norm $E_{P_X}[\|f(X)\|_1]$.
Hence, since $\bigotimes_{i=1}^r L^1(P_X)$ is complete we can conclude that there is an $\tilde f \in \bigotimes_{i=1}^r L^1(P_X)$ such that 
\begin{equation}\label{lm:closedR4}
\lim_{n\to \infty} E_{P_X}[\|\Omega^\dagger \Omega f_n(X) - \tilde f(X)\|_1] = 0.
\end{equation}
Therefore, the same manipulations as in result \eqref{lm:closedR2} and $\Omega = \Omega \Omega^\dagger \Omega$ imply that 
\begin{multline*}
E_{\dom}[|\ell(T^\star,X) - A(\tilde f)(T^\star,X)|] \leq \lim_{n\to \infty} E_{\dom}[|A(f_n - \tilde f)(T^\star,X)|] \\
\leq \lim_{n\to \infty}E_{P_X}[\|\Omega f_n(X) - \Omega \tilde f(X)\|_1] \leq \|\Omega\|_{o,1} E_{P_X}[\|\Omega^\dagger \Omega f_n(X) - \tilde f(X)\|_1] = 0,
\end{multline*}
where the final inequality holds for $\|\cdot\|_{o,1}$ the operator norm of $\Omega :\mathbf R^{r}\to\mathbf R^{d^\star}$ when both the range and domain are endowed with $\|\cdot\|_1$, while the final equality follows by \eqref{lm:closedR4}.
We can thus conclude the range of $A$ is closed, which establishes the lemma. \qed \\


\noindent {\Large {\bf A.2 ~ Proofs for Sections \ref{subsec:inftype} and \ref{subsec:infout}}}\\


\noindent \emph{Proof of Theorem \ref{th:typenorm}.} We begin by noting that by definition of $\hat \lambda$ and $\hat \psi_k$ in \eqref{eq:inftype4} we may apply Lemma \ref{lm:auxtypeinf} with $W_i$ satisfying $P(W_i = 1)=1$ to obtain that
\begin{equation}\label{th:typenorm1}
\hat \lambda = \frac{1}{n}\sum_{i=1}^n (\psi(T_i,Z_i,X_i) + \lambda_{\tr}) + o_P(\frac{\sigma}{\sqrt n}),
\end{equation}
which establishes the first equality in \eqref{th:typenormdisp}.
Since $\|\psi\|_\infty \lesssim B$ by Assumption \ref{ass:regtype}(i) and $\sigma^2 = \text{Var}_P\{\psi(T,Z,X)\}$ by definition, Theorem 1.1 in \cite{zhai2018high} further yields
\begin{equation}\label{th:typenorm2}
\frac{1}{\sqrt n \sigma}\sum_{i=1}^n (\psi(T_i,Z_i,X_i) - E_P[\psi(T,Z,X)]) = \mathbb Z + O_P(\frac{B\log(n)}{\sigma\sqrt n})
\end{equation}
for $\mathbb Z$ a standard normal random variable possibly depending on $n$.
The theorem follows from \eqref{th:typenorm1}, \eqref{th:typenorm2}, Lemma \ref{lm:auxtypebias}, and $B\log(n) = o(\sigma\sqrt n)$ by Assumption \ref{ass:regtype}(i). \qed

\noindent \emph{Proof of Theorem \ref{th:typeboot}.} First note that Lemma \ref{lm:auxtypeinf}, $(\hat \lambda - \lambda_{\tr}) = O_P(\sigma/\sqrt n)$ by Theorem \ref{th:typenorm}, and $\sum_i W_i/n = o_P(1)$ due to $E[W]=0$ together allow us to conclude that
\begin{align}\label{th:typeboot1}
\frac{\sqrt n}{\sigma}(\hat \lambda^* - \hat \lambda) & = \frac{1}{\sqrt n \sigma} \sum_{i=1}^n W_i \psi(T_i,Z_i,X_i) + \frac{1}{n} \sum_{i=1}^n W_i \times \frac{\sqrt n}{\sigma}(\lambda_{\tr}-\hat \lambda) + o_P(1) \notag \\ & = \frac{1}{\sqrt n \sigma} \sum_{i=1}^n W_i \psi(T_i,Z_i,X_i) + o_P(1).
\end{align}
Next note that $\sigma^2 \equiv \text{Var}_P\{\psi(T,Z,X)\}$ by definition, $\|\psi\|_\infty \lesssim B$ and $B\log(n) = o(\sigma \sqrt n)$ by Assumption \ref{ass:regtype}(i), and Bernstein's inequality (see, e.g., Lemma 2.2.9 in \cite{vandervaart:wellner:1996}), yield that for $n$ sufficiently large
\begin{equation}\label{th:typeboot2}
P(|\frac{1}{ n\sigma}\sum_{i=1}^n (\psi(T_i,Z_i,X_i) - E_P[\psi(T,Z,X)])| > \frac{\log(n)}{\sqrt n})\\
\leq 2 \exp\{-\frac{\log^2(n)}{4}\}.
\end{equation}
In particular, note that result \eqref{th:typeboot2} together with Lemma \ref{lm:auxtypebias} imply that we have
\begin{equation}\label{th:typeboot3}
|\frac{1}{n}\sum_{i=1}^n \{\psi^2(T_i,Z_i,X_i) - (\psi(T_i,Z_i,X_i)-E_P[\psi(T,Z,X)])^2\}| = o_P(\frac{\sigma^2\log(n)}{n}).
\end{equation}
Moreover, Markov's inequality, $\|\psi\|_\infty \lesssim B$, and $\sigma^2 \equiv \text{Var}_P\{\psi(T,Z,X)\}$ further imply
\begin{multline}\label{th:typeboot4}
P(|\frac{1}{n}\sum_{i=1}^n \frac{(\psi(T_i,Z_i,X_i) - E_P[\psi(T,Z,X)])^2}{\sigma^2} - 1|>\epsilon)\\
\leq \frac{1}{n\epsilon^2 \sigma^4}\text{Var}_P\{(\psi(T,Z,X) - E_P[\psi(T,Z,X)])^2\} \lesssim \frac{B^2 \sigma^2}{n \sigma^4}
\end{multline}
for any $\epsilon >0$.
Therefore, setting $\hat \sigma^2 \equiv \sum_i \psi^2(T_i,Z_i,X_i)/n$, we can conclude from results \eqref{th:typeboot3} and \eqref{th:typeboot4} together with $B\log(n)=o(\sigma \sqrt n)$ by Assumption \ref{ass:regtype}(i) that
\begin{equation}\label{th:typeboot5}
\frac{\hat \sigma^2}{\sigma^2} = \frac{1}{n}\sum_{i=1}^n \frac{(\psi(T,Z,X) - E_P[\psi(T,Z,X)])^2}{\sigma^2} + o_P(1) = 1+o_P(1).
\end{equation}
To conclude, note that $\{W_i\}_{i=1}^n$ being i.i.d.\ standard normal random variables and $\{W_i\}_{i=1}^n$ being independent of the sample $\{Y_i,T_i,Z_i,X_i\}_{i=1}^n$ imply that the variable
\begin{equation*}
\mathbb Z^* \equiv \frac{1}{\sqrt n \hat \sigma}\sum_{i=1}^n W_i\psi(T_i,Z_i,X_i) 
\end{equation*}
satisfies $\mathbb Z^*\sim N(0,1)$ conditionally on $\{Y_i,T_i,Z_i,X_i\}_{i=1}^n$ and hence $\mathbb Z^*$ is independent of $\{Y_i,T_i,Z_i,X_i\}_{i=1}^n$. 
The theorem therefore follows from \eqref{th:typeboot1} and \eqref{th:typeboot5}. \qed

\noindent \emph{Proof of Theorem \ref{th:outnorm}.} The proof follows by identical arguments as those employed in Theorem \ref{th:typenorm} but relying on Lemmas \ref{lm:auxoutinf} and \ref{lm:auxoutbias} in place of \ref{lm:auxtypeinf} and \ref{lm:auxtypebias}. \qed

\noindent \emph{Proof of Theorem \ref{th:outboot}.} The proof follows by identical arguments as those employed in Theorem \ref{th:typenorm} but relying on Theorem \ref{th:outnorm} and Lemmas \ref{lm:auxoutinf} and \ref{lm:auxoutbias} in place of Theorem \ref{th:typenorm} and Lemmas \ref{lm:auxtypeinf} and \ref{lm:auxtypebias}. \qed

\begin{lemma}\label{lm:auxtypeinf}
Let Assumptions \ref{ass:proptype}(i)(iii) and \ref{ass:regtype}(i)(ii)(v) hold, $\{W_i\}_{i=1}^n$ be an i.i.d.\ sequence independent of $\{Y_i,T_i,Z_i,X_i\}_{i=1}^n$ satisfying $E[W^2] < \infty$, $\psi$ and $\hat \psi_k$ be as defined in \eqref{eq:inftype3} and \eqref{eq:inftype4} respectively, and $\sigma^2 \equiv \text{\rm Var}_P\{\psi(T,Z,X)\}$. 
Then it follows that
$$\frac{1}{n}\sum_{k=1}^K\sum_{i\in I_k} W_i \{\hat \psi_k(T_i,Z_i,X_i) + \hat \lambda_n\} = \frac{1}{n}\sum_{i=1}^n W_i\{\psi(T_i,Z_i,X_i) + \lambda_{\tr}\} + o_P(\frac{\sigma}{\sqrt n}).$$
\end{lemma}

\noindent \emph{Proof.} Let $\hat \Delta_{t,k}^\gamma \equiv (\hat \gamma_{t,k} - \gamma_t)$, $\hat \Delta_{t,k}^\beta \equiv (\hat \beta_{t,k} - \beta_t)$, and note that for $1\leq k \leq K$ we have
\begin{align}
    \frac{1}{n}\sum_{i\in I_k} W_i  & \{(\hat \psi_k(T_i,Z_i,X_i) + \hat \lambda) - (\psi(T_i,Z_i,X_i) + \lambda_{\tr})\} \notag \\&  =  \frac{1}{n}\sum_{i\in I_{k}} \sum_{t\in \mathbf T} W_i(1\{T_i=t\} - b(Z_i,X_i)^\prime \beta_{t})b(Z_i,X_i)^\prime \hat \Delta_{t,k}^\gamma \notag \\ 
    &+ \frac{1}{n}\sum_{i\in I_{k}} \sum_{t\in \mathbf T} W_i\{E_{\mu_{Z|X}}[\nu_j(t,Z,X_i)b(Z,X_i)^\prime] - b(Z_i,X_i)^\prime \gamma_t b(Z_i,X_i)^\prime\}\hat \Delta_{t,k}^\beta \notag \\ 
    & - \frac{1}{n}\sum_{i\in I_{k}} \sum_{t\in \mathbf T} W_i(\hat \Delta^\gamma_{t,k})^\prime b(Z_i,X_i)b(Z_i,X_i)^\prime \hat \Delta^\beta_{t,k} \label{eq:auxtypeinf3}.
\end{align}
Next observe that by definition of $\beta_t$, it must satisfy the following first order condition 
\begin{equation}\label{eq:auxtypeinf4}
E_P[(1\{T=t\}-b(Z,X)^\prime \beta_t)b(Z,X)] = 0.
\end{equation}
Hence, since $\hat \gamma_{t,k}$ is computed using the observations in $I_k^c$, which are independent of the observations in $I_k$, and $\{W_i\}_{i=1}^n$ is independent of $\{Y_i,T_i,Z_i,X_i\}_{i=1}^n$, \eqref{eq:auxtypeinf4} yields
\begin{equation*}
E[\frac{1}{n}\sum_{i\in I_k}\sum_{t\in \mathbf T} W_i(1\{T_i = t\}-b(Z_i,X_i)^\prime \beta_t)b(Z_i,X_i)^\prime \hat \Delta_{t,k}^\gamma|\{\hat \gamma_{t,k}\}_{t\in \mathbf T}] =0.
\end{equation*}
Therefore, by employing that the observations within $I_k$ are i.i.d., $|I_k|\asymp n$ by Assumption \ref{ass:regtype}(v), $\|b^\prime \beta_t\|_\infty$ being bounded by Assumption \ref{ass:regtype}(i), and $E[W^2] < \infty$ we obtain
\begin{multline}\label{eq:auxtypeinf6}
\text{Var}\{\frac{1}{n}\sum_{i\in I_k}\sum_{t\in \mathbf T} W_i (1\{T_i = t\}-b(Z_i,X_i)^\prime \beta_t)b(Z_i,X_i)^\prime \hat \Delta_{t,k}^\gamma|\{\hat \gamma_{t,k}\}_{t\in \mathbf T}\}\\
\lesssim \frac{1}{n} \sum_{t\in \mathbf T} E_P[(b(Z,X)^\prime \hat\Delta_{t,k}^\gamma)^2].
\end{multline}
Moreover, by identical arguments but relying on the first order condition for $\gamma_t$ yields
\begin{align}\label{eq:auxtypeinf7}
\text{Var} & \{\frac{1}{n} \sum_{i\in I_k}\sum_{t\in \mathbf T} W_i\{E_{\mu_{Z|X}}[\nu_j(t,Z,X_i)b(Z,X_i)^\prime] - b(Z_i,X_i)^\prime \gamma_t b(Z_i,X_i)^\prime\}\hat \Delta_{t,k}^\beta|\{\hat \beta_{t,k}\}_{t\in \mathbf T}\}\notag \\
& \lesssim \frac{1}{n}\sum_{t\in \mathbf T} \{E_{P}[(E_{\mu_{Z|X}}[\nu_j(t,Z,X)b(Z,X)^\prime \hat \Delta^\beta_{t,k}])^2] + E_P[(b(Z,X)^\prime \gamma_t)^2(b(Z,X)^\prime \hat \Delta^\beta_{t,k})^2] \notag \\ & \lesssim  \frac{B^2}{n}\sum_{t\in \mathbf T} E_P[(b(Z,X)^\prime \hat \Delta_{t,k}^\beta)^2],
\end{align}
where in the final inequality we employed Jensen's inequality, Assumptions \ref{ass:proptype}(iii) and \ref{ass:regtype}(i), and $d\mu_{Z|X}/dP_{Z|X} = 1/\pi$.
Finally, note that the Cauchy-Schwarz inequality yields
\begin{multline}\label{eq:auxtypeinf8}
|\frac{1}{n}\sum_{i\in I_{k}} \sum_{t\in \mathbf T} W_i(\hat \Delta^\gamma_{t,k})^\prime b(Z_i,X_i)b(Z_i,X_i)^\prime \hat \Delta^\beta_{t,k}|\\
\leq \{\frac{1}{n}\sum_{i\in I_k} W_i^2( b(Z_i,X_i)^\prime \hat \Delta_{t,k}^\gamma)^2\}^{1/2} \times \{\frac{1}{n}\sum_{i\in I_k} (b(Z_i,X_i)^\prime \hat \Delta_{t,k}^\beta )^2\}^{1/2}.
\end{multline} 
Therefore, combining results \eqref{eq:auxtypeinf3}, \eqref{eq:auxtypeinf6}, \eqref{eq:auxtypeinf7}, and \eqref{eq:auxtypeinf8}, $E[W_i^2] < \infty$, $\{W_i\}_{i=1}^n$ being independent of $\{Y_i,T_i,Z_i,X_i\}_{i=1}^n$, and Markov's inequality we obtain that
\begin{multline*}
\frac{1}{n}\sum_{i\in I_k} W_i  \{(\hat \psi_k(T_i,Z_i,X_i) + \hat \lambda_n) - (\psi(T_i,Z_i,X_i) + \lambda_{\tr})\} \\
= O_P(\sum_{t\in \mathbf T} \frac{r_{t}^\gamma + r_{t}^\beta B}{\sqrt n} + r_{t}^\beta r_{t}^\gamma) = o_P(\frac{\sigma}{\sqrt n}) ,
\end{multline*}
where the final equality follows from Assumption \ref{ass:regtype}(ii). \qed

\begin{lemma}\label{lm:auxtypebias}
If Assumptions \ref{ass:proptype}(ii), \ref{ass:regtype}(iii)(iv) hold, and $\psi$ is as in  \eqref{eq:inftype3}, then
$$|E_P[\psi(T,Z,X)]| =   o(\frac{\sigma}{\sqrt n}).$$
\end{lemma}

\noindent \emph{Proof.} First note that the first order condition implied by the definition of $\gamma_t$ yields
\begin{equation}\label{lm:auxtypebias1}
E_P[(b(Z,X)^\prime \gamma_t )(b(Z,X)^\prime \beta_t)] - E_{P_X}[E_{\mu_{Z|X}}[\nu_j(T,Z,X) b(Z,X)^\prime \beta_t]] = 0.
\end{equation}
Hence, by combining the definition of $\psi$, the first order condition in \eqref{lm:auxtypebias1}, the law of iterated expectations, and $\kappa_j = \nu_j/\pi$ we are able to conclude that
\begin{align}\label{lm:auxtypebias2}
|E_P & [\psi(T,Z,X)] - (E_P[\kappa_j(T,Z,X)]-\lambda_{\tr})| \notag \\
& = |\sum_{t\in \mathbf T} E_P[(b(Z,X)^\prime\gamma_t - \kappa_j(t,Z,X))1\{T=t\}]| \notag \\
& = |\sum_{t\in \mathbf T} (E_P[(b(Z,X)^\prime \gamma_t - \kappa_j(t,Z,X))(P(T=t|Z,X)-b(Z,X)^\prime \beta_t)]|.
\end{align}
Hence, result \eqref{lm:auxtypebias2}, the Cauchy-Schwarz inequality, and Assumptions \ref{ass:regtype}(iii)(iv) yield
\begin{equation*}
|E_P[\psi(T,Z,X)]| \leq |E_P[\kappa_j(T,Z,X) - \lambda_{\tr}| + O(\sum_{t\in \mathbf T} \delta_{t}^\beta \delta_{t}^\gamma) =  o(\frac{\sigma}{\sqrt n}),
\end{equation*}
which establishes the claim of the lemma. \qed

\begin{lemma}\label{lm:auxoutinf}
Let Assumptions \ref{ass:propout}(i)(iii) and \ref{ass:regout}(i)(ii)(v) hold, $\{W_i\}_{i=1}^n$ be an i.i.d.\ sequence independent of $\{Y_i,T_i,Z_i,X_i\}_{i=1}^n$ satisfying $E[W^2] < \infty$, $\psi$ and $\hat \psi_k$ be as defined in \eqref{eq:infout3} and \eqref{eq:infout4} respectively, and $\sigma^2 \equiv \text{\rm Var}_P\{\psi(Y,T,Z,X)\}$. 
Then it follows that
$$\frac{1}{n}\sum_{k=1}^K\sum_{i\in I_k} W_i \{\hat \psi_k(Y_i,T_i,Z_i,X_i) + \hat \lambda\} = \frac{1}{n}\sum_{i=1}^n W_i\{\psi(Y_i,T_i,Z_i,X_i) + \lambda_{\tr}\} + o_P(\frac{\sigma}{\sqrt n}).$$
\end{lemma}

\noindent \emph{Proof.} The proof follows from identical arguments to those in Lemma \ref{lm:auxtypeinf}. \qed

\begin{lemma}\label{lm:auxoutbias}
If Assumptions \ref{ass:propout}(ii), \ref{ass:regout}(iii)(iv) hold, and $\psi$ is as in \eqref{eq:infout3}, then
$$|E_P[\psi(Y,T,Z,X)]| =   o(\frac{\sigma}{\sqrt n}).$$
\end{lemma}

\noindent \emph{Proof.} The proof follows from identical arguments to those in Lemma \ref{lm:auxtypebias}. \qed

\begin{lemma}\label{lm:delta}
Let $V\equiv (Y,T,Z,X)$, $\{V_i\}_{i=1}^n$ be i.i.d., $\{W_i\}_{i=1}^n$ be i.i.d.\ with $W\sim N(0,1)$ independent of $\{V_i\}_{i=1}^n$, and $\phi:\mathbf R^q \to \mathbf R^p$ be differentiable at $(\lambda_{\tr 1},\ldots, \lambda_{\tr q})^\prime \equiv \lambda_{\tr} \in \mathbf R^q$ with derivative $\phi^\prime_{\lambda_{\tr}}$.
Suppose $\hat \lambda \equiv (\hat \lambda_1,\ldots, \hat \lambda_q)^\prime $ and $\hat \lambda^* \equiv (\hat \lambda_1^*,\ldots, \hat \lambda_q^*)^\prime $ satisfy
$$\frac{\sqrt n}{\sigma_{j}}(\hat \lambda_{j} - \lambda_{\tr j}) =  \frac{1}{\sqrt n \sigma_{j}} \sum_{i=1}^n \psi_j(V_i) + o_P(1) \hspace{0.2 in} \frac{\sqrt n}{\sigma_{j}}(\hat \lambda_{j}^* - \hat \lambda_{j})  = \frac{1}{\sqrt n \sigma_{j}} \sum_{i=1}^n W_i\psi_j(V_i) + o_P(1) $$
with $\sigma_j^2 \equiv \text{\rm Var}_P\{\psi_j(V)\}$ and let $\bar \sigma \equiv \max_{1\leq j \leq q} \sigma_{j}$.  
If $B \equiv \max_{1\leq j \leq q} \|\psi_j\|_\infty < \infty$, $E_P[\psi_j(V)] = o(\bar \sigma/\sqrt n)$, $B\log(n) = o(\bar \sigma \sqrt n)$, and $\bar \sigma = o(\sqrt n)$, then it follows
\begin{align}
\frac{\sqrt n}{\bar \sigma}(\phi(\hat \lambda) - \phi(\lambda_{\tr})) & = \phi^\prime_{\lambda_{\tr}}(\mathbb G) + o_P(1) \label{lm:deltadisp1} \\
\frac{\sqrt n}{\bar \sigma}(\phi(\hat \lambda^*) - \phi(\hat \lambda)) & = \phi^\prime_{\lambda_{\tr}}(\mathbb G^*) + o_P(1) ,\label{lm:deltadisp2}
\end{align}
where $\mathbb G$ and $\mathbb G^*$ have the same distribution and $\mathbb G^*$ is independent of $\{V_i\}_{i=1}^n$.
\end{lemma}

\noindent \emph{Proof.} First set $\psi(V) \equiv (\psi_1(V),\ldots, \psi_q(V))^\prime$ and note that $\sigma_{j}/\bar \sigma \leq 1$ by definition of $\bar \sigma$ and our requirements on $(\hat \lambda_j-\lambda_{\tr j})$ together allow us to conclude that
\begin{multline}\label{lm:delta1}
 \frac{\sqrt n}{\bar \sigma} (\hat \lambda - \lambda_{\tr}) = \frac{1}{\sqrt n \bar \sigma}\sum_{i=1}^n \psi(V_i) + o_P(1) \\
 = \frac{1}{\sqrt n \bar \sigma} \sum_{i=1}^n (\psi(V_i) - E_P[\psi(V)]) + o_P(1) = \mathbb G + o_P(1),
 \end{multline}
where the second equality holds due to $E_P[\psi_j(V)] = o(\bar \sigma/\sqrt n)$ by hypothesis, and the final equality holds for some Gaussian $\mathbb G\sim N(0,\text{Var}_P\{\psi(V)\}/\bar \sigma^2)$ by Theorem 1.1 in \cite{zhai2018high} and $B\log(n)/\bar \sigma \sqrt n = o(1)$ by hypothesis.
In particular, note that since the variance of each coordinate of $\mathbb G$ is bounded by one, we must have $\|\mathbb G\| = O_P(1)$ and therefore \eqref{lm:delta1} implies $\|\hat \lambda - \lambda_{\tr}\| = O_P(\bar \sigma/\sqrt n)$.
Hence, $\phi : \mathbf R^q \to \mathbf R^p$ being differentiable by assumption together with result \eqref{lm:delta1} allow us to conclude
\begin{equation*}
\frac{\sqrt n}{\bar \sigma}(\phi(\hat \lambda) - \phi(\lambda_{\tr})) = \frac{\sqrt n}{\bar \sigma} \phi^\prime_{\lambda_{\tr}}(\hat \lambda - \lambda_{\tr}) + \frac{\sqrt n}{\bar \sigma}\times o(\|\hat \lambda - \lambda_{\tr}\|) = \phi^\prime_{\lambda_{\tr}}(\mathbb G) + o_P(1),
\end{equation*}
where in the final result we employed that $\bar \sigma/\sqrt n = o(1)$ by hypothesis and $\|\hat \lambda - \lambda_{\tr}\| = O_P(\bar \sigma/\sqrt n)$ as already shown. 
Thus, claim \eqref{lm:deltadisp1} holds.

To establish claim \eqref{lm:deltadisp2} we first note that since $E_P[\psi_j(V)] = o(\bar \sigma/\sqrt n)$ by hypothesis and $\sqrt n \bar W_n = O_P(1)$ due to $W\sim N(0,1)$, we can conclude that
\begin{multline*}
\frac{1}{\sqrt n \bar \sigma} \sum_{i=1}^n W_i \times \frac{1}{n}\sum_{i=1}^n \psi_j(V_i)
\\ = \frac{1}{\sqrt n \bar \sigma} \sum_{i=1}^n W_i \times \frac{1}{n}\sum_{i=1}^n (\psi_j(V_i)-E_P[\psi_j(V)]) + o_P(1) =o_P(1)
\end{multline*}
where the final equality holds due to $\sqrt n \bar W_n = O_P(1)$, Chebychev's inequality, and $\sigma_{j}/\bar \sigma \leq 1$.
Therefore, it follows from our condition on $(\hat \lambda^*-\hat \lambda)$ that we must have
\begin{align}\label{lm:delta4}
 \frac{\sqrt n}{\bar \sigma} (\hat \lambda^* - \hat\lambda) & = \frac{1}{\sqrt n \bar \sigma}\sum_{i=1}^n W_i (\psi(V_i) -\frac{1}{n}\sum_{k=1}^n \psi(V_k)) + o_P(1) \notag \\
& = \mathbb G^* + o_P(1),
 \end{align}
where the final equality holds for some $\mathbb G^*\sim N(0,\text{Var}_P\{\psi(V)/\bar \sigma\})$ independent of the data by Theorem S.7.1 in \cite{cns} -- to apply said theorem set, in their notation, $f_{n,P}^{d_n}(V) = (\psi(V) - E_P[\psi(V)])/\bar \sigma$ and note that then $C_n = O(1)$ due to $\sigma_{j}/\bar \sigma \leq 1$, $K_n \asymp B/\bar \sigma$, $J_{1n} = 0$, and $J_{2n} = O(1)$.
Claim \eqref{lm:deltadisp2} of the lemma then follows by identical arguments to those employed in showing claim \eqref{lm:deltadisp1} but relying on result \eqref{lm:delta4} in place of result \eqref{lm:delta1}. \qed


\vspace{0.2 in}

\noindent {\Large {\bf A.3 ~ Proofs for Section \ref{subsec:eff}}}\\


\noindent \emph{Proof of Theorem \ref{th:effmain}.} 
Let $\eta \mapsto Q_{\eta,g}$ be a submodel with $Q_{0,g} = Q\in \Theta_0$ inducing a path $\eta \mapsto P_{\eta,s}$. 
Then note that by Proposition 4 in \cite{le1988preservation} we have
\begin{equation}\label{th:effmain1}
s(Y,T,Z,X) = E_Q[g(Y^\star,T^\star,Z,X)|Y,T,Z,X].
\end{equation}
Next define $T_1(Q) \equiv \{g \in L^2(Q_{Y^\star T^\star X}) : E_Q[g(Y^\star,T^\star,X)|Z,X] = 0\}$ and note  Lemma \ref{lm:auxTQ} implies that $g = g_1 + g_2$ for some $g_1\in T_1(Q)$ and $g_2\in L^2_0(P_{ZX})$. 
Further set
\begin{equation*}
\Delta(Y,T,Z,X)  \equiv \kappa(Y,T,Z,X) - E_P[\kappa(Y,T,Z,X)|Z,X] + E_P[\kappa(Y,T,Z,X)|X] 
\end{equation*}
and note that the law of iterated expectations and the definition of $T_1(Q)$ yield that
\begin{align}\label{th:effmain3}
E_Q[\kappa(Y,T,Z,X)g_1(Y^\star,T^\star,X)] & = E_Q[\Delta(Y,T,Z,X)g_1(Y^\star,T^\star,X)]\notag \\
E_P[E_P[\kappa(Y,T,Z,X)|X]g_2(Z,X)] & = E_P[\Delta(Y,T,Z,X)g_2(Z,X)].
\end{align}
Next observe that $g = g_1 + g_2$ and Lemma F.1 in \cite{chen2018overidentification} yield that
\begin{equation}\label{th:effmain4}
\frac{\partial}{\partial \eta} \lambda_{Q_{\eta,g}}\Big|_{\eta = 0} = E_{Q}[\ell(Y^\star,T^\star,X)(g_1(Y^\star,T^\star,X) + g_2(Z,X))].
\end{equation}
Therefore, $\dom(\Upsilon(\kappa) = \ell)= 1$, $Q\ll \dom$ for any $Q\in \Theta_0$, Corollary \ref{cor:cond}, results \eqref{th:effmain3} and \eqref{th:effmain4}, the law of iterated expectations, and $(Y^\star,T^\star)\indep Z|X$ under $Q$ imply
\begin{align}\label{th:effmain5}
\frac{\partial}{\partial \eta} \lambda_{Q_{\eta,g}}\Big|_{\eta = 0}&  = E_Q[E_Q[\kappa(Y,T,Z,X)|Y^\star,T^\star,X](g_1(Y^\star,T^\star,X) + g_2(Z,X))] \notag \\
 & = E_Q[\Delta(Y,T,Z,X)(g_1(Y^\star,T^\star,X) + g_2(Z,X))] \notag \\
 & = E_P[\Delta(Y,T,Z,X)s(Y,T,Z,X)],
\end{align}
where the final equality follows from $Q$ inducing $P$ due to $Q\in \Theta_0$, the law of iterated expectations, and result \eqref{th:effmain1}.
For any closed linear subspace $V$ of a Hilbert space $\mathbf H$ and $f\in \mathbf H$ we let $\text{Proj}\{f|V\}$ denote the projection of $f$ onto $V$ (understood to be with respect to the norm $\|\cdot\|_{\mathbf H}$ of $\mathbf H$).
Then note that for $T(P)$ the tangent set and $\bar T(P)$ the tangent space (as defined in, e.g., Theorem \ref{th:eff}) result \eqref{th:effmain5} yields that
\begin{multline}\label{th:effmain6}
\sup_{Q_{\cdot,g}} I^{-1}(Q_{\cdot,g}) =  \sup_{0\neq s\in T(P)}\frac{(E_P[\Delta(Y,T,Z,X)s(Y,T,Z,X)])^2}{E_P[s^2(Y,T,Z,X)]} \\
     = \sup_{0\neq s\in \bar T(P)} \frac{(E_P[\Delta(Y,T,Z,X) s(Y,T,Z,X)])^2}{E_P[s^2(Y,T,Z,X)]} = \|\text{Proj}\{\Delta|\bar T(P)\}\|_{P,2}^2    ,
\end{multline}
where the second equality follows by continuity of the objective in $s$ under $\|\cdot\|_{P,2}$.
Moreover, employing that $\bar T(P) = [N(\Eff)]^\perp \oplus L^2_0(P_{ZX})$ by Theorem \ref{th:eff} we obtain
\begin{align}\label{th:effmain7}
\text{Proj}\{\kappa|\bar T(P)\} 
& = \text{Proj}\{\kappa|[N(\Eff)]^\perp\} + \text{Proj}\{\kappa|L^2_0(P_{ZX})\} \notag \\
& = \varphi(Y,T,Z,X) + E_P[\kappa(Y,T,Z,X)|Z,X]  - E_P[\kappa(Y,T,Z,X)],
\end{align}
where the second equality follows from the definition of $\varphi$ and $\text{Proj}\{\kappa|L^2_0(P_{ZX})\} = E_P[\kappa(Y,T,Z,X)|Z,X] - E_P[\kappa(Y,T,Z,X)]$.
Since $L^2_0(P_{ZX})\subseteq \bar T(P)$ and the law of iterated expectations implies $E_P[\kappa(Y,T,Z,X)|X] - E_P[\kappa(Y,T,Z,X)|Z,X] \in L^2_0(P_{ZX})$,  we thus obtain from result \eqref{th:effmain7} and the definition of $\Delta$ that
\begin{equation}\label{th:effmain8}
\text{Proj}\{\Delta|\bar T(P)\} = \varphi(Y,T,Z,X) + E_P[\kappa (Y,T,Z,X)|X] - E_P[\kappa(Y,T,Z,X)].
\end{equation}
Part (i) of the theorem therefore follows from results \eqref{th:effmain6}, \eqref{th:effmain8}, and $[N(\Eff)]^\perp$ and $L^2_0(P_{ZX})$ being orthogonal subspaces of $L^2_0(P)$.

In order to establish part (ii), we first construct a smaller set of submodels that contains the ``least favorable" paths.
To this end, note that Lemma \ref{lm:auxadj}(ii) implies we may be view $\Eff$ as a map from $L^2(P)$ to $T_1(\dom)$.
Letting $\bar R(\Eff,\dom)$ denote the $\|\cdot\|_{\dom,2}$-closure of $R(\Eff,\dom) \equiv \{g \in L^2(\dom) : g = \Eff(f) \text{ for some } f\in L^2(P)\}$, then set
\begin{equation*}
\mathcal P \equiv \{ Q_{\cdot,g} : Q_{\cdot,g} \text{ is a submodel,}~ Q_{\eta,g} \ll \dom, ~ Q_{0,g} = \dom, ~ g\in \bar R(\Eff,\dom) \oplus L^2_0(P_{ZX})\};
\end{equation*}
i.e.\ $\mathcal P$ consists of the submodels passing through $\dom$ whose score belongs to the subspace $\bar R(\Eff,\dom) \oplus L^2_0(P_{ZX})$. 
Further note that since $R(\Eff,\dom) \subseteq T_1(\dom)$ and $T_1(\dom)$ is closed under $\|\cdot\|_{\dom,2}$, it follows that $\bar R(\Eff,\dom) \subseteq T_1(\dom)$.
Hence, Lemma \ref{lm:auxTQ} implies that
\begin{equation}\label{th:effmain10}
T(\dom,\mathcal P) \equiv \{g : Q_{\cdot,g} \in \mathcal P\} = \bar R(\Eff,\dom)\oplus L^2_0(P_{ZX})  .
\end{equation}
Next, define $\Eff_{\dom}^\prime(g) \equiv E_{\dom}[g(Y^\star,T^\star,Z,X)|Y,T,Z,X]$ for any $g \in T(\dom,\mathcal P)$, and note that by Jensen's inequality we may view $\Eff_{\dom}^\prime$ as a map from $T(\dom,\mathcal P)$ to $L^2(P)$.
Similarly set
\begin{equation}\label{th:effmain11}
\mathcal E(f) \equiv \Eff(f) + E_P[f(Y,T,Z,X)|Z,X]  - E_P[f(Y,T,Z,X)]  
\end{equation}
for any $f\in L^2(P)$, and note that $\mathcal E(f) \in T(\dom,\mathcal P)$ by result \eqref{th:effmain10}.
Moreover, for any $f\in L^2(P)$ and $g \in T(\dom,\mathcal P)$ we obtain from result \eqref{th:effmain10} implying that $g = g_1+g_2$ for some $g_1\in \bar R(\Eff,\dom)$ and $g_2 \in L^2_0(P_{ZX})$, and the law of iterated expectations that
\begin{equation}\label{th:effmain12}
\langle f,\Eff_{\dom}^\prime(g)\rangle_P = \langle f,g_1 + g_2\rangle_{\dom}  = \langle \mathcal E (f),g\rangle_{\dom},  
\end{equation}
where the final equality follows by definition \eqref{th:effmain11} and the same arguments employed in \eqref{th:effmain3}.
In particular, result \eqref{th:effmain12} implies $\mathcal E : L^2(P)\to T(\dom, \mathcal P)$ is the adjoint of $\Eff_{\dom}^\prime : T(\dom,\mathcal P) \to L^2(P)$.
Thus, since Proposition 4 in \cite{le1988preservation} implies that the set of paths $\mathcal P$ fit the framework of Section 3 in \cite{van1991differentiable}, Theorem 4.1 in \cite{van1991differentiable} (applied with $A = \Eff_{\dom}^\prime$ and $A^* = \mathcal E$), $\mathcal P$ being a subset of all submodels, and result \eqref{th:effmain4} yields that $I^{-1}$ being finite implies
\begin{equation}\label{th:effmain13}
\dom(\mathcal E(f_0) = \text{Proj}\{\ell|T(\dom,\mathcal P)\}) = 1 \text{ for some }    f_0 \in L^2(P).
\end{equation}

To conclude the proof, we aim to show that if condition \eqref{th:effmain13} holds, then we must have $\|\ell - \Id(\kappa)\|_{\dom,2} = 0$ for some $\kappa \in L^2(P)$. 
To this end, suppose \eqref{th:effmain13} holds and note that since $\mathcal E(f) = \mathcal E(f+c)$ for any $c\in \mathbf R$, we may assume without loss of generality that $f_0 \in L^2_0(P)$.
Moreover, result \eqref{th:effmain10} and the definition of $\mathcal E$ imply
\begin{equation}\label{th:effmain14}
\text{Proj}\{\ell|L^2_0(P_{ZX})\} = \text{Proj}\{f_0|L^2_0(P_{ZX})\} \hspace{0.5 in}\Eff(f_0) = \text{Proj}\{\ell|\bar R(\Eff,\dom)\}.
\end{equation}
In particular, result \eqref{th:effmain14}, the law of iterated expectations, $\dom$ inducing $P$ due to $\dom \in \Theta_0$, and $f_0 \in L^2_0(P)$ implying $E_P[f_0(Y,T,Z,X)|Z,X] = \text{Proj}\{f_0|L^2_0(P_{ZX})\}$ yield
\begin{equation}\label{th:effmain15}
E_{\dom}[(\ell(Y^\star,T^\star,X) - E_{\dom}[f_0(Y,T,Z,X)|Y^\star,T^\star,X])E_P[g(Y,T,Z,X)|X]] \\
= 0
\end{equation}
for any $g\in L^2_0(P)$.
Thus, result \eqref{th:effmain15} and Corollary \ref{cor:cond} yield, for any $g\in L^2_0(P)$, that
\begin{equation}\label{th:effmain16}
\langle \ell - \Id(f_0),\Id(g)\rangle_{\dom} =     \langle \ell - \Id(f_0),\Eff(g)\rangle_{\dom} = \langle \ell - \Eff(f_0),\Eff(g)\rangle_{\dom} = 0
\end{equation}
where the second equality holds by the law of iterated expectations and Corollary \ref{cor:cond},
and the final equality by \eqref{th:effmain14}.
Setting $\kappa = f_0 + \lambda_{\tr}$, then note that $\Id(c) = c$ for any $c\in \mathbf R$, the law of iterated expectations, result \eqref{th:effmain16}, and $\lambda_{\tr} = E_{\dom}[\ell(Y^*,T^*,X)]$ due to $\lambda_{\tr}$ being identified imply that for any $g\in L^2_0(P)$ and $c\in \mathbf R$ we have
\begin{equation}\label{th:effmain16p5}
0 = \langle \ell - \Id(f_0),\Id(g)\rangle_{\dom} = \langle \ell - \Id(\kappa),\Id(g)\rangle_{\dom} = \langle \ell - \Id(\kappa),\Id(g+ c)\rangle_{\dom}.
\end{equation}
It follows from \eqref{th:effmain16p5} that $\Id(\kappa)$ equals the projection of $\ell$ onto the $\|\cdot\|_{\dom,2}$-closure of $\Id(L^2(P))$, and therefore that $\Id(\kappa)$ is bounded by hypothesis.
Also note that Assumption \ref{ass:genpar} holds due to $\ell$ being bounded and Assumption \ref{ass:genpar}(iii) holding by hypothesis.
Hence, $\lambda_{\tr}$ being identified, Theorem \ref{th:genmain}, and Lemma \ref{lm:seqclos} applied with $\mathcal C = \mathcal R$ yield
\begin{equation}\label{th:effmain17}
0 = \inf_{f\in L^1(P)}\|\ell - \Id(f)\|_{\dom,1} = \sup_{\|f\|_{\dom,\infty} \leq 1} \langle f,\ell\rangle_{\dom} \text{ s.t. } \langle f,\Id(g) \rangle_{\dom} = 0 \text{ for all } g\in L^1(P),
\end{equation}
where the second equality follows from Theorem 5.8.1 in \cite{luenberger:1969}.
Moreover, since we have shown $\|\Id(\kappa)\|_{\dom,\infty} < \infty$ and $\ell$ is bounded, we next suppose without loss of generality that $\|\ell - \Id(\kappa)\|_{\dom,\infty} > 0$ (because otherwise the theorem trivially follows).
By result \eqref{th:effmain16p5}, we then obtain that $(\ell - \Id(\kappa))/\|\ell- \Id(\kappa)\|_{\dom,\infty}$ satisfies the constraints in the maximization problem on the right hand side of \eqref{th:effmain17}, and hence we can conclude
\begin{equation*}
0 \geq \langle \ell - \Id(\kappa),\ell\rangle_{\dom} = \|\ell - \Id(\kappa)\|_{\dom,2}^2 + \langle \ell - \Id(\kappa),\Id(\kappa)\rangle_{\dom} = \|\ell - \Id(\kappa)\|_{\dom,2}^2,
\end{equation*}
where the final equality follows $\Id(\kappa)$ equaling the projection of $\ell$ onto the $\|\cdot\|_{\dom,2}$ closure of $\Id(L^2(P))$. 
Thus, \eqref{th:effmain13} implies $\dom(\Id(\kappa) = \ell) = 1$ for some $\kappa \in L^2(P)$ and since \eqref{th:effmain13} is a necessary condition for $I^{-1}$ to be finite, part (ii) of the theorem follows. \qed

\noindent \emph{Proof of Corollary \ref{cor:justid}.} First note that the conditions of part (i) and Lemma \ref{lm:auxnull} imply $N(\Eff) = L^2(P_{ZX})$.
Hence, since $L^2(P) = L^2(P_{ZX}) \oplus [L^2(P_{ZX})]^\perp$, it follows that the projection of $\kappa$ onto $[N(\Eff)]^\perp$ equals $\kappa(Y,T,Z,X)-E_P[\kappa(Y,T,Z,X)|Z,X]$.
The first claim of the corollary thus follows from Theorem \ref{th:effmain}(i), Lemma \ref{lm:auxtypeff}, and $E_P[\kappa(Y,T,Z,X)] = \lambda_{\tr}$ due to $\Id(\kappa) = \ell$ and Lemma \ref{lm:warmup}.

Next, let $N(\Id) \equiv \{s\in L^2(P) : \|\Id(s)\|_{\dom,2} = 0\}$ and for any $s\in N(\Id)$ let $\tilde s_t$ equal
\begin{equation}\label{cor:justid1}
\tilde s_t(Z,X) \equiv E_{\dom_{Y^\star(t)|X}}[s(Y^\star(t),t,Z,X)].
\end{equation}
Then note that $\dom \in \Theta_0$, Jensen's inequality, and the definition of $\dom^{\rm io}$ imply that
\begin{align}\label{cor:justid2}
E_P[1\{T=t\}\tilde s_t^2(Z,X)] & = E_{\dom_{T^\star ZX}}[1\{T^\star(Z)=t\}(E_{\dom_{Y^\star(t)|X}}[s(Y^\star(t),t,Z,X)])^2] \notag \\
& \leq E_{\dom^{\rm io}}[1\{T^\star(Z) = t\}s^2(Y^\star(t),t,Z,X)] \notag \\
& \lesssim E_{\dom}[1\{T^\star(Z)=t\}s^2(Y^\star(t),t,Z,X)] \notag \\
& \leq E_P[s^2(Y,T,Z,X)],
\end{align}
where the second inequality follows from $d\dom^{\rm io}/d\dom$ being bounded by Assumption \ref{ass:4out}(ii), and the final follows from $\dom \in \Theta_0$.
In particular, since $s\in L^2(P)$, result \eqref{cor:justid2} implies $\tilde s_t \delta_t \in L^2(P)$ as well.
Moreover, $s\in N(\Id)$ and $d\dom^{\rm io}/d\dom$ being bounded yield
\begin{multline}\label{cor:justid3}
0 = \|\Id(s)\|_{\dom,2}^2 \gtrsim \|\Id(s)\|_{\dom^{\rm io},2}^2 =  \|(\sum_{t\in \mathbf T} \Id(\delta_t(s - \tilde s_t)) + \Id(\sum_{t\in \mathbf T} \delta_t \tilde s_t)\|_{\dom^{\rm io}}^2 \\ = \sum_{t\in \mathbf T} \|\Id(\delta_t(s-\tilde s_t))\|_{\dom^{\rm io},2}^2 + \|\Id(\sum_{t\in \mathbf T}(\delta_t \tilde s_t)\|^2_{\dom^{\rm io},2},
\end{multline}
where the final equality follows by noting that $\langle \Id(\delta_{t_1}(s - \tilde s_{t_1}),\Id(\delta_{t_2}s)\rangle_{\dom^{\rm io}} = 0$ and $\langle \Id(\delta_{t_1}(s - \tilde s_{t_1}),\Id(\delta_{t_2}\tilde s_{t_2})\rangle_{\dom^{\rm io}} = 0$ for any $t_1\neq t_2$ by definition of $\tilde s_t$ and $\dom^{\rm io}$.
Since result \eqref{cor:justid3}, $\delta_t \tilde s_t \in L^2(P)$, and $\dom \ll \dom^{\rm io}$ imply $\|\Id(\delta_t(s-\tilde s_t)\|_{\dom,2} = 0$, it follows from the hypotheses of part (ii) of the corollary that $\|\delta_t(s-\tilde s_t)\|_{\dom,2} = 0$.
Hence, we obtain that $s = \sum_{t\in \mathbf T} \delta_t \tilde s_t$ and since $s\in N(\Id)$ was arbitrary, we can conclude that $N(\Id)\subseteq L^2(P_{TZX})$.
However, by hypothesis $N(\Id)\cap L^2(P_{TZX}) \subseteq L^2(P_{ZX})$, and therefore Lemma \ref{lm:auxnull} implies that $N(\Eff) = L^2(P_{ZX})$, which together with the same arguments employed in part (i) yields the second claim of the corollary. \qed

\begin{theorem}\label{th:eff}
Let Assumptions  \ref{ass:setup}, \ref{ass:model} hold, $\mu$ be known, and define the tangent set
$$T(P) \equiv \{s \in L^2(P) : \eta \mapsto P_{\eta,s} \text{ is induced by some submodel } \eta \mapsto Q_{\eta,g} \}.
$$
Then, the tangent space satifies $\bar T(P) = [N(\Eff)]^\perp \oplus L^2_0(P_{ZX})$, where $\bar T(P)$ denotes the $\|\cdot\|_{P,2}$-closure of $T(P)$ and $L^2_0(P_{ZX}) \equiv \{f\in L^2(P_{ZX}) : E_P[f(Z,X)] = 0\}$,
\end{theorem}

\noindent \emph{Proof.} First set $T_1(Q) \equiv \{g\in L^2(Q_{Y^\star T^\star X}) : E_Q[g(Y^\star,T^\star,X)|Z,X] = 0\}$ for any $Q\in \Theta_0$ and define a linear map $\Eff_Q^\prime : L^2(Q)\to L^2(P)$ to be given by
\begin{equation*}
\Eff^\prime_Q(g) \equiv E_Q[g(Y^\star,T^\star,Z,X)|Y,T,Z,X].
\end{equation*}
Next set $N(\Eff,Q) \equiv \{s \in L^2(P) : \|\Eff(s)\|_{Q,2} = 0\}$ noting that $N(\Eff,\dom) = N(\Eff)$ for $N(\Eff)$ as defined in the main text -- for ease of exposition we omitted the dependence on $Q$ from the main text, but we make such dependence explicit in this proof to enhance the clarity of the arguments that follow.
Setting $[ N(\Eff,Q)]^\perp \equiv \{s \in L^2(P) : \langle s, s^\prime\rangle_P = 0 \text{ for all } s^\prime \in N(\Eff,Q)\}$ and letting $\text{cl}\{A\}$ denote the $\|\cdot\|_{P,2}$ closure of any set $A\subseteq L^2(P)$, then observe that Lemma \ref{lm:auxadj}(ii) and Theorem 6.7.3 in \cite{luenberger:1969} imply
\begin{equation}\label{th:eff2}
\Eff_Q^\prime(T_1(Q)) \subseteq \text{cl}\{\Eff_Q^\prime(T_1(Q))\} = [N(\Eff,Q)]^\perp \subseteq [N(\Eff,\dom)]^\perp,
\end{equation}
where the final set inclusion follows from $Q\ll \dom$ for any $Q\in \Theta_0$ implying that $N(\Eff,\dom)\subseteq N(\Eff,Q)$.
Further note that, by direct calculation, it is possible to verify that $\dom(\Eff(s) = 0) = 1$ for any $s\in L^2(P_{ZX})$ and therefore it follows that $[N(\Eff,\dom)]^\perp$ and $L^2_0(P_{ZX})$ are orthogonal. 
Hence, if $\{s_n\}$ is a sequence in $[N(\Eff,\dom)]^\perp + L^2_0(P_{ZX})$, then writing $s_n = s_{1n} + s_{2n}$ for some $\{s_{1n}\}\subset [N(\Eff,\dom)]^\perp $ and $\{s_{2n}\}\subset L^2_0(P_{ZX})$ we obtain from the orthogonality of $[N(\Eff,\dom)]^\perp$ and $L^2_0(P_{ZX})$ that $\|s_n\|_{P,2}^2 = \|s_{1n}\|_{P,2}^2 + \|s_{2n}\|_{P,2}^2$.
Therefore, if $\{s_n\}$ is a Cauchy sequence, then so must be $\{s_{1n}\}$ and $\{s_{2n}\}$ and hence, since $[N(\Eff,\dom)]^\perp$ and $L^2_0(P_{ZX})$ are complete, we can conclude that $\{s_n\}$ has a limit in $[N(\Eff,\dom)]^\perp + L^2_0(P_{ZX})$.
In particular, it follows that $[N(\Eff,\dom)]^\perp + L^2_0(P_{ZX})$ is closed, which together with result \eqref{th:eff2} and Lemma \ref{lm:auxadj}(i) implies that
\begin{equation}\label{th:eff3}
\bar T(P) \subseteq \text{cl}\{[N(\Eff,\dom)]^\perp + L^2_0(P_{ZX})\} = [N(\Eff,\dom)]^\perp + L^2_0(P_{ZX}).
\end{equation}
Conversely, note that Lemma \ref{lm:auxadj}(ii) and Theorem 6.7.3 in \cite{luenberger:1969} yield
\begin{equation}\label{th:eff4}
[N(\Eff,\dom)]^\perp + L^2_0(P_{ZX}) = \text{cl}\{\Eff^\prime_{\dom}(T_1(\dom))\} + L^2_0(P_{ZX})   \subseteq \bar T(P),    
\end{equation}
where the final set inclusion follows from Lemma \ref{lm:auxadj}(i).
The theorem therefore follows from \eqref{th:eff3}, \eqref{th:eff4}, and the orthogonality of $[N(\Eff,\dom)]^\perp$ and $L^2_0(P_{ZX})$. \qed

\begin{lemma}\label{lm:auxadj}
Let Assumptions  \ref{ass:setup} and \ref{ass:model}(i)(ii) hold, $\mu$ be known, for any $Q\in \Theta_0$ let $T_1(Q) \equiv \{g\in L^2(Q_{Y^\star T^\star X}) : E_Q[g(Y^\star, T^\star, X)|Z,X] = 0\}$, and for any $g\in L^2(Q)$ set $$\Eff^\prime_Q(g) \equiv E_Q[g(Y^\star,T^\star,Z,X)|Y,T,Z,X].$$
Then: (i) $T(P) = \bigcup_{Q\in \Theta_0} \Eff_Q^\prime(T_1(Q)) +  L^2_0(P_{ZX})$ with $L^2_0(P_{ZX}) \equiv \{f\in L^2(P_{ZX}) : E_P[f(Z,X)] = 0\}$; and (ii) $\Eff$ (as in \eqref{def:Eff}) is the adjoint of $\Eff^\prime_Q : T_1(Q) \to L^2(P)$.
\end{lemma}

\noindent \emph{Proof.} For any $Q\in \Theta_0$ let $T(Q) \equiv \{g\in L^2(Q) : \eta \mapsto Q_{\eta,g} \text{ is a submodel  with } Q_{0,g} = Q\}$ and note Lemma \ref{lm:auxTQ}, $Q_{ZX} = P_{ZX}$, and the linearity of $\Eff^\prime_Q:L^2(Q)\to L^2(P)$ imply 
\begin{equation}\label{lm:auxadj1}
\Eff_Q^\prime(T(Q)) = \Eff_Q^\prime(T_1(Q)) + \Eff_Q^\prime(L^2_0(P_{ZX})) = 
\Eff_Q^\prime(T_1(Q)) + L^2_0(P_{ZX}).
\end{equation}
To establish part (i), then note that Proposition 4 in \cite{le1988preservation} implies that any submodel $\eta \mapsto Q_{\eta,g}$ induces a path $\eta \mapsto P_{\eta,s}$ with score $s = \Eff_{Q_{0,g}}^\prime(g)$.
Therefore part (i) of the lemma follows from \eqref{lm:auxadj1} and the definition of $T(P)$.

In order to establish part (ii), first note that for any $s \in L^2(P)$ and $Q\in \Theta_0$ we can conclude from the definition of $\Eff$ in \eqref{def:Eff} and Corollary \ref{cor:cond} that
\begin{equation}\label{lm:auxadj2}
    \Eff(s) = E_Q[s(Y,T,Z,X)|Y^\star,T^\star,X] - E_P[s(Y,T,Z,X)|X].
\end{equation}
Hence, since $(Y^\star,T^\star)\indep X|Z$ under $Q$ and $Q$ induces $P$ due to $Q\in \Theta_0$, result \eqref{lm:auxadj2} and the law of iterated expectations imply that $\Eff(s)\in T_1(Q)$ for any $s\in L^2(P)$.
Next, let $g\in T_1(Q)$ and $s\in L^2(P)$ be arbitrary, and note that the definition of $\Eff_Q^\prime$, the law of iterated expectations, and $g\in T_1(Q)$ allow us to conclude that
\begin{equation*}
    \langle \Eff_Q^\prime(g),s\rangle_P  = E_Q[g(Y^\star,T^\star,X)(s(Y,T,Z,X) - E_P[s(Y,T,Z,X)|Z,X]]   = \langle g,\Eff(s)\rangle_Q,
\end{equation*}
where the final equality follows from the law of iterated expectations, $(Y^\star,T^\star)\indep Z|X$ under $Q$, and \eqref{lm:auxadj2}.
Hence, $\Eff : L^2(P)\to T_1(Q)$ is indeed the adjoint of $\Eff_Q^\prime : T_1(Q)\to L^2(P)$, which establishes part (ii) of the lemma. \qed

\begin{lemma}\label{lm:auxTQ}
Let Assumptions  \ref{ass:setup} and \ref{ass:model}(i)(ii) hold, $\mu$ be known, $Q\in \Theta_0$, and set
\begin{align*}
 T_1(Q) & \equiv \{g \in L^2(Q_{Y^\star T^\star X}) : E_Q[g(Y^\star,T^\star,X)|Z,X] = 0\} \\    
 T(Q) & \equiv \{g \in L^2(Q) : \eta \mapsto Q_{\eta,g} \text{ is a submodel with } Q_{0,g} = Q\}.
\end{align*}
Then $T(Q) = T_1(Q) + L^2_0(P_{ZX})$, where $L^2_0(P_{ZX}) \equiv \{f\in L^2(P_{ZX}) : E_P[f(Z,X)] = 0\}$.
Moreover, the lemma also holds if when defining $T(Q)$ we require $Q_{\eta,g} \ll Q_{0,g}$ for all $\eta$.
\end{lemma}

\noindent \emph{Proof.} Fix  $g\in T(Q)$ and note that Lemma \ref{lm:auxrep} implies that $g = g_1 + g_2$, where
\begin{align*}
g_1(Y^\star,T^\star,X) & = E_Q[g(Y^\star,T^\star,Z,X)|Y^\star,T^\star,X] - E_Q[g(Y^\star,T^\star,Z,X)|X]  \\
g_2(Z,X) & = E_Q[g(Y^\star,T^\star,Z,X)|Z,X]  .
\end{align*}
Moreover, since $(Y^\star,T^\star)\indep Z|X$ under $Q$ and $Q_{ZX} = P_{ZX}$ because $Q\in \Theta_0$, it follows from the law of iterated expectations and $E_Q[g(Y^\star,T^\star,Z,X)] = 0$ that $g_1\in T_1(Q)$ and $g_2 \in L^2_0(P_{ZX})$.
It thus follows that $T(Q) \subseteq T_1(Q)+L^2_0(P_{ZX})$.
In order to establish the reverse inclusion, we rely on a construction from Example 3.2.1 in \cite{bickel:klaassen:ritov:wellner}.
Specifically, let $g_1\in T_1(Q)$ and $g_2\in L^2_0(P_{ZX})$ be arbitrary and set
\begin{equation}\label{lm:auxTQ2}
\frac{dQ_{\eta}}{d\mu} \equiv \frac{dQ}{d\mu} \frac{\Psi(\eta g_1)\Psi(\eta g_2)}{c(\eta)} \hspace{0.5 in} c(\eta) \equiv \int \Psi(\eta g_1)\Psi(\eta g_2) dQ
\end{equation}
where $\Psi : \mathbf R \to (0,\infty)$ is any continuously differentiable function with $\Psi(0) = \Psi^{\prime}(0) = 1$ and $\Psi$, $\Psi^\prime$, and $\Psi^\prime/\Psi$ bounded.
Next define $\pi(\eta,X) \equiv E_Q[\Psi(\eta g_1(Y^\star,T^\star,X))|X]$ and note that \eqref{lm:auxTQ2}, the law of iterated expectations, and $(Y^\star,T^\star)\indep Z|X$ under $Q$ yield
\begin{equation}\label{lm:auxTQ3}
E_{Q_\eta}[1\{(Z,X)\in A\}] = E_Q[1\{(Z,X)\in A\} \Psi(\eta g_2(Z,X))\frac{\pi(\eta,X)}{c(\eta)}]
\end{equation}
for any measurable  $A$.
In particular,  \eqref{lm:auxTQ3} implies $Q_{\eta,ZX} \ll Q_{ZX}$ and $dQ_{\eta,ZX}/dQ_{ZX} = \Psi(\eta g_2)\pi(\eta,\cdot)/c(\eta)$.
Moreover, for any $h\in L^\infty(Q_{\eta,ZX})$ and $f\in L^1(Q_{\eta,Y^\star T^\star X})$, definition \eqref{lm:auxTQ2}, the law of iterated expectations, $(Y^\star,T^\star)\indep Z|X$ under $Q$, and \eqref{lm:auxTQ3} yield
\begin{align}\label{lm:auxTQ4}
E_{Q_\eta}& [h(Z,X) f(Y^\star,T^\star,X)] \notag\\
& = E_{Q}[h(Z,X)\frac{\Psi(\eta g_2(Z,X))}{c(\eta)} E_Q[f(Y^\star,T^\star,X)\Psi(\eta g_1(Y^\star,T^\star,X))|X]] \notag\\
& = E_{Q_{\eta}}[h(Z,X)\frac{E_Q[f(Y^\star,T^\star,X)\Psi(\eta g_1(Y^\star,T^\star,X))|X]}{\pi(\eta,X)}].
\end{align}
Hence, since \eqref{lm:auxTQ4} holds for any bounded $h$, it follows for any $f\in L^1(Q_{\eta,Y^\star T^\star X})$ that
\begin{align*}
E_{Q_{\eta}}[f(Y^\star,T^\star,X)|Z,X] & = E_{Q_{\eta}}[f(Y^\star,T^\star,X)|X] \\
&= \frac{E_Q[f(Y^\star,T^\star,X)\Psi(\eta g_1(Y^\star,T^\star,X))|X]}{\pi(\eta,X)};
\end{align*}
see, e.g., Definition 10.1.1 in \cite{bogachev2:2007}.
Therefore, since $f\in L^1(Q_{\eta,Y^\star T^\star X})$ was arbitrary, we can conclude that $(Y^\star,T^\star)\indep Z|X$ under $Q_\eta$.
Finally, note that if $g_1 = g_2 =0$, then trivially $g_1+g_2 \in T(Q)$. 
On the other hand, if either $g_1$ or $g_2$ do not equal zero, then Proposition 2.1.1 in \cite{bickel:klaassen:ritov:wellner} implies $\eta \mapsto dQ_\eta/d\mu$ is a regular parametric model in a neighborhood of zero. Moreover, by direct calculation 
\begin{equation*}
\frac{d}{d\eta} \log(\frac{dQ_\eta}{d\mu}) \Big|_{\eta = 0} = g_1 + g_2
\end{equation*}
due to $\Psi(0) = \Psi^\prime(0) = 1$ and therefore $g_1+g_2 \in T(Q)$. Thus, we can conclude $T_1(Q)+L^2_0(P_{ZX})\subseteq T(Q)$, and the claim of the lemma follows. \qed

\begin{lemma}\label{lm:auxrep}
Let Assumptions \ref{ass:setup} and \ref{ass:model}(i)(ii) hold, $\mu$ be known, $\eta \mapsto Q_{\eta,g}$ be a submodel with $Q_{0,g} = Q \in \Theta_0$, and let $V\equiv (Y^\star,T^\star,Z,X)$. Then, it follows  that:
$$g(V) = E_{Q}[g(V)|Y^\star, T^\star, X] + E_{Q}[g(V)|Z, X] - E_{Q}[g(V)|X].$$
\end{lemma}

\noindent \emph{Proof.} For notational simplicity we first define the function $\Delta_Q\in L^2(Q)$ to be given by
\begin{equation}\label{lm:auxrep1}
\Delta_Q(V) \equiv g(V) - E_Q[g(V)|Y^\star,T^\star,X] - E_Q[g(V)|Z,X] + E_Q[g(V)|X].    
\end{equation}
Next note that since $Q\in \Theta_0$ we must have $(Y^\star,T^\star)\indep Z|X$ under $Q$ and therefore
\begin{align}\label{lm:auxrep2}
    E_Q[h(Z,X)-E_Q[h(Z,X)|X]|Y^\star,T^\star,X] & = 0 \notag \\
    E_Q[f(Y^\star,T^\star,X) - E_Q[f(Y^\star,T^\star,X)|X]|Z,X] & = 0
\end{align}
for any bounded functions $h$ and $f$.
In particular, definition \eqref{lm:auxrep1}, result \eqref{lm:auxrep2}, the law of iterated expectations, and Lemma \ref{lm:auxortho} imply that
\begin{equation}\label{lm:auxrep3}
    E_Q[\Delta_Q(V)(h(Z,X)-E_{Q}[h(Z,X)|X])(f(Y^\star,T^\star,X)-E_Q[f(Y^\star,T^\star,X)|X])] = 0.
\end{equation}
Moreover, the law of iterated expectations and $(Y^\star,T^\star)\indep Z|X$ under $Q$ also yield that
\begin{equation}\label{lm:auxrep4}
    E_Q[\Delta_Q(V)|Y^\star,T^\star,X] = E_Q[\Delta_Q(V)|Z,X] = 0.
\end{equation}
Therefore, results \eqref{lm:auxrep3} and \eqref{lm:auxrep4} imply that for any bounded $f$ and $h$ we have
\begin{equation}\label{lm:auxrep5}
    E_Q[\Delta_Q(V)h(Z,X)f(Y^\star,T^\star,X)] = 0.
\end{equation}

We next establish the lemma by showing that result \eqref{lm:auxrep5} implies that $\Delta_Q(V) = 0$.
To this end, we let $\mathcal F$ denote the $\sigma$-field generated by $(\mathcal F_{Y^\star}\times \mathcal F_{T^\star}\times \mathcal F_{Z}\times \mathcal F_{X})$ which, as in the rest of the literature, we assume equals the $\sigma$-field on which $Q$ is defined (here, $\mathcal F_U$ denotes the $\sigma$-field on which $Q_U$ is defined).
We also define the class of sets
\begin{equation*}
\mathcal A \equiv \{A\in \mathcal F : E_Q[1\{(Y^\star,T^\star,Z,X)\in A\}\Delta_Q(V)] = 0\}    
\end{equation*}
and note \eqref{lm:auxrep5} implies $\mathbf Y^\star\times \mathbf T^\star\times \mathbf X \times \mathbf Z \in \mathcal A$.
Also, if $A_1,A_2\in \mathcal A$ and $A_1\subseteq A_2$ then 
\begin{multline*}
E_Q[\Delta_Q(V)1\{(Y^\star,T^\star,Z,X) \in A_2\setminus A_1\}] \\ =     E_Q[\Delta_Q(V)(1\{(Y^\star,T^\star,Z,X) \in A_2\}-1\{(Y^\star,T^\star,Z,X) \in A_1\})] = 0,
\end{multline*}
which implies $A_2\setminus A_1\in \mathcal A$.
Similarly, if $\{A_i\}_{i=1}^\infty \subset \mathcal A$ is a sequence of pairwise disjoint sets, then
the dominated convergence theorem implies that
\begin{multline}\label{lm:auxrep8}
E_Q[\Delta_Q(V)1\{(Y^\star,T^\star,Z,X)\in \bigcup_{i=1}^\infty A_i\}] = \lim_{n\to \infty}E_Q[\Delta_Q(V)1\{(Y^\star,T^\star,Z,X)\in \bigcup_{i=1}^n A_i\}]\\
=\lim_{n\to \infty}\sum_{i=1}^n E_Q[\Delta_Q(V)1\{(Y^\star,T^\star,Z,X)\in A_i\}] = 0,
\end{multline}
where the second and third equalities follow from $\{A_i\}_{i=1}^\infty$ being disjoint and $A_i\in \mathcal A$.
Result \eqref{lm:auxrep8} implies $\bigcup_{i=1}^\infty \in \mathcal A$ and therefore that $\mathcal A$ is a $\lambda$-system.
On the other hand, if $A_{Y^\star}\in \mathcal F_{Y^\star}$, $A_{T^\star}\in \mathcal F_{T^\star}$, $A_Z\in \mathcal F_Z$, and $A_X\in \mathcal F_X$, then setting $h(Z,X) = 1\{(Z,X)\in A_Z\times A_X\}$ and $f(Y^\star,T^\star,X) = 1\{(Y^\star,T^\star)\in A_{Y^\star}\times A_{T^\star}\}$ in \eqref{lm:auxrep5} yields
\begin{multline*}
    E_Q[\Delta_Q(V)1\{(Y^\star,T^\star,Z,X)\in A_{Y^\star}\times A_{T^\star} \times A_Z\times A_X\}] \\ = E_Q[\Delta_Q(V)1\{(Y^\star,T^\star)\in  A_{Y^\star}\times A_{T^\star}\}1\{(Z,X)\in  A_Z\times A_X\}] = 0.
\end{multline*}
In particular, we obtain that $(\mathcal F_{Y^\star}\times \mathcal F_{T^\star}\times \mathcal F_{Z}\times \mathcal F_{X})\subseteq \mathcal A$.
Hence, since $(\mathcal F_{Y^\star}\times \mathcal F_{T^\star}\times \mathcal F_{Z}\times \mathcal F_{X})$ is a $\pi$-system and $\mathcal F$ is generated by $(\mathcal F_{Y^\star}\times \mathcal F_{T^\star}\times \mathcal F_{Z}\times \mathcal F_{X})$, the $\pi-\lambda$ theorem (see, e.g., Theorem 2.38 in \cite{pollard2002user}) yields that $\mathcal A = \mathcal F$.
Thus, we obtain
\begin{equation*}
E_Q[|\Delta_Q(V)|] = E_Q[\Delta_Q(V)1\{\Delta_Q(V) \geq 0\}] - E_Q[\Delta_Q(V)1\{\Delta_Q(V) < 0\}] = 0,
\end{equation*}
which establishes the claim of the lemma. \qed

\begin{lemma}\label{lm:auxortho}
Let Assumptions \ref{ass:setup} and \ref{ass:model}(i)(ii) hold, $\mu$ be known, and $\eta \mapsto Q_{\eta,g}$ be a submodel with $Q_{0,g} = Q\in \Theta_0$. Then, for any $h\in L^\infty(\mu_{ZX})$ and $f \in L^\infty(\mu_{Y^\star T^\star X})$:
$$E_{Q}[g(Y^\star,T^\star,Z,X)(h(Z,X)-E_{Q}[h(Z,X)|X])(f(Y^\star,T^\star,X)-E_Q[f(Y^\star,T^\star,X)|X])]=0.$$
\end{lemma}

\noindent \emph{Proof.}
In what follows we write $E_{\eta}$ in place of $E_{Q_{\eta,g}}$ and $E$ in place of $E_Q$.
Next note that $f$ and $h$ being bounded and Lemma F.1 in \cite{chen2018overidentification} imply
\begin{multline}\label{lm:auxortho1}
\lim_{\eta \downarrow 0} \frac{1}{\eta}\{E_\eta[h(Z,X)f(Y^\star,T^\star,X)] - E[h(Z,X)f(Y^\star,T^\star,X)]\} \\ = E[h(Z,X)f(Y^\star,T^\star,X)g(Y^\star,T^\star,Z,X)].
\end{multline}
Moreover, a second application of Lemma F.1 in \cite{chen2018overidentification} also yields 
\begin{align}\label{lm:auxortho2}
\lim_{\eta \downarrow 0} \frac{1}{\eta}\{ & E_\eta[h(Z,X)E_{\eta}[f(Y^\star,T^\star,X)|X]] - E[h(Z,X)E_\eta[f(Y^\star,T^\star,X)|X]]\} \notag \\ 
& =
\lim_{\eta \downarrow 0} E[h(Z,X)E_\eta[f(Y^\star,T^\star,X)|X]g(Y^\star,T^\star,Z,X)] \notag \\ & = E[h(Z,X)E[f(Y^\star,T^\star,X)|X]g(Y^\star,T^\star,Z,X)]
\end{align}
where the final result follows from the Cauchy-Schwarz inequality and Lemma \ref{lm:auxcondlim}.
Next note that the law of iterated expectations and similar arguments also yield
\begin{align}\label{lm:auxortho3}
\lim_{\eta \downarrow 0} \frac{1}{\eta}\{ & E_\eta[E[h(Z,X)|X]f(Y^\star,T^\star,X)] - E[E[h(Z,X)|X]E_\eta[f(Y^\star,T^\star,X)|X]]\} \notag \\ 
& =
\lim_{\eta \downarrow 0} E[E[h(Z,X)|X]E_\eta[f(Y^\star,T^\star,X)|X]g(Y^\star,T^\star,Z,X)] \notag \\ & = E[E[h(Z,X)|X]E[f(Y^\star,T^\star,X)|X]g(Y^\star,T^\star,Z,X)],
\end{align}
while a final application of Lemma F.1 in \cite{chen2018overidentification} further implies that
\begin{multline}\label{lm:auxortho4}
    \lim_{\eta \downarrow 0} \frac{1}{\eta}\{ E_\eta[E[h(Z,X)|X]f(Y^\star,T^\star,X)] - E[E[h(Z,X)|X]f(Y^\star,T^\star,X)]\} \\ =  E[E[h(Z,X)|X]f(Y^\star,T^\star,X)g(Y^\star,T^\star,Z,X)].
\end{multline}
To conclude, note that since $(Y^\star,T^\star)\indep Z|X$ under $Q_{\eta,g}$ for all $\eta \geq 0$ we must have 
\begin{equation}\label{lm:auxortho5}
E_{\eta}[h(Z,X)E_{\eta}[f(Y^\star,T^\star,X)|X]] = E_{\eta}[h(Z,X)f(Y^\star,T^\star,X)]
\end{equation}
for any $\eta \geq 0$. 
In particular, since $Q_{0,g} = Q$, result \eqref{lm:auxortho5} allows us to conclude that
\begin{multline}\label{lm:auxortho6}
\lim_{\eta \downarrow 0} \frac{1}{\eta}\{E_\eta[h(Z,X)E_\eta[f(Y^\star,T^\star,X)|X]] - E[h(Z,X)E[f(Y^\star,T^\star,X)|X]]\}\\
= \lim_{\eta \downarrow 0} \frac{1}{\eta}\{E_\eta[h(Z,X)f(Y^\star,T^\star,X)] - E[h(Z,X)f(Y^\star,T^\star,X)]\}.
\end{multline}
The claim of the lemma therefore follows from combining the equality in \eqref{lm:auxortho6} with results \eqref{lm:auxortho1}, \eqref{lm:auxortho2}, \eqref{lm:auxortho3} and \eqref{lm:auxortho4}. \qed

\begin{lemma}\label{lm:auxcondlim}
If $\mu$ is known, $\eta \mapsto Q_{\eta,g}$ is a path, and $f\in L^\infty(\mu)$, then it follows that
$$\lim_{\eta \downarrow 0} E_{Q_{0,g}}[(E_{Q_{\eta,g}}[f(Y^\star,T^\star,Z,X)|X] - E_{Q_{0,g}}[f(Y^\star,T^\star,Z,X)|X])^2] = 0.$$
\end{lemma}

\noindent \emph{Proof.} Set $V \equiv (Y^\star,T^\star,Z,X)$ for notational simplicity and define the sets $A_\eta^+ \equiv \{X : E_{Q_{\eta,g}}[f(V)|X] \geq E_{Q_{0,g}}[f(V)|X]\}$ and $A_\eta^{-} \equiv \{X : E_{Q_{\eta,g}}[f(V)|X] < E_{Q_{0,g}}[f(V)|X]\}$. 
Then note that since $f$ is bounded by hypothesis we can conclude that
\begin{multline}\label{lm:auxcondlim1}
     E_{Q_{0,g}}[(E_{Q_{\eta,g}}[f(V)|X] - E_{Q_{0,g}}[f(V)|X])^2] 
    \lesssim  E_{Q_{0,g}}[|E_{Q_{\eta,g}}[f(V)|X] - E_{Q_{0,g}}[f(V)|X]|]\\
    =   E_{Q_{0,g}}[(1\{X\in A_\eta^+\} - 1\{X\in A_\eta^-\}) (E_{Q_{\eta,g}}[f(V)|X] - f(V))],
\end{multline}
where the equality follows from the definitions of $A_\eta^+$ and $A_\eta^{-}$ and the law of iterated expectations.
However, by Lemma F.1 in \cite{chen2018overidentification} we have that
\begin{multline}\label{lm:auxcondlim2}
    \lim_{\eta \downarrow 0} E_{Q_{0,g}}[(1\{X\in A_\eta^+\} - 1\{X\in A_\eta^-\}) (E_{Q_{\eta,g}}[f(V)|X] - f(V))] \\
    = \lim_{\eta \downarrow 0}E_{Q_{\eta,g}}[(1\{X\in A_\eta^+\} - 1\{X\in A_\eta^-\}) (E_{Q_{\eta,g}}[f(V)|X] - f(V))] = 0,
\end{multline}
where the final equality follows from the law of iterated expectations.
Results \eqref{lm:auxcondlim1} and \eqref{lm:auxcondlim2} together establish the claim of the lemma. \qed

\begin{lemma}\label{lm:auxnull}
Let Assumptions \ref{ass:setup}, \ref{ass:model} hold, $N(\Id)\equiv \{s\in L^2(P) : \|\Id(s)\|_{\dom,2} = 0\}$, and $[L^2(P_{ZX})]^\perp \equiv \{s\in L^2(P) : \langle s,\tilde s\rangle_P = 0 \text{ for all } \tilde s \in L^2(P_{ZX})\}$. Then it follows that
$$N(\Eff)=(N(\Id) \cap [L^2(P_{ZX})]^\perp) \oplus L^2(P_{ZX}).$$
\end{lemma}

\noindent \emph{Proof.} Let $s_1 \in N(\Id) \cap [L^2(P_{ZX})]^\perp$ and $s_2 \in L^2(P_{ZX})$ be arbitrary and note that
\begin{multline*}
\Eff(s_1 + s_2) = \Eff(s_1) = - E_P[s_1(Y,T,Z,X)|X] 
\\ = - E_{\dom}[E_{\dom}[s_1(Y,T,Z,X)|Y^\star,T^\star,X]|X] = 0   
\end{multline*}
where in the first equality we used that $\Eff(s_2) = 0$ for any $s_2\in L^2(P_{ZX})$, the second equality follows from $s_1\in N(\Id)$, the third inequality follows from $\dom \in \Theta_0$ and the law of iterated expectations, and the final equality from Corollary \ref{cor:cond} and $s_1\in N(\Id)$.
Thus, we must have $ (N(\Id) \cap [L^2(P_{ZX})]^\perp) \oplus L^2(P_{ZX}) \subseteq N(\Eff)$.
For the reverse inclusion let $s \in N(\Eff)$ be arbitrary and set $s_1 \equiv s - s_2$ with $s_2$ given by 
\begin{equation*}
s_2(Z,X) \equiv E_P[s(Y,T,Z,X)|Z,X].    
\end{equation*}
Note that $s_2\in L^2(P_{ZX})$, $s_1 \in [L^2(P_{ZX})]^\perp$, and by the law of iterated expectations 
\begin{equation*}
\Id(s_2) = \sum_{t\in \mathbf T} E_{P_{Z|X}}[s_2(Z,X)1\{T^\star(Z)= t\}] = E_P[s(Y,T,Z,X)|X].
\end{equation*}
Thus, since $s\in N(\Eff)$ we obtain that $\Id(s_1) = \Id(s) - \Id(s_2) = \Eff(s) = 0$, which implies $s_1 \in N(\Id)\cap[L^2(P_{ZX})]^\perp$.
Hence, we conclude $N(\Eff) \subseteq (N(\Id) \cap [L^2(P_{ZX})]^\perp) \oplus L^2(P_{ZX})$ and the claim of the lemma follows. \qed

\begin{lemma}\label{lm:auxtypeff}
Suppose that the conditions of Theorem \ref{th:typenorm} (resp. Theorem \ref{th:outnorm}) hold with $\max_t \delta_t^\beta \vee \delta_t^\gamma = o(1)$ (resp. $\delta^\beta \vee \delta^\gamma = o(1))$, the conditions of Theorem \ref{th:effmain}(i) hold with a $\kappa$ satisfying Assumption \ref{ass:proptype}(ii) (resp. Assumption \ref{ass:propout}(ii)), and define
$$\tilde \psi(Y,T,Z,X) \equiv \kappa(Y,T,Z,X) - E_P[\kappa(Y,T,Z,X)|Z,X] + E_P[\kappa(Y,T,Z,X)|X] - \lambda_{\tr}. $$
Then, it follows that the estimator $\hat \lambda$ of Section \ref{subsec:inftype} (resp. Section \ref{subsec:infout}) satisfies 
$$\sqrt n\{\hat \lambda - \lambda_{\tr}\} \stackrel{d}{\rightarrow} N(0,\text{\rm Var}_P\{\tilde \psi(Y,T,Z,X)\}).$$
\end{lemma}

\noindent \emph{Proof.} We only establish the claim concerning Section \ref{subsec:inftype}, since the claim concerning Section \ref{subsec:infout} follows by identical arguments.
First note that by Assumption \ref{ass:proptype}(iii), $\kappa \in L^\infty(P_{TZX})$ and hence $\kappa = \nu/\pi$, and the law of iterated expectations yield
\begin{multline}\label{lm:auxtypeff1}
\tilde \psi(Y,T,Z,X) = \sum_{t\in \mathbf T}\kappa(t,Z,X)(1\{T=t\}-P(T=t|Z,X)) \\
+ \sum_{t\in \mathbf T} E_{\mu_{Z|X}}[\nu(t,Z,X)P(T=t|Z,X)] -\lambda_{\tr}.
\end{multline}
For $\psi$ as in \eqref{eq:inftype3}, we then obtain from \eqref{lm:auxtypeff1}, $\max_{t} \|b^\prime \beta_t\|_\infty = O(1)$ by Assumption \ref{ass:regtype}(i), $\|\kappa\|_\infty \vee \|\nu\|_\infty < \infty$ by Assumptions \ref{ass:proptype}(ii)(iii), and Jensen's inequality that
\begin{multline}\label{lm:auxtypeff2}
E_P[(\tilde \psi(Y,T,Z,X) - \psi(Y,T,Z,X))^2] \lesssim
\sum_{t\in \mathbf T} E_P[(b(Z,X)^\prime \beta_t - P(T=t|Z,X))^2 \\ + \sum_{t\in \mathbf T} (b(Z,X)^\prime \gamma_t - \kappa(t,Z,X))^2 ] = o(1),
\end{multline}
where the final equality follows from $\max_t \delta_t^\beta \vee \delta_t^\gamma = o(1)$ by hypothesis.
Setting $\tilde \sigma^2 \equiv \text{Var}_P\{\tilde \psi(Y,T,Z,X)\}$ and $\sigma^2 \equiv \text{Var}_P\{\psi(Y,T,Z,X)\},$ it then follows from \eqref{lm:auxtypeff2} that $\sigma^2 = \tilde \sigma^2 + o(1)$ and hence that $\sigma^2 = O(1)$ due to $\|\kappa\|_\infty < \infty$.
Thus, $E[\tilde \psi(Y,T,Z,X)] = 0$ due to $\Id(\kappa) = \ell$ and Lemma \ref{lm:warmup}, Lemma \ref{lm:auxtypebias}, $\sigma^2 = O(1)$, and \eqref{lm:auxtypeff2} imply
\begin{multline}\label{lm:auxtypeff3}
E[(\frac{1}{\sqrt n}\sum_{i=1}^n \psi(Y_i,T_i,Z_i,X_i) - \tilde \psi(Y_i,T_i,Z_i,X_i))^2] \\
= E[(\psi(Y,T,Z,X)-\tilde \psi(Y,T,Z,X))^2] +o(1) = o(1).
\end{multline}
Thus, Theorem \ref{th:typenorm}, $\sigma^2 = O(1)$, result \eqref{lm:auxtypeff3}, and Markov's inequality yield that
\begin{equation*}
\sqrt n\{\hat \lambda - \lambda_{\tr}\} = \frac{1}{\sqrt n}\sum_{i=1}^n \tilde \psi(Y_i,T_i,Z_i,X_i) + o_P(1),
\end{equation*}
which together with the central limit theorem establishes the claim of the lemma. \qed


\vspace{0.2 in}

\noindent {\Large {\bf A.4 ~ Additional Details for Section \ref{sec:mto}}}\\


The MTO experiment offered incentives to households that were socially disadvantaged, encouraging them to relocate from economically deprived areas to more affluent neighborhoods. 
The experiment was conducted over a period of four years, from June 1994 to July 1998, as documented by \cite{Orr_etal_2003}. 
Eligible households were those that belonged to the low-income group and had children under the age of 18,
residing in the most impoverished housing projects of five major US cities, namely Baltimore,
Boston, Chicago, Los Angeles, and New York. 
The majority of these households, i.e.\ 75\%, relied on welfare, while only a third had completed high school. African Americans comprised the majority of the sample, constituting 62\%, followed by Hispanics at 30\%. Female-headed households made up 92\% of the participants.

Our dataset comprises 3039 families residing in high-poverty neighborhoods at the onset of the intervention. These families were randomly assigned to either the control group, consisting of 1310 families, or the experimental group, comprising of 1729 families. 
The experimental group received a rent-subsidizing voucher that incentivized families to relocate from the high-poverty public housing they lived in to low-poverty communities, namely, neighborhoods
where less than 10\% of households were living below the poverty line according to the 1990 US
Census. 
Families in the control group did not receive any voucher.
The Department of Housing and Urban Development (HUD) set the subsidy amount and unit eligibility based on the Applicable Payment Standard (APS). 
Landlords could not discriminate against a voucher recipient, and leases were automatically renewed. Families that decided to use the experimental voucher were required to live in the low-poverty neighborhood for a year but could move afterward. 
HUD paid rent directly to the landlord and required that households pay 30\% of their monthly adjusted gross income to offset the cost of rent and utilities.
A total of 818 out of the 1,729 experimental families agreed to use the voucher to relocate to low-poverty neighborhoods. 
Experimental families that did not use the voucher and control families were also allowed to move to low-poverty neighborhoods. 

We investigate labor market outcomes surveyed at the MTO interim evaluation in 2002 \citep{Orr_etal_2003}.
For control variables $X$ we follow the literature in employing:
\begin{enumerate}
    \item Experimental site indicators.
    \item Indicator for whether a household member had a disability.
    \item Indicator for no teens (ages 13-17) in the household at the onset of the intervention.
    \item Indicator for whether the family had previously applied for a Section 8 voucher.
    \item Indicator for whether the family had moved more than three times in the five years prior to the onset of the intervention.
    \item Indicator for whether respondent reported not having friends in the neighborhood.
    \item Indicator for whether respondent was very likely to tell a neighbor if he/she saw a neighbor's child getting into trouble.
    \item Indicator for whether a family member had been assaulted during the six months preceding the baseline survey.
    \item Assessment of whether the streets near home were very unsafe at night.
    \item Baseline respondent's primary or secondary reason for wanting to move was to get away from gangs or drugs.
\end{enumerate}

All our estimates rely on the person-level weights, denoted by $\{\omega_i\}_{i=1}^n$, for the adult survey of the interim analyses, as described in the MTO Interim Impacts Evaluation manual, 2003, Appendix B. 
We drop any observations with a missing value for any of the outcomes of interest, treatment status, or baseline characteristics.

\vspace{0.2 in}

\noindent {\large {\bf A.4.1 ~ Additional Details for Section \ref{subsec:mtot}}} \\

All functionals about types fall within the framework of Section \ref{subsec:inftype}.
Moreover, since the instrument $Z\in \{0,1\}$ and treatment $T = (D,M)\in \{0,1\}\times \{0,1\}$ are discrete, the estimation algorithm may be implemented in the manner discussed in Example \ref{ex:disc}.
To map this problem into the notation of Example \ref{ex:disc} simply interpret $T^* \equiv (D^*(0),D^*(1),M^*(0),M^*(1))$ as a vector in $\mathbf R^4$ and require that $\mu(T^*\in \mathbf R^*) = 1$ where
\begin{equation*}
\mathbf R^* \equiv \left\{\left(\begin{array}{c}0\\0\\0\\0\end{array}\right), \left(\begin{array}{c}0\\0\\1\\1\end{array}\right), \left(\begin{array}{c}0\\1\\0\\0\end{array}\right), \left(\begin{array}{c}0\\1\\0\\1\end{array}\right), \left(\begin{array}{c}0\\1\\1\\1\end{array}\right), \left(\begin{array}{c}1\\1\\0\\0\end{array}\right), \left(\begin{array}{c}1\\1\\1\\1\end{array}\right)\right\}
\end{equation*}

Due to the sample size, we do not employ sample splitting -- a modification to the algorithm of Section \ref{subsec:inftype} that is justified under appropriate sparsity assumptions. 
We additionally incorporate the weights $\{\omega_i\}_{i=1}^n$ in estimation by proceeding as follows:

\noindent {\sc Step A.1.} Set $b(Z,X)\in \mathbf R^p$ to consist of the functions generated by interacting $Z$ and $(1-Z)$ with every coordinate of the baseline covariates $X$. \qed

\noindent {\sc Step A.2.} For each treatment value $t \in \mathbf T$ we estimate the following LASSO regression
$$\hat \beta_{t} \in \arg\min_{\beta \in \mathbf R^p} \sum_{i =1}^n \omega_i (1\{T_i = t\} - b(Z_i,X_i)^\prime \beta)^2 + \alpha \|\beta\|_1,$$
where the penalty $\alpha$ is chosen through leave-one-out cross validation.
We also compute
$$\hat \gamma_{t} \in \arg\min_{\gamma \in \mathbf R^p} \sum_{i=1}^n \omega_i\{\frac{1}{2}(b(Z_i,X_i)^\prime \gamma)^2 - E_{\mu_{Z|X}} [\nu(t,Z,X_i)b(Z,X_i)^\prime \gamma]\} + \alpha \|\gamma\|_1, 
 $$
where $\alpha$ is again chosen by leave-one-out cross validation and, since $p < n$, we follow Remark \ref{rm:lasso} below to compute $\hat \gamma_t$ through a LASSO regression. \qed

\noindent {\sc Step A.3.} We estimate $\lambda_{\tr} = E_{\tr}[\ell(T^*,X)]$ by employing the plug-in estimator
$$\hat \lambda \equiv  \sum_{i =1}^n \omega_i\{\sum_{t\in \mathbf T} b(Z_i,X_i)^\prime \hat \gamma_{t}(1\{T_i = t\} - b(Z_i,X_i)^\prime \hat \beta_{t}) + E_{\mu_{Z|X}}[\nu(t,Z,X_i)b(Z,X_i)^\prime \hat \beta_{t}]\}.$$
Recall $\hat \lambda$ is simply a sample analogue to the moment condition in \eqref{eq:inftype2}. \qed

By applying the algorithm with $\ell(T^*,X) = 1\{T^* = t^*\}$ for each possible type $t^*$ we obtain estimates of the type probabilities.
The standard error $\hat \sigma$ for $\hat \lambda$ then satisfies
$$\hat \sigma^2 = \sum_{i=1}^n \omega_i^2 (\sum_{t\in \mathbf T} b(Z_i,X_i)^\prime \hat \gamma_{t}(1\{T_i = t\} - b(Z_i,X_i)^\prime \hat \beta_{t}) + E_{\mu_{Z|X}}[\nu(t,Z,X_i)b(Z,X_i)^\prime \hat \beta_{t}] -\hat \lambda)^2$$
To estimate the expectation of a coordinate $X^{(j)}$ of the baseline covariates $X$ conditional on type $t^*$ (as in Table \ref{tab:type}), we simply rely on the equality
$$E_{\tr}[X^{(j)}|T^* = t^*] = \frac{E_{\tr}[X^{(j)}1\{T^*=t^*\}]}{E_{\tr}[1\{T^* =t^*\}]}$$
and construct a plug-in estimator by applying the preceding algorithm with $\ell(T^*,X) = X^{(j)}1\{T^*=t^*\}$ and $\ell(T^*,X) = 1\{T^* = t^*\}$.
Standard errors for these estimators are obtained via the Delta method.

\begin{remark}\label{rm:lasso} \rm
Whenever the dimension $p$ of $b(Z,X)$ is smaller than $n$, the estimator $\hat \gamma_t$ can be computed through a LASSO regression.
Specifically, by setting 
$$\tilde Y_i \equiv b(Z_i,X_i)^\prime (\sum_{j=1}^n \omega_j b(Z_j,X_j) b(Z_j,X_j)^\prime )^{-1}\sum_{j=1}^n \omega_j  E_{\mu_{Z|X}}[\nu(t,Z,X_j)b(Z,X_j)],$$
it is possible to show that $\hat \gamma_t$ also equals the solution to the LASSO regression
$$\min_{\gamma \in \mathbf R^p} \sum_{i=1}^n \omega_i(\tilde Y_i - b(Z_i,X_i)^\prime \gamma)^2 + \alpha \|\gamma\|_1.$$
This observation is helpful for computational purposes, because it allows us to rely on readily available LASSO routines to compute the estimator $\hat \gamma_t$. \qed
\end{remark}

\vspace{0.2 in}

\noindent {\large {\bf A.4.1 ~ Additional Details for Section \ref{subsec:mtoo}}} \\

All parameters examined in Section \ref{subsec:mtoo} depend on expectations with the structure
\begin{equation}\label{supp:mto1}
E_{\tr}[\rho(Y^*(t))\ell(T^*,X)]
\end{equation}
and on type probabilities.
Moreover, recall that a necessary and sufficient condition for identification of \eqref{supp:mto1} is that there exist a function $\kappa$ satisfying 
\begin{equation}\label{supp:mto2}
E_{\tr}[\sum_{z\in \mathbf Z} 1\{T^*(z)=t\}\kappa(z,X)P(Z=z|X)|V^*(t),X] = E_{\tr}[\ell(T^*,X)|V^*(t),X],
\end{equation}
where $V^*(t) = T^*$ if $t\in\{(0,1),(1,0)\}$, $V^*((0,0)) = T^*1\{T^*\notin\{CN,CC\}\}$, and $V^*((1,1)) = T^*1\{T^*\notin\{CA,CC\}\}$. 
For certain choices of functions $\ell(T^*,X)$, the identifying equation in \eqref{supp:mto2} has the structure assumed in Section \ref{subsec:infout}.
In particular, 
\begin{equation}\label{supp:mto3}
E_{\tr}[\rho(Y^*(t))1\{T^* \in A\}] \text{ with } \left\{\begin{array}{cl}
t = (0,0) & \text{ and } A = \{NN\} \text{ or } \{CN,CC\} \\
t = (0,1) & \text{ and } A = \{NA\} \text{ or } \{CA\} \\
t = (1,0) & \text{ and } A = \{CN\} \text{ or } \{AN\} \\
t = (1,1) & \text{ and } A = \{CC,CA\} \text{ or } \{AA\}\end{array}\right.
\end{equation}
are identified by $E_P[\rho(Y)1\{T=t\}\kappa(Z,X)]$ with $\kappa(Z,X) = \nu(Z,X)/\pi(Z,X)$ for some known function $\nu$ that may be found by proceeding as in our discussion of Example \ref{ex:disc}.
For expectations that fall within the scope of \eqref{supp:mto3} we therefore employ the algorithm in Section \ref{subsec:infout} but without sample splitting and with the inclusion of person-level weights:

\noindent {\sc Step B.1.} Set $b(Z,X)\in \mathbf R^p$ to consist of the functions generated by interacting $Z$ and $(1-Z)$ with every coordinate of the baseline covariates $X$. \qed

\noindent {\sc Step B.2.} Compute the following two estimators through LASSO regressions
\begin{align*}
\hat \beta & \in \arg\min_{\beta \in \mathbf R^p} \sum_{i =1}^n \omega_i (\rho(Y_i)1\{T_i = t\} - b(Z_i,X_i)^\prime \beta)^2 + \alpha \|\beta\|_1 \\
\hat \gamma & \in \arg\min_{\gamma \in \mathbf R^p} \sum_{i=1}^n \omega_i\{\frac{1}{2}(b(Z_i,X_i)^\prime \gamma)^2 - E_{\mu_{Z|X}}[\nu(Z,X_i)b(Z,X_i)^\prime \gamma]\} + \alpha \|\gamma\|_1,
\end{align*}
where the penalty $\alpha$ is chosen through leave-one-out cross validation and in computing $\hat \gamma$ we rely on Remark \ref{rm:lasso}. \qed

\noindent {\sc Step B.3.} We estimate $\lambda_{\tr} = E_{\tr}[\rho(Y^*(t))\ell(T^*,X)]$ employing the plug-in estimator
$$\hat \lambda \equiv  \sum_{i =1}^n \omega_i\{b(Z_i,X_i)^\prime \hat \gamma(\rho(Y_i)1\{T_i = t\} - b(Z_i,X_i)^\prime \hat \beta) + E_{\mu_{Z|X}}[\nu(Z,X_i)b(Z,X_i)^\prime \hat \beta]\}.$$

Certain parameters that are relevant for our analysis, however, fall outside the scope of Section \ref{subsec:infout} and the preceding algorithm.
These parameters have the structure
$$E_{\tr}[\rho(Y^*(t))1\{T^* = t^*\}] \text{ with } \left\{\begin{array}{cl} t = (0,0) & \text{ and } t^* = CN \\ t = (1,1) & \text{ and } t^* = CA\end{array}\right.$$
and remain identified by the expectation $E_P[\rho(Y)1\{T=t\}\kappa(Z,X)],$ but the relevant $\kappa$ no longer satisfies $\kappa(Z,X) = \nu(Z,X)/\pi(Z,X)$ for some known $\nu$.
For instance, for identifying $E_{\tr}[\rho(Y^*(0,0))1\{T^*=CN\}]$ equation \eqref{supp:mto2} implies the relevant $\kappa$ solves
$$E_{P_{Z|X}}  [1\{t^*(Z)=(0,0)\}\kappa(Z,X)] = 0 \text{ for all }  t^*\notin \{CN,CC\}$$
and
$$E_{\tr}  [1\{T^*(Z)=(0,0)\}\kappa(Z,X)|T^*\in \{CN,CC\},X] = \tr(T^*=CN|T^*\in \{CN,CC\},X).$$
Based on these observations, it is then possible to obtain an orthogonal score for estimating $E_{\tr}[\rho(Y^*(0,0))1\{T^*=CN\}]$.
Specifically, defining the nuisance parameters
\begin{align*}
    m(Z,X) & \equiv E[\rho(Y)1\{T=(0,0)\}|Z,X]  \\
    p_{00}(Z,X) & \equiv P(T=(0,0)|Z,X) \\
    p_{10}(Z,X) & \equiv P(T=(1,0)|Z,X) \\
    \kappa_c(Z,X) & \equiv (\frac{1\{Z=0\}}{P(Z=0|X)}-\frac{1\{Z=1\}}{P(Z=1|X)}) \\
    u(X) & \equiv \tr(T^*\in \{CC,CN\}|X) \\
    c(X) & \equiv \tr(T^* = CN|T^*\in \{CN,CC\},X)     ,
\end{align*}
it is possible to show that the orthogonal score for $E_{\tr}[\rho(Y^*(0,0))1\{T^*=CN\}]$ equals
\begin{align*}
E [\rho(Y^*(0,0)) & 1\{T^*=CN\}] \notag \\
& = E[(Y1\{T=(0,0)\} - m(Z,X))c(X)\kappa_c(Z,X)] \notag \\
& + E[(m(0,X)-m(1,X))c(X)]\notag \\
& - E[\frac{(m(0,X)-m(1,X))}{u(X)}(1\{T=(1,0)\}-p_{10}(Z,X))\kappa_c(Z,X)]\notag \\
& - E[\frac{(m(0,X)-m(1,X))}{u(X)}(p_{10}(0,X)- p_{10}(1,X)] \notag \\
& - E[\frac{(m(0,X)-m(1,X))c(X)}{u(X)}(1\{T=(0,0)\}-p_{00}(Z,X))\kappa_c(Z,X)] \notag \\
& - E[\frac{(m(0,X)-m(1,X))c(X)}{u(X)}(p_{00}(0,X)-p_{00}(1,X))] .
\end{align*}

Given this orthogonal score, we then obtain an estimator by proceding as follows:

\noindent {\sc Step C.1.} Set $b(Z,X)\in \mathbf R^p$ to consist of the functions generated by interacting $Z$ and $(1-Z)$ with every coordinate of the baseline covariates $X$ and compute
\begin{align*}
\hat \beta_{m} & \in \arg\min_{\beta \in \mathbf R^p} \sum_{i =1}^n \omega_i(\rho(Y_i)1\{T_i = (0,0)\} - b(Z_i,X_i)^\prime \beta)^2 + \alpha \|\beta\|_1 \\
\hat \beta_{00} & \in \arg\min_{\beta \in \mathbf R^p} \sum_{i =1}^n \omega_i(1\{T_i = (0,0)\} - b(Z_i,X_i)^\prime \beta)^2 + \alpha \|\beta\|_1 \\
\hat \beta_{10} & \in \arg\min_{\beta \in \mathbf R^p} \sum_{i =1}^n 
\omega_i(1\{T_i = (1,0)\} - b(Z_i,X_i)^\prime \beta)^2 + \alpha \|\beta\|_1 \end{align*}
through LASSO regression and the penalty $\alpha$ selected by leave-one-out cross validation.
Similarly, by relying on Remark \ref{rm:lasso} we also compute the estimator
$$
\hat \gamma_{\kappa}  \in \arg\min_{\gamma \in \mathbf R^p} \sum_{i =1}^n\omega_i \{\frac{1}{2}(b(Z_i,X_i)^\prime \gamma)^2 - (b(0,X_i)-b(1,X_i))^\prime \gamma \} + \alpha \|\gamma\|_1,
$$
through a LASSO regression and select $\alpha$ through leave-one-out cross validation. \qed

\noindent {\sc Step C.2.} Set $f(X)\in \mathbf R^q$ to equal $X$ and compute the following penalized estimators
\begin{align*}
    \hat \pi_{cc} & \in \arg\min_{\pi \in \mathbf R^q}\sum_{i=1}^n \omega_i(1\{T_i \in \{(0,0),(1,0)\}\}b(Z_i,X_i)^\prime \hat \gamma_\kappa - f(X_i)^\prime \pi)^2 + \alpha \|\pi\|_1 \\
    \hat \pi_{cn} & \in \arg\min_{\pi \in \mathbf R^q} \sum_{i=1}^n \omega_i(1\{T_i = (1,0)\}b(Z_i,X_i)^\prime \hat \gamma_\kappa + f(X_i)^\prime \pi)^2 + \alpha \|\pi\|_1,
\end{align*}
where the penalties $\alpha$ are selected by leave-one-out cross validation. \qed

\noindent {\sc Step C.3.} Using the estimators from Steps 1 and 2 define the following estimators
\begin{align*}
    \hat m(Z_i,X_i) & \equiv b(Z_i,X_i)^\prime \hat \beta_m \hspace{0.5 in}
    \hat p_{00}(Z_i,X_i)  \equiv b(Z_i,X_i)^\prime \hat \beta_{00} \\
    \hat p_{10}(Z_i,X_i) & \equiv b(Z_i,X_i)^\prime \hat \beta_{10} \hspace{0.5 in}
    \hat \kappa_c(Z_i,X_i)  \equiv b(Z_i,X_i)^\prime \hat \gamma_{\kappa}\\
    \hat u(X_i) & \equiv  f(X_i)^\prime (\hat \pi_{cc} + \hat \pi_{cn}) \hspace{0.5 in}
    \hat c(X_i)  \equiv \frac{f(X_i)^\prime \hat \pi_{cn}}{\hat u(X_i)}.
\end{align*}
Employing these estimators we put them all together into the orthogonal score by setting
\begin{align*}
\hat \psi(Y_i,T_i,Z_i,X_i) & = (\rho(Y_i)1\{T_i=(0,0)\} - \hat m(Z_i,X_i))\hat c(X_i)\hat \kappa_c(Z_i,X_i) \notag \\
& + (\hat m(0,X_i)-\hat m(1,X_i))\hat c(X_i) \notag \\
& - \frac{(\hat m(0,X_i)-\hat m(1,X_i))}{\hat u(X_i)}(1\{T_i=(1,0)\}-\hat p_{10}(Z_i,X_i))\hat\kappa_c(Z_i,X_i) \notag \\
& -\frac{(\hat m(0,X_i)-\hat m(1,X_i))}{\hat u(X_i)}(\hat p_{10}(0,X_i)- \hat p_{10}(1,X_i)) \notag \\
& - \frac{(\hat m(0,X_i)-\hat m(1,X_i))\hat c(X_i)}{\hat u(X_i)}(1\{T_i=(0,0)\}-\hat p_{00}(Z_i,X_i))\hat \kappa_c(Z_i,X_i) \notag \\
& -\frac{(\hat m(0,X_i)-\hat m(1,X_i))\hat c(X_i)}{\hat u(X_i)}(\hat p_{00}(0,X_i)-\hat p_{00}(1,X_i)).  
\end{align*}
Our estimator for $E_{\tr}[Y^*(0,0)1\{T^*=CN\}]$ then equals $\hat \lambda = \sum_i \omega_i \hat \psi(Y_i,T_i,Z_i,X_i)$. \qed

An estimator for $E_{\tr}[Y^*(1,1)1\{T^*= CA\}]$ can be obtained through similar steps:

\noindent {\sc Step D.1.} Set $b(Z,X)\in \mathbf R^p$ to consist of the functions generated by interacting $Z$ and $(1-Z)$ with every coordinate of the baseline covariates $X$ and compute 
\begin{align*}
\hat \beta_{m} & \in \arg\min_{\beta \in \mathbf R^p} \sum_{i =1}^n (\rho(Y_i)1\{T_i = (1,1)\} - b(Z_i,X_i)^\prime \beta)^2 + \alpha \|\beta\|_1 \\
\hat \beta_{11} & \in \arg\min_{\beta \in \mathbf R^p} \sum_{i =1}^n(1\{T_i = (1,1)\} - b(Z_i,X_i)^\prime \beta)^2 + \alpha \|\beta\|_1 \\
\hat \beta_{01} & \in \arg\min_{\beta \in \mathbf R^p} \sum_{i =1}^n (1\{T_i = (0,1)\} - b(Z_i,X_i)^\prime \beta)^2 + \alpha \|\beta\|_1 \\
\hat \gamma_{\kappa} & \in \arg\min_{\gamma \in \mathbf R^p} \sum_{i =1}^n\{\frac{1}{2}(b(Z_i,X_i)^\prime \gamma)^2 - (b(1,X_i)-b(0,X_i))^\prime \gamma\} + \alpha \|\gamma\|_1,
\end{align*}
with $\alpha$ selected through leave-one-out cross validation. \qed

\noindent {\sc Step D.2.} Set $f(X)\in \mathbf R^q$ to equal $X$ and compute the following penalized estimators
\begin{align*}
    \hat \pi_{cc} & \in \arg\min_{\pi \in \mathbf R^q}\sum_{i=1}^n (1\{T_i = \{(0,0),(1,0)\}\}b(Z_i,X_i)^\prime \hat \gamma_\kappa + f(X_i)^\prime \pi)^2 + \alpha \|\pi\|_1 \\
    \hat \pi_{ca} & \in \arg\min_{\pi \in \mathbf R^q} \sum_{i=1}^n (1\{T_i = (0,1)\}b(Z_i,X_i)^\prime \hat \gamma_\kappa + f(X_i)^\prime \pi)^2 + \alpha \|\pi\|_1
\end{align*}
with $\alpha$ selected through leave-one-out cross validation. \qed

\noindent {\sc Step D.3.} Using the estimators from Steps 1 and 2 define the following estimators
\begin{align*}
    \hat m(Z_i,X_i) & \equiv b(Z_i,X_i)^\prime \hat \beta_m \hspace{0.5 in}
    \hat p_{11}(Z_i,X_i)  \equiv b(Z_i,X_i)^\prime \hat \beta_{11} \\
    \hat p_{01}(Z_i,X_i) & \equiv b(Z_i,X_i)^\prime \hat \beta_{01} \hspace{0.5 in}
    \hat \kappa_c(Z_i,X_i)  \equiv b(Z_i,X_i)^\prime \hat \gamma_{\kappa}\\
    \hat u(X_i) & \equiv  f(X_i)^\prime (\hat \pi_{cc}+\hat \pi_{ca}) \hspace{0.5 in}
    \hat c(X_i)  \equiv \frac{f(X_i)^\prime \hat \pi_{ca}}{\hat u(X_i)}.
\end{align*}
Employing these estimators we put them all together into the orthogonal score by setting
\begin{align*}
\hat \psi(Y_i,T_i,Z_i,X_i) & =    (\rho(Y_i)1\{T_i=(1,1)\} - \hat m(Z_i,X_i))\hat c(X_i)\hat \kappa_c(Z_i,X_i) \notag \\
& + (\hat m(1,X_i)-\hat m(0,X_i))\hat c(X_i) \notag \\
& - \frac{(\hat m(1,X_i)-\hat m(0,X_i))}{\hat u(X_i)}(1\{T_i=(0,1)\}-\hat p_{01}(Z_i,X_i))\hat\kappa_c(Z_i,X_i) \notag \\
& -\frac{(\hat m(1,X_i)-\hat m(0,X_i))}{\hat u(X_i)}(\hat p_{01}(1,X_i)- \hat p_{01}(0,X_i)) \notag \\
& - \frac{(\hat m(1,X_i)-\hat m(0,X_i))\hat c(X_i)}{\hat u(X_i)}(1\{T_i=(1,1)\}-\hat p_{11}(Z_i,X_i))\hat \kappa_c(Z_i,X_i) \notag \\
& -\frac{(\hat m(1,X_i)-\hat m(0,X_i))\hat c(X_i)}{\hat u(X_i)}(\hat p_{11}(1,X_i)-\hat p_{11}(0,X_i)).  
\end{align*}
Our estimator for $E_{\tr}[Y^*(1,1)1\{T^*=CA\}]$ then equals $\hat \lambda = \sum_i \omega_i \hat \psi(Y_i,T_i,Z_i,X_i)$. \qed

All the parameters in Section \ref{subsec:mtoo} can be computed by employing plug-in estimators based on the preceding algorithms.
For instance, to estimate $\text{CDE}_0$ we employ that
$$\text{CDE}_0 = \frac{E_{\tr}[Y^*(1,0)1\{T^*=CN\}] - E_{\tr}[Y^*(0,0)1\{T^*=CN\}]}{\tr(T^* = CN)}$$
and estimate $E_{\tr}[Y^*(1,0)1\{T^*=CN\}]$ using Steps B.1-B.3, $E_{\tr}[Y^*(0,0)1\{T^*=CN\}]$ using Steps C.1-C.3, and $\tr(T^*=CN)$ using Steps A.1-A.3. 
Similarly, noting
$$\text{CDE}_1 = \frac{E_{\tr}[Y^*(1,1)1\{T^*=CA\}] - E_{\tr}[Y^*(0,1)1\{T^*=CA\}]}{\tr(T^* = CA)}$$
we obtain a plug-in estimator by employing Steps D.1-D.3 to estimate $E_{\tr}[Y^*(1,1)1\{T^*=CA\}]$, Steps B.1-B.3 to estimate $E_{\tr}[Y^*(0,1)1\{T^*=CA\}]$, and Steps A.1-A.3 to estimate $\tr(T^*=CA)$.
Finally, to estimate $\text{CTE}$ we observe that
\begin{multline*}
\text{CTE} = \frac{E_{\tr}[Y^*(1,1)1\{T^*\in \{CA,CC\}\}]-E_{\tr}[Y^*(1,1)1\{T^*=CA\})]}{\tr(T^*=CC)}\\ - \frac{E_{\tr}[Y^*(0,0)1\{T^*\in \{CN,CC\}\}]-E_{\tr}[Y^*(0,0)1\{T^*=CN\}]}{\tr(T^*=CC)},
\end{multline*}
and compute $E_{\tr}[Y^*(1,1)1\{T^*\in \{CA,CC\}\}]$ and $E_{\tr}[Y^*(0,0)1\{T^*\in\{CN,CC\}\}]$ employing Steps B.1-B.3, $E_{\tr}[Y^*(1,1)1\{T^*=CA\}]$ and $E_{\tr}[Y^*(0,0)1\{T^*=CN\}]$ employing Steps D.1-D.3 and C.1-C.3 respectively, and $\tr(T^*=CC)$ employing Steps A.1-A.3.
All standard errors are obatined through the Delta method.


\phantomsection
\addcontentsline{toc}{section}{References}

{\small
\singlespacing
\bibliography{references}
}

\end{document}